\documentclass[pra,
   reprint,
   superscriptaddress,
   showpacs,
   preprintnumbers,
   ]{revtex4-1}
\usepackage{amsfonts}
\usepackage{amssymb}
\usepackage{amsmath}
\usepackage{color}
\usepackage{graphicx}
\usepackage{epsfig}
\usepackage{subfigure}
\usepackage{dcolumn}
\usepackage{time}
\usepackage[pdftex,colorlinks=true,linkcolor=red,citecolor=blue]{hyperref}
\definecolor{dark}{gray}{.5}
\definecolor{light}{gray}{.75}
\definecolor{darkmagenta}{rgb}{.5,0,.5}
\newcommand{\Veff}{\mathcal{V}_{\text{eff}}}             % effective potential
\newcommand{\VeffR}{\mathcal{V}_{\text{eff}}^{\text{R}}} % effective potential regulated
\newcommand{\Veffzero}{\mathcal{V}_0}                    % zero effective potential
\newcommand{\VeffzeroR}{\mathcal{V}_{\text{R}}}          % zero effective potential regulated
\newcommand{\Grand}{\Omega}                              % grand potential
\newcommand{\lambdaR}{\lambda_{\text{}}}                % lambda regulated
\newcommand{\muR}{\mu_{\text{}}}                        % mu regulated
\newcommand{\rD}{\text{D}}

\newcommand{\bk}{\mathbf{k}}

\newcommand{\br}{\mathbf{r}}

\newcommand{\calD}{\mathcal{D}}

\newcommand{\calG}{\mathcal{G}}

\newcommand{\calL}{\mathcal{L}}
\newcommand{\calM}{\mathcal{M}}
\newcommand{\calN}{\mathcal{N}}

\newcommand{\calR}{\mathcal{R}}

\newcommand{\calV}{\mathcal{V}}

\newcommand{\Real}[1]{\ensuremath{\mathcal{R}e \{ \, #1 \, \} }}

\newcommand{\kb}{k_{\text{B}}}                 % Bose momentum
\newcommand{\QED}%
   {$\mathcal{Q\kern-.1em \lower.6ex\hbox{$\mathcal{E}$}\kern-.1667em D}$}
\newcommand{\QCD}%
   {$\mathcal{Q\kern-.1em \lower.6ex\hbox{$\mathcal{C}$}\kern-.1667em D}$}
\newcommand{\SUSYext}%
   {$\mathcal{S \kern-0.08em \lower 0.5ex \hbox{$\mathcal{U}$}
   \kern-0.05em S \kern-0.2em \lower 0.5ex
   \hbox{$\mathcal{Y}$}}\kern-0.05em{}_{\text{ext}}$}
\newcommand{\Qquad}[1]{\qquad\text{#1}\qquad}      % big spacer
\newcolumntype{d}[1]{D{.}{.}{#1}}
\newcommand{\Set}[1]{ \bigl ( \, #1 \, \bigr )}    % Set
\newcommand{\rd}{\mathrm{d}}

\newcommand{\Partial}[4]%                          % partial derivatives
   {\Bigl ( \frac{\partial #1 }{\partial #2 } \Bigr )_{\! #3, #4 }}
\newcommand{\Intk}{\int% 
   \!\!\frac{ \mathrm{d}^3 k }{ (2\pi)^3 } }       % int d^3 k / (2\pi)^3
\newcommand{\IntkR}{\int^{\Lambda}% 
   \!\!\frac{ \mathrm{d}^3 k }{ (2\pi)^3 } }       % int^{\Lambda} d^3 k / (2\pi)^3

\newcommand{\Det}[1]{\det [ \, #1 \, ]}

\newcommand{\Ln}[1]{\ln [ \, #1 \, ]}
\newcommand{\bra}[1]%
   {\ensuremath{\langle \, #1 \, |}}
\newcommand{\Bra}[1]%
   {\ensuremath{\langle \, #1 \, |}}
\newcommand{\bigbra}[1]%
   {\ensuremath{\Bigl \langle \, #1 \, \Bigr |}}
\newcommand{\ket}[1]%
   {\ensuremath{| \, #1 \, \rangle}}
\newcommand{\Ket}[1]%
   {\ensuremath{| \, #1 \, \rangle}}
\newcommand{\bigket}[1]%
   {\ensuremath{\Bigl | \, #1 \, \Bigr \rangle}}
\newcommand{\braket}[2]%
   {\ensuremath{\langle \, #1 \, | \, #2 \, \rangle}}
\newcommand{\Braket}[2]%
   {\ensuremath{\langle \, #1 \, | \, #2 \, \rangle}}
\newcommand{\matrixelement}[3]%
   {\ensuremath{\langle \, #1 \, | \, #2 \, | \, #3 \, \rangle}}
\newcommand{\MatEl}[3]%
   {\ensuremath{\langle \, #1 \, | \, #2 \, | \, #3 \, \rangle}}
\newcommand{\pbra}[1]%
   {\ensuremath{( \, #1 \, |}}
\newcommand{\pket}[1]%
   {\ensuremath{| \, #1 \, )}}    
\newcommand{\pbraket}[2]%
   {\ensuremath{( \, #1 \, | \, #2 \, )}}
\newcommand{\braV}[1]%
   {\ensuremath{\langle \, #1 \, \Vert}}
\newcommand{\ketV}[1]%
   {\ensuremath{\Vert \, #1 \, \rangle}}
\newcommand{\Comm}[2]%
   {\ensuremath{[ \, #1, #2 \, ]}}
\newcommand{\AntiComm}[2]%
   {\ensuremath{\{ \, #1, #2 \, \}}}
\newcommand{\Pbracket}[2]%
   {\ensuremath{\{ \, #1, #2 \, \} }}
\newcommand{\PBracket}[2]%
   { \ensuremath{ \{ \, #1, #2 \, \}_{\lower1.0ex\hbox{\scriptsize \text{PB}}} } }
\newcommand{\wedgeComm}[2]
   {\ensuremath{[ \, #1, #2 \, ]_{\lower1.0ex\hbox{\scriptsize $\wedge$}} }}
\newcommand{\Expect}[1]%                           % Expect
   {\ensuremath{\langle \, #1 \,  \rangle}}
\newcommand{\expect}[1]%
   {\ensuremath{\langle \, #1 \,  \rangle}}
\newcommand{\expectbig}[1]%
   {\ensuremath{\Bigl \langle \, #1 \, \Bigr \rangle}}
\newcommand{\expectc}[2]%
   {\ensuremath{\langle \, \{ \, #1 , #2 \, \} \, \rangle}}
\newcommand{\expectq}[2]%
   {\ensuremath{\langle \, [ \, #1 , #2 \, ] \, \rangle}}
\newcommand{\expectT}[1]%
   {\ensuremath{\langle \, \mathcal{T} \{ \, #1 \, \} \, \rangle}}
\newcommand{\expectaT}[1]%
   {\ensuremath{\langle \, \mathcal{T}^{\ast} \{ \, #1 \, \} \, \rangle}}
\newcommand{\expectTbig}[1]%
   {\ensuremath{\biggl \langle \, \mathcal{T}  \biggl \{ \, #1 \,%
\biggr \} \, \biggr \rangle}}
\newcommand{\expectTabig}[1]%
   {\ensuremath{\biggl \langle \, \mathcal{T}^{\ast} \biggl \{ \, #1 \,%
\biggr \} \, \biggr \rangle}}
\newcommand{\Tproduct}[1]%
   {\ensuremath{\mathcal{T} \{ \, #1 \, \} } }
\newcommand{\aTproduct}[1]%
   {\ensuremath{\mathcal{T}^{\ast} \{ \, #1 \, \} } }
\newcommand{\Nproduct}[1]%
   {\ensuremath{\mathcal{N} \{ \, #1 \, \} } }
\newcommand{\ctpTproduct}[1]%
   {\ensuremath{\mathcal{T}_{\mathcal{C}} \{ \, #1 \, \} }}
\newcommand{\tauordered}[1]%
   {\ensuremath{\mathcal{T}_{\tau} \{ \, #1 \, \} }}
\newcommand{\Tr}[1]{\mathrm{Tr} [ \, #1 \, ]}

\newcommand{\expectTproduct}[1]%
   {\ensuremath{\langle \, \mathcal{T} \{ \, #1 \, \} \, \rangle}}
\newcommand{\expectTCproduct}[1]%
   {\ensuremath{\langle \, \mathcal{T}_{\mathcal{C} \{ \, #1 \, \} \, \rangle}}}
\newcommand{\expectComm}[2]%
   {\ensuremath{\langle \, [ \, #1 , #2 \, ] \, \rangle}}
\newcommand{\expectPbracket}[2]%
   {\ensuremath{\langle \, \{ \, #1 , #2 \, \} \, \rangle}}
\newcommand{\threej}[6]%
{\begin{pmatrix} #1 & #2 & #3 \\ #4 & #5 & #6 \end{pmatrix}}
\newcommand{\sixj}[6]%
{\begin{Bmatrix} #1 & #2 & #3 \\ #4 & #5 & #6 \end{Bmatrix}}
\newcommand{\ninej}[9]%
{\begin{Bmatrix} #1 & #2 & #3 \\ #4 & #5 & #6 \\%
 #7 & #8 & #9 \end{Bmatrix}}
\newcommand{\reducedme}[3]%
{\langle \, #1 \, \Vert \, #2 \, \Vert \, #3 \, \rangle }
\begin{document}
%
%%%%%%%%%%%%%%%%%%%%%%%%%%%%%%%%%%%%%%%%%%%%%%%%%%%%%%%%%%%%%%%%%%%%%%
%
% titlepage
%
\preprint{LA-UR-11-01238}
%\preprint{LA-UR-11-01238; PR-BEC-v6; \today,\ \now}
%
\title[BEC]
   {Auxiliary field approach to dilute Bose gases with tunable interactions}

\author{Fred Cooper}
\email{cooper@santafe.edu}
\affiliation{Santa Fe Institute,
   Santa Fe, NM 87501}
\affiliation{Theoretical Division,
   Los Alamos National Laboratory,
   Los Alamos, NM 87545}

\author{Bogdan Mihaila}
\email{bmihaila@lanl.gov}
\affiliation{Materials Science and Technology Division,
   Los Alamos National Laboratory,
   Los Alamos, NM 87545}

\author{John F. Dawson}
\email{john.dawson@unh.edu}
\affiliation{Department of Physics,
   University of New Hampshire,
   Durham, NH 03824}

\author{Chih-Chun Chien}
\email{chinchun@lanl.gov}
\affiliation{Theoretical Division
   Los Alamos National Laboratory,
   Los Alamos, NM 87545}

\author{Eddy Timmermans}
\email{eddy@lanl.gov}
\affiliation{Center for Nonlinear Studies,
   Los Alamos National Laboratory,
   Los Alamos, NM 87545}

\date{\today, \now \ EST}


\begin{abstract}
We rewrite the Lagrangian for a dilute Bose gas in terms of auxiliary fields related to the normal and anomalous condensate densities.  We derive the loop expansion of the effective action in the composite-field propagators.  The lowest-order auxiliary field (LOAF) theory is a conserving mean-field approximation consistent with the Goldstone theorem without some of the difficulties plaguing approximations such as the Hartree and Popov approximations.  LOAF predicts a second-order phase transition.  We give a set of Feynman rules for improving results to any order in the loop expansion in terms of composite-field propagators.  We compare results of the LOAF approximation with those derived using the Popov approximation. LOAF allows us to explore the critical regime for all values of the coupling constant and we determine various parameters in the unitarity limit. 
\end{abstract}
\maketitle
%\tableofcontents
%\listoffigures

%
%%%%%%%%%%%%%%%%%%%%%%%%%%%%%%%%%%%%%%%%%%%%%%%%%%%%%%%%%%%%%%%%%%%%%%
%
\section{\label{s:intro}Introduction}

In 1911, Kamerlingh Onnes found that liquid $^{4}$He, when cooled below $2.2$ K began to expand rather than contract\cite{r:Kamerling:1911fk}. ÊThe transition, later named the lambda-transition was recognized in 1938 as the onset of superfluidity\cite{r:Kapitza:1938kx,r:Allen:1938vn}. ÊThe connection with Bose-Einstein condensation (BEC), first argued by F. London on the basis of the near identical values of the lambda transition temperature $T_{c}$ and the critical temperature $T_{c}^{0}$ for BEC of noninteracting bosons\cite{r:London:1938ys,r:London:1938zr} sparked a series of weakly interacting BEC studies when Bogoliubov\cite{r:Bogoliubov:1947ys} pointed out  that the BEC elementary excitations satisfy the Landau criterion for superfluidity\cite{r:Landau:1941ly}. ÊIn the weakly interacting limit, the interactions can be characterized by a single parameter\cite{r:Lee:1957ve} --- the scattering length $a$ --- giving the results a powerful, general applicability.  The hope of studying bosons with short-range inter-particle interactions of a strength that can be tuned all the way from weakly interacting ($\rho^{1/3}a \ll 1$) to universality ($\rho^{1/3}a \gg 1$), appeared thwarted when it was found that the three-body loss-rate in cold atom traps scales as $\propto a^{4}$ near a Feshbach resonance\cite{r:Fedichev:1996hc,r:Esry:1991ij}.  In cold atom traps, only fermions have been obtained in the strongly interacting, quantum degenerate regime in equilibrium\cite{r:Shin:2007oq}, in which case three-body loss is reduced by virtue of the Pauli exclusion principle.  Recently, however, it was pointed out\cite{r:Daley:2009bs} that three-body losses can be strongly suppressed in an optical lattice when the average number of bosons per lattice site is two or less.  The development of novel cold atom technology\cite{r:Henderson:2006fv,r:Henderson:2009dz} leads to  the prospect of studying finite temperature properties, such as the BEC transition temperature, $T_{c}$, superfluid to normal fluid ratio, depletion, and specific heat, at fixed particle density~$\rho$. 

At finite temperature the description of BEC's even in the weakly interacting regime
remains a challenge. ÊStandard approximations such as the Hartree-Fock-Bogoliubov (HFB) and the Popov schemes, generally fall within the Hohenberg and Martin classification\cite{r:Hohenberg:1965fu} of conserving and gapless approximations which imply that they either violate Goldstone's (or Hugenholz-Pines) theorem or general conservation laws\cite{r:Hohenberg:1965fu}.  Both these approximations predict the BEC-transition to be a first-order transition, whereas we expect the transition to be second-order\cite{r:Andersen:2004uq}.  The calculation of $T_c$, first undertaken by Toyoda\cite{r:Toyoda:1982pi} to explain the difference between the lambda-transition temperature $T_{c}$ ($2.2$ K) and the $T_{c}^{0}$ ($3.1$ K) of the noninteracting BEC at the same density, exemplifies the difficulties of understanding the theory near $T_c$: whereas Toyoda found a $T_{c}$ -decrease with increasing scattering length, K.~Huang later pointed out that the calculation had a sign error, giving an increasing value of $T_{c}$ \cite{r:Huang:1999qa}. ÊHowever, Baym and collaborators\cite{r:Baym:1999mi,r:Baym:2000fk} noted that the Toyoda expansion involves an expansion in a large parameter. ÊTheir calculation found a linear increase of $(T_{c}-T^{0}_{c})/T^{0}_{c}$ with $\rho a^{3}$. ÊThe fact that the helium lambda transition temperature falls below $T^{0}_{c}$ may be explained by quantum Monte-Carlo calculations\cite{r:Baym:1999mi}, which found that the critical temperature of a hard-sphere boson gas increases at low values of $\rho a^{3}$, then turns over and drops below $T^{0}_{c}$ near $\rho a^{3} \approx 0.1$.

In this paper, we discuss in detail a theoretical description that we introduced recently\cite{r:Cooper:2010fk} to describe a large interval of $\rho^{1/3}a$ values, satisfies Goldstone's theorem, yields a Bose-Einstein transition that is second-order, gives the same critical temperature variation found in Refs.~\onlinecite{r:Baym:1999mi,r:Baym:2000fk} but at a lower order of the calculation, while also predicting reasonable values for the depletion. ÊThis method then resolves many of the main challenges in describing boson physics over a large temperature and $\rho^{1/3}a$ regime and it's predictions will be available for experimental testing in the near future.  The approach we present here is different from other resummation schemes such as the large-$N$ expansion (which is a special case of this expansion), in that it treats the normal and anomalous densities on an equal footing.  

In the following, we will discuss the general features that arise when rewriting the original theory in terms of composite fields.  One aspect of this approach is that one can systematically calculate corrections to the mean-field results presented earlier\cite{r:Cooper:2010fk} in a loop expansion in the composite-field propagators.  We derive the Feynman rules for such an expansion using the propagators and vertices of the mean-field approximation.  At each level of this loop expansion one maintains the features that the results are both gapless and conserving.  The broken $U(1)$ symmetry Ward identities guarantee Goldstone's theorem order-by-order in the loop expansion in terms of auxiliary-field propagators\cite{r:Bender:1977bh}.  

In our auxiliary field formalism, we introduce two auxiliary fields related to the normal and anomalous densities by means of the Hubbard-Stratonovitch transformation\cite{r:Hubbard:1959kx,r:Stratonovich:1958vn}, utilizing methods discussed in the quantum field theory community~\cite{r:Bender:1977bh,r:Coleman:1974ve,r:Root:1974qf}. This transformation has already been shown to be quite useful in discussing the properties of the BCS-BEC crossover in the analogous 4-fermi theory for the BCS phase~\cite{r:Melo:1993vn,r:Engelbrecht:1997fk,r:Floerchinger:2008kxx}.  The path integral formulation of the grand canonical partition function can be found in Negele and Orland~\cite{r:Negele:1988fk}.  The Hubbard Stratonovich transformation is used to replace the original quartic interaction with an interaction quadratic in the original fields. 
%When  related to the Bogoliubov transformation\cite{r:Bogoliubov:1947ys} used to diagonalize the Hamiltonian for many-body systems.  
An excellent review of previous use of path integral methods to study dilute Bose gases is found in the review article of Andersen\cite{r:Andersen:2004uq}.  The use of path integral methods to study various topics in dilute gases began with the work of Braaten and Nieto~\cite{r:Braaten:1997uq}. Path integral methods have recently been used to study static and dynamical properties of the dilute Bose gases~\cite{r:ReyHuCalzettaRouraClark03,r:gasenzer:2005,r:gasenzer:2006,r:gasenzer:2007,PhysRevB.81.235108,r:Floerchinger:2008kx,r:Floerchinger:2008kxx}. An excellent summary of this approach and its connection to the more traditional Hamiltonian approach is to be found in the recent book by Calzetta and Hu~\cite{r:Calzetta:2008pb}.  We also point out that the 1/N expansion, which is a special case of the method being proposed here, has a long history of use in high-energy and condensed matter physics~\cite{r:brezin:1993,r:Moshe:2003uq}.  It has been used to calculate the critical temperature by Baym, Blaizot and Zinn-Justin~\cite{r:Baym:2000fk}.  This calculation gives the same result for $T_c$ as the method we are describing here. However, our approach can be used at all temperatures. Corrections to the 1/N result to calculating $T_c$ were obtained by Arnold and Tomasik~\cite{PhysRevA.62.063604}.

The paper is organized as follows:  In Sec.~\ref{s:auxfield} we discuss the auxiliary-field formalism and rewrite the Lagrangian for weakly interacting Bosons in terms of two auxiliary fields.  In Sec.~\ref{s:Seff} we derive the loop expansion by performing the path integral over the original fields $\phi_i$ and then performing the resulting path integral over the auxiliary fields by stationary phase.  In Sec.~\ref{s:effpotcon} we find the leading-order loop expansion in the auxiliary fields (LOAF) for the action. In Sec.~\ref{s:theta} we set the auxiliary-field parameter $\theta$ and discuss the leading-order effective potential for both the ground state and at finite temperature.  In Sec.~\ref{s:meanfield} we discuss related mean-field approximations.  In Sec.~\ref{ss:results} we discuss numerical results for the theory at finite temperature and varying dimensionless coupling constant $\rho^{1/3}a$.  We compare the LOAF approximation to the Popov approximation in detail.  
We conclude in Sec.~\ref{s:conclusions}. 
Finally, in App.~\ref{s:renorm} we discuss the connection between regularization of the effective potential and renormalization of the parameters.  In App.~\ref{s:building} we give the rules for determining all the Feynman graphs for the expansion using the mean-field propagators and vertices.  

%
%%%%%%%%%%%%%%%%%%%%%%%%%%%%%%%%%%%%%%%%%%%%%%%%%%%%%%%%%%%%%%%%%%%%%%
%
\section{\label{s:auxfield}The auxiliary-field formalism}

The classical action $S[\, \phi,\phi^{\ast} \, ]$ is given by
\begin{equation}\label{af.e:actionI}
   S[\, \phi,\phi^{\ast} \, ]
   =
   \int \! [\rd x] \> 
   \calL[ \, \phi,\phi^{\ast} \, ] \>,
\end{equation}
where $[\rd x] \equiv \rd t \, \rd^3 x$ and where the Lagrangian density is
\begin{align}
   \calL[ \, \phi,\phi^{\ast} \, ]
   &=
   \frac{i \hbar}{2} \, 
   [ \, 
      \phi^{\ast}(x) \, ( \partial_t \, \phi(x) )
      -
      ( \partial_t \, \phi^{\ast}(x) ) \, \phi(x) \,
   ]
   \label{af.e:LagI} \\
   & \quad
   -
   \phi^{\ast}(x) \, 
   \Bigl \{ \,
      -
      \frac{\hbar^2\nabla^2}{2m}
      -
      \mu_0 \,
   \Bigr \} \, 
   \phi(x)
   -
   \frac{\lambda_0}{2} \, | \, \phi(x) |^4 \>.
   \notag
\end{align}
Here $\mu_0$ and $\lambda_0$ are the bare (unrenormalized) chemical potential and contact interaction strength respectively.  We introduce two auxiliary fields, a real field, $\chi(x)$, and a complex field, $A(x)$, by means of the Hubbard-Stratonovitch transformation\cite{r:Hubbard:1959kx,r:Stratonovich:1958vn}, utilizing methods discussed in Refs.~\onlinecite{r:Bender:1977bh,r:Coleman:1974ve,r:Root:1974qf}.  In our case, the auxiliary-field Lagrangian density takes the form
\begin{align}
   &\calL_{\text{aux}}[\phi,\phi^{\ast},\chi,A,A^{\ast}]
   =
   \frac{1}{2 \lambda_0} \,
   \bigl [ \,
      \chi(x) - \lambda_0 \, \cosh\theta \, | \phi(x) |^2 \,
   \bigr ]^2
   \notag \\
   & \qquad
   -
   \frac{1}{2 \lambda_0} \,
   \bigl | \,
      A(x) 
      - 
      \lambda_0 \, \sinh\theta \, \phi^{2}(x) \,
   \bigr |^2
   \>,
   \label{af.e:Laux}
\end{align}
which we add to Eq.~\eqref{af.e:LagI}.  Here $\theta$ is a parameter which provides a mixing between the normal and anomalous densities.  In Sec.~\ref{s:meanfield}, we will see that choosing $\theta = 0$ leads to the usual large-$N$ expansion which has only the auxiliary field $\chi$  \cite{r:Coleman:1974ve,r:Root:1974qf}.   In lowest order, $\theta = 0$ gives a gapless solution very similar to the free Bose gas in the condensed phase.  If instead we choose $\theta$ such that $\sinh \theta = 1$, then in the weak coupling limit  our results agree with the Bogoliubov theory\cite{r:Bogoliubov:1947ys,r:Andersen:2004uq}, which represents the leading-order low-density expansion.  Of course all values of $\theta$ lead to a complete resummation of the original theory in terms of different combinations of the composite fields.  

For an arbitrary parameter $\theta$,  the action is given by
\begin{align}
   &S[\Phi,J] 
   =
   \label{af.e:actionII} \\
   &
   - 
   \frac{1}{2} \, 
   \iint  [\rd x] \, [\rd x'] \,
   \phi_a(x) \, \calG^{-1}{}^a{}_b[\chi](x,x') \, \phi^b(x')
   \notag \\
   &
   +
   \int \rd x \,
   \Bigl \{ \,
      \frac{\chi_i(x) \, \chi^i(x)}{2\lambda_0}
      +
      \Phi_{\alpha}(x) \, J^{\alpha}(x) \,
   \Bigr \} \>.
   \notag
\end{align}
with
\begin{align}
   &\calG^{-1}{}^a{}_b[\chi]
   = 
   \delta(x,x') \,
   \bigl \{ \,
      G^{-1}_0{}^a{}_b  
      +
      V^{a}{}_{b}[\chi](x) \,
   \bigr \}  \>,
   \label{af.e:G0invV0def} \\
   &G^{-1}_0{}^a{}_b
   =
   \begin{pmatrix}
      h - \mu_0 & 0 \\[3pt]
      0 & h^{\ast} - \mu_0 
   \end{pmatrix} \>,
   \quad
   h
   =
   -
   \frac{\hbar^2 \nabla^2}{2m}
   -
   i \hbar \frac{\partial}{\partial t} \>,   
   \notag \\
   &V^{a}{}_{b}[\chi](x)
   =
   \begin{pmatrix}
      \chi(x) \cosh\theta & - A(x) \sinh\theta \\
      - A^{\ast}(x) \sinh\theta & \chi(x) \cosh\theta
   \end{pmatrix} \>.
   \notag
\end{align}
Here we have introduced a two-component notation using Roman indices $a,b,c,\dotsb$ for the fields $\phi(x)$ and $\phi^{\ast}(x)$ and currents $j(x)$ and $j^{\ast}(x)$,
\begin{subequations}\label{af.e:notation}
\begin{alignat}{2}
   \phi^a(x)
   &=
   \Set{ \phi(x), \phi^{\ast}(x) } \>,
   &\quad
   \phi_a(x)
   &=
   \Set{ \phi^{\ast}(x), \phi(x) } \>,
   \\
   j^a(x)
   &=
   \Set{ j(x), j^{\ast}(x) } \>,
   &\quad
   j_a(x)
   &=
   \Set{ j^{\ast}(x), j(x) } \>,
\end{alignat}
\end{subequations}
for $a=1,2$, and a three-component notation using Roman indices $i,j,k,\dotsb$ for the fields $\chi(x)$, $A(x)$, and $A^{\ast}(x)$,
\begin{align}
   \chi^{i}(x)
   &=
   \bigl ( \,
      \chi(x), A(x)/\sqrt{2}, A^{\ast}(x)/\sqrt{2} \,
   \bigr ) \>,
   \label{af.e:chiSupperdefs} \\
   S^{i}(x)
   &=
   \bigl ( \,
      s(x), S(x)/\sqrt{2}, S^{\ast}(x)/\sqrt{2} \,
   \bigr ) \>,
   \notag
\end{align}
and
\begin{align}
   \chi_{i}(x)
   &=
   \bigl ( \,
      \chi(x), - A^{\ast}(x)/\sqrt{2}, - A(x)/\sqrt{2} \,
   \bigr ) \>,
   \label{af.e:chiSlowerdefs} \\
   S_{i}(x)
   &=
   \bigl ( \,
      s(x), -S^{\ast}(x)/\sqrt{2}, -S(x)/\sqrt{2} \,
   \bigr ) \>,
   \notag
\end{align}
for $i=1,2,3$.  For convenience, we also define five-component fields with Greek indices $\Phi^{\alpha}(x) = \Set{\phi^a(x),\chi^i(x)}$ and currents $J^{\alpha}(x) = \Set{j^a(x),S^i(x) }$.  These definitions define a metric $\eta_{\alpha,\beta}$ for raising and lowering indices.  We use this notation throughout this paper.

The action is invariant under a global $U(1)$ transformation, $\phi(x) \mapsto e^{i\alpha} \phi(x)$, $A(x) \rightarrow e^{2 i \alpha} A(x)$, and $\chi(x) \mapsto \chi(x)$.  In components, the equations of motion are
\begin{gather}
   [ \  h - \mu_0 + \chi(x) \cosh\theta \, ]\, \phi(x) 
   -
   A(x) \, \phi^{\ast}(x) \, \sinh\theta
   =
   j(x) \>,
   \notag \\
   \chi(x) / \lambda_0
   =
   | \, \phi(x) \, |^2 \, \cosh\theta
   -
   s(x) \>,
   \notag \\
   A(x) / \lambda_0
   =
   \phi^2(x) \, \sinh\theta
   -
   S(x) \>.
   \label{af.e:alleom}
\end{gather}   
We note that substituting $\chi(x)$ and $A(x)$ from the last two lines of Eqs.~\eqref{af.e:alleom} (for zero currents) into the first line of Eqs.~\eqref{af.e:alleom} gives the equation of motion for the field $\phi(x)$ with \emph{no} auxiliary fields\cite{r:Andersen:2004uq}.    

Parametrizing the Green function $\calG$ as
\begin{equation}\label{af.e:Green}
   \calG(x,x')
   =
   \begin{pmatrix}
      G(x,x') & K(x,x') \\
      K^{\ast}(x,x') & G^{\ast}(x,x')
   \end{pmatrix} \>,
\end{equation}
and using
\begin{equation}\label{af.e:invert}
   \int [\rd x'] \, 
   \calG^{-1}(x,x') \, \calG(x',x'')
   =
   \delta(x,x'') \>,
\end{equation}
we obtain the equations
\begin{subequations}\label{af.e:GKeom}
\begin{align}
   &
   \bigl [ \,
      h_0
      - \mu +
      \chi(x) \cosh \theta  \,
   \bigr ]  \, G(x,x')
   \label{af.e:Geom} \\
   & \qquad\qquad
   -
   A(x) \, K^{\ast}(x,x') \sinh \theta
   =
   \delta(x,x') \>,
   \notag \\
   &
   \bigl [ \,
      h_0 - \mu
      +
      \chi(x) \cosh \theta \,
   \bigr ] \, K(x,x')
   \label{af.e:Keom} \\
   & \qquad\qquad
   -
   A(x) \, G^{\ast}(x,x') \sinh \theta
   =
   0 \>,
   \notag
\end{align}
\end{subequations}
and the complex conjugates.  Here, $G(x,x')$ and $K(x,x')$ are the normal and anomalous correlation functions.

%
%%%%%%%%%%%%%%%%%%%%%%%%%%%%%%%%%%%%%%%%%%%%%%%%%%%%%%%%%%%%%%%%%%%%%%
%
\section{\label{s:Seff}Auxiliary-field loop expansion}

\begin{figure}[t]
   \centering
   \includegraphics[width=0.9\columnwidth]{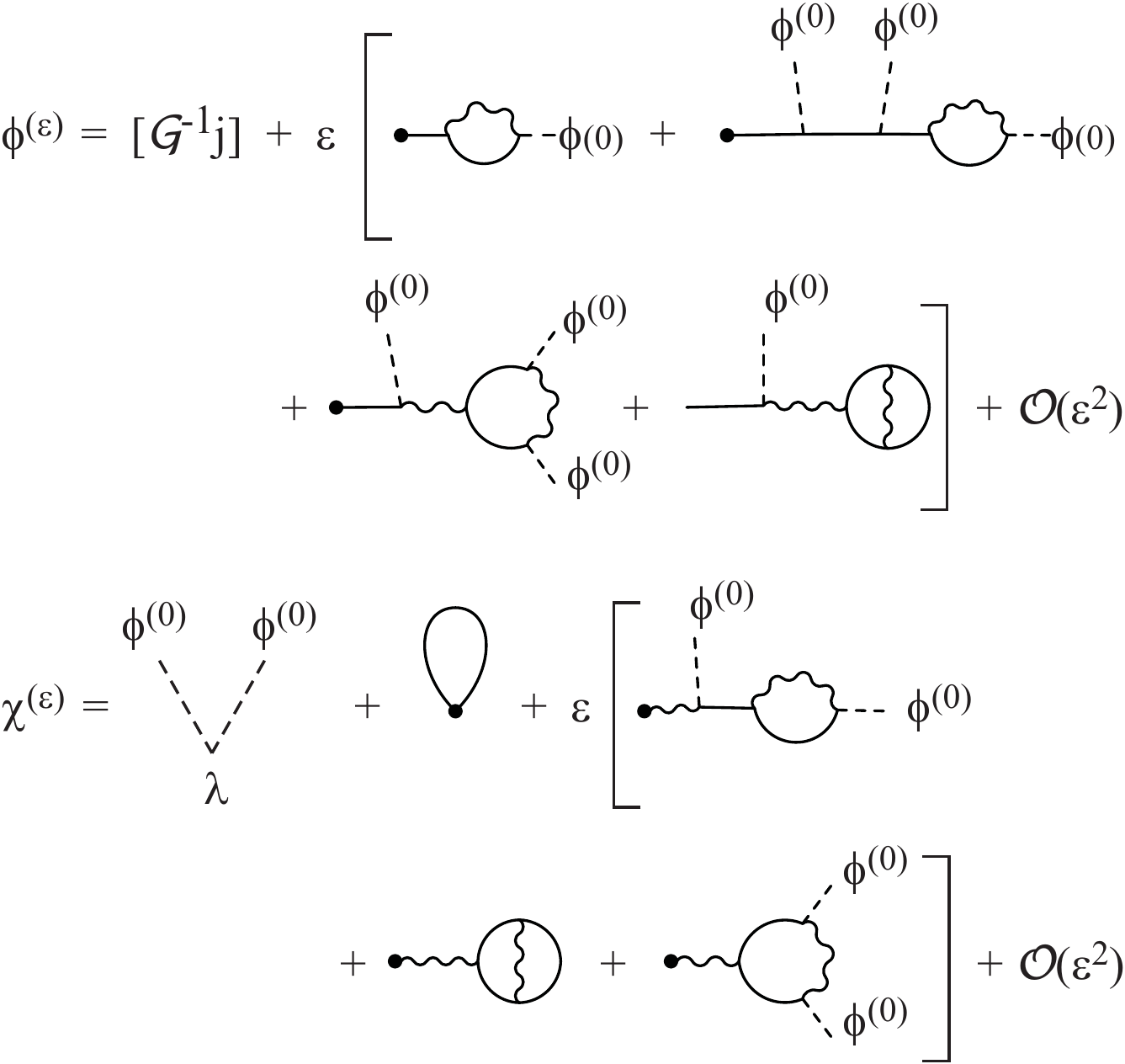}
   \caption{\label{f:phi_chi} Feynman diagrams for $\phi$ and $\chi$ to first order in~$\epsilon$. Solid and wavy lines correspond to the propagators of~$\phi$ and~$\chi$. Dashed lines denote the zeroth-order $\phi^{(0)}$.}
\end{figure}

The generating functional for connected graphs is
\begin{equation}\label{afle.e:Z}
   Z[J]
   =
   e^{i W[J] / \hbar}
   = 
   \calN
   \int \rD \Phi \>
   e^{ i S[\Phi,J] / \hbar } \>,
\end{equation}
with $S[\Phi,J]$ given by Eq.~\eqref{af.e:actionII}.  
Average values of the fields are given by
\begin{equation}\label{afle.e:avePhis}
   \Expect{\Phi^{\alpha}(x) }
   =
   \frac{\hbar}{i} \,
   \frac{1}{Z[J]} \, \frac{\delta Z[J]}{\delta J_{\alpha}(x)} \Big |_{J=0}
   =
   \frac{\delta W[J]}{\delta J_{\alpha}(x)} \Big |_{J=0} \>.
\end{equation}
If we integrate out the auxiliary fields $A(x)$ and $\chi(x)$, we obtain the  path integral for the original Lagrangian of Eq.~\eqref{af.e:LagI}.  The strategy we will use here is to reverse the order of integration and first do the path integral over the fields $\phi^a(x)$ exactly and then perform the path integration over the auxiliary fields by stationary phase to obtain a loop expansion in the auxiliary fields.  Performing the path integral over the fields $\phi^a$, we obtain 
\begin{equation}
   Z[J] 
   = 
   \calN' \int \rD \chi \, e^{i S_{\text{eff}}[ \chi,J ] / (\epsilon \hbar) }
   \>,
   \label{eq:Z}
\end{equation}
where the effective action is given by
\begin{align}
   S_{\text{eff}}[ \chi,J ]
   &=
   \frac{1}{2} \iint [\rd x] \, [\rd x'] \,
   j_{a}(x) \, \calG[\chi]^a{}_b(x,x') \, j^a(x)
   \notag \\
   & \quad
   +
   \int \! [\rd x] \,
   \Bigl \{ 
      \frac{\chi_i(x) \, \chi^{i}(x)}{2\lambda_0}
      +
      \chi_{i}(x) \, S^{i}(x)  
      \notag \\
      & \qquad\qquad\qquad
      - 
      \frac{\hbar}{2i}  
      \Tr{ \Ln{ \calG^{-1}(x,x) } } \,
   \Bigr \} \>.
   \label{afle.e:Seffxx}
\end{align}
Here $\chi^i(x)$ is defined in Eq.~\eqref{af.e:chiSupperdefs}.
As shown in Ref.~\onlinecite{r:Bender:1977bh}, the dimensionless parameter $\epsilon$ (which we eventually set equal to one) in Eq.~\eqref{eq:Z} allows us to count loops for the auxiliary-field propagators in the effective theory in analogy with $\hbar$ which counts loops for the $\phi$-propagator for the original Lagrangian.  The stationary point $\chi_0^{i}(x)$ of the effective action are defined by $\delta S_{\text{eff}}[ \chi,J ] /  \delta \chi_{i}(x) = 0$, i.e
\begin{align}
   \frac{\chi_0(x)}{\lambda_0}
   &=
   \bigl \{ \,
      | \phi_0(x) |^2
      +
      \hbar\, \Real{G(x,x)} / i \,
   \bigr \} \, \cosh\theta
   - 
   s(x)
  \notag \\
  \frac{A_0(x)}{\lambda_0}
   &=
   \bigl \{ \,
       \phi^2_0(x) 
      +
      \hbar \, K(x,x) / i \,
   \bigr \} \, \sinh\theta
   - 
   S(x) \>,
   \label{afle.e:chi0A0eqs}
\end{align}
where $\phi^a_0(x)$ is given by
\begin{equation}\label{afle.e:phi0def}
   \phi^a_0[\chi_0,J](x)
   =
   \int [\rd x'] \, \calG[\chi_0]^a{}_b(x,x') \, j^b(x') \>.
\end{equation}
Both $\chi_0(x)$ and $A_0(x)$ include self consistent fluctuations and are functionals of all the currents $J^{\alpha}(x)$.  Expanding the effective action about the stationary point, we find
\begin{align}\label{afle.e:Seffexpand}
   S_{\text{eff}}[ \chi,J ]
   &=
   S_{\text{eff}}[ \chi_0,J ]
   \\ & \qquad
   +
   \frac{1}{2} \iint [\rd x] \, [\rd x'] \,
   \calD^{-1}_{ij}[\chi_0](x,x')
   \notag \\
   & \qquad 
   \times
   ( \chi^i(x) - \chi^i_0(x) ) \,
   ( \chi^j(x') - \chi^j_0(x') )
   +
   \dotsb
   \notag
\end{align}
where $\calD_{ij}^{-1}[\chi_0](x,x')$ is given by the second-order derivatives
\begin{align}
   \calD^{-1}_{ij}[\chi_0](x,x')
   &=
   \frac{ \delta^2 \, S_{\text{eff}}[ \chi^a] }
        { \delta \chi^i(x) \, \delta \chi^j(x') } \, \bigg |_{\chi_0} 
   \label{afle.e:Dinverse} \\
   &=
   \frac{1}{\lambda_0} \eta_{ij} \, \delta(x,x')
   +
   \Pi_{ij}[\chi_0](x,x') \>,
   \notag
\end{align}
evaluated at the stationary points.  Here $\Pi_{ij}[\chi_0](x,x')$ is the polarization and is calculated in App.~\ref{s:building}.  We perform the remaining gaussian path integral over the fields $\chi_i$ by saddle point methods, obtaining
\begin{align}
   \epsilon \, W[J]
   &=
   S_0
   +
   S_{\text{eff}} [\chi_0,J]
   \label{afle.e:WtoSD} \\
   & \qquad
   -
   \frac{\epsilon \hbar}{2i} 
   \int \! [\rd x] \, \Tr{ \Ln{ \calD^{-1} [\chi_i, J](x,x) } } \>,
   \notag
\end{align}
where $S_0$ is a normalization constant.  
From this we calculate the order $\epsilon$ corrections to $\phi^a = \delta W / \delta j_a$ and $\chi^i = \delta W / \delta S_i$.  Schematically, these one-point functions are shown in Fig.~\ref{f:phi_chi}.  
%
%%%%%%%%%%%%%%%%%%%%%%%%%%%%%%%%%%%%%%%%%%%%%%%%%%%%%%%%%%%%%%%%%%%%%%
%
%
%%%%%%%%%%%%%%%%%%%%%%%%%%%%%%%%%%%%%%%%%%%%%%%%%%%%%%%%%%%%%%%%%%%%%%
%
The vertex function $\Gamma[\Phi]$ is constructed by a Legendre tranformation (see for example Ref.~\onlinecite{ref:ItzyksonZuber}) by 
\begin{equation}  \label{afle.e:vertexfctdef}
  \Gamma[\Phi]
   =
   \int [\rd x] \, J_{\alpha}(x) \, \Phi^{\alpha}(x)
   -
   W[J] \>.
\end{equation}
Here $\Gamma[\Phi]$ is the generator of the one-particle-irreducible (1-PI) graphs of the theory\cite{r:LW,r:Baym62,r:CJT}, with
\begin{equation}\label{afle.e:dGammadphidchi}
   \frac{\delta \Gamma[\Phi]}{\delta \Phi_{\alpha}(x)}
   =
   J^{\alpha}(x) \>.
\end{equation}
%So the currents vanish at the minimum of $\Gamma[\Phi]$.  
Keeping only the gaussian fluctuations in $W[J]$, we find
\begin{align}
   &\epsilon \, \Gamma[\Phi]
   =
   \frac{1}{2} \iint [\rd x] \, [\rd x'] \,
   \phi_a(x) \, \calG^{-1}[\chi]^{a}{}_{b}(x,x') \, \phi^b(x')
   \notag \\
   & \quad
   -
   \int [\rd x] \,
   \Bigl \{ \,
      \frac{\chi_{i}(x) \, \chi^{i}(x)}{2\lambda_0}
      -
      \frac{\hbar}{2i} \,
      \Tr{ \Ln{ \calG^{-1}[\chi](x,x) } }
      \notag \\
      &\quad
      -
      \frac{\epsilon \hbar}{2i} \,
      \Tr{ \Ln{ \calD^{-1}[\Phi]}(x,x) } \, 
   \Bigr \}
   +
   \epsilon \, \Gamma_0
   +
   \dotsb
   \>,
\label{gamma}  
\end{align}
which is the negative of the classical action plus self consistent one loop corrections in the $\phi^a$ and $\chi^i$ propagators. 
Here, $\Gamma_0$ is an adjustable constant used to set the minimum of the effective potential to have finite reference energy.
The effective potential $\Veff[\Phi]$ is defined for static fields $\Phi$ by
\begin{align}
   &\Veff[\Phi]
   =
   \frac{\epsilon \Gamma[\Phi]}{V T} 
   =
   \Veffzero
   +
   \frac{1}{2} \,
   \phi_a \, V[\chi]^{a}{}_{b} \, \phi^b 
   -  
   \frac{\chi_{i} \, \chi^{i}}{2\lambda_0}
   \label{afle.e:Veff} \\
   & \qquad
   - 
   \frac{\hbar}{2i V T} \,  
   \Tr{ \Ln{ \calG^{-1}[\chi](x,x) } } 
   \notag \\
   & \qquad
   -
   \frac{\epsilon \hbar}{2i V T} \,
   \Tr{ \Ln{ \calD^{-1}[\Phi](x,x) } } 
   + 
   \text{O}(\epsilon^2) \>,
   \notag
\end{align}
where
\begin{equation}\label{afle.e:Vabdef}
   V[\chi]^a{}_b
   =
   \begin{pmatrix}
      \chi \cosh\theta - \mu & 
      -A \sinh\theta \\
      -A^{\ast} \sinh\theta & 
      \chi \cosh\theta - \mu
   \end{pmatrix} \>.
\end{equation}
We will see below that for the static case, $\calG^{-1}[\chi](x,x)$ and $\calD^{-1}[\Phi](x,x)$ are independent of $x$.

For a system in equilibrium at temperature $T$, we Wick rotate the time variable to Euclidian time $\tau$ according to the Matsubara prescription, $t \rightarrow -i \hbar \tau$.  Then the effective potential becomes the grand potential $\Grand[\Phi]$ per unit volume, $\Veff[\Phi] \rightarrow \Grand[\Phi] / V$. (Details of the Matsubara formalism can be found for example in Ref.~\onlinecite{r:Negele:1988fk}.)
So to leading order in $\epsilon$, the thermal effective potential is given by
\begin{align}
   \Veff[\Phi] 
   &=
   \Veffzero
   +
   \frac{1}{2} \, \phi_a \, V[\chi]^{a}{}_{b} \, \phi^b  
   -  
   \frac{\chi_{i} \, \chi^{i}}{2\lambda_0}
   \label{afle.e:leading} \\
   & \qquad
   - 
   \frac{1}{2 \beta V}  
   \Tr{ \Ln{ \calG^{-1}[\chi](x,x) } } \>,
   \notag
\end{align}
and where $\Veffzero$ is a normalization constant.  At the next order we have the additional term
\begin{equation}  
   \Veff^{(1)}[\Phi]  
   =  
   - 
   \frac{\epsilon}{2 \beta V}  
   \Tr{ \Ln{ \calD^{-1}[\Phi](x,x) } } \>.
\end{equation}
Here and throughout this section, we suppress the dependence of quantities on $\theta$ and the thermodynamic variables $\Set{T,\mu,V}$.  The thermodynamic effective potential $\Veff[\Phi_0]$ is obtained by evaluating the effective potential at zero currents.  From \eqref{afle.e:dGammadphidchi}, this is when the fields $\Phi_0$ satisfy
\begin{equation}\label{afle.e:dGammadphidchiII}
   \frac{\delta \, \Veff[\Phi_0]}{\delta \Phi_{\alpha}(x)}
   =
   0 \>,
   \Qquad{for $\alpha = 1,\dotsb,5$.}
\end{equation}
We call these the ``gap equations'' in analogy with the corresponding equations in BCS theory.

The Green functions are periodic with Matsubara frequency $\omega_n = 2 \pi n/\beta$ with $\beta = 1 / ( \kb T )$, and are expanded in a Fourier series,
\begin{equation}
   \calG[\chi](x,x')
   = 
   \frac{1}{\beta} 
   \sum_{\bk,n} 
   \tilde{\calG}[\chi](\bk,n) \,
   e^{i [ \, k \cdot (\br-\br') - \omega_n(\tau-\tau') \, ]} \>.
   \label{afle.e:Fourierexpand}
\end{equation}   
Writing the Green function equation in $\bk$-$n$ space as
\begin{equation}\label{afle.e:GG}
   \tilde{\calG}^{-1}[\chi](\bk,n) \,
   \tilde{\calG}[\chi](\bk,n)
   =
   1 \>,
\end{equation}
we find
\begin{align}
   &\tilde{\calG}^{-1}[\chi](\bk,n)
   \label{afle.e:tGinverse} \\
   & \quad
   =
   \begin{pmatrix}
      \xi_k + \chi \cosh\theta - i \omega_n & 
      -A \sinh\theta \\
      -A^{\ast} \sinh\theta & 
      \xi_k + \chi \cosh\theta + i \omega_n
   \end{pmatrix} \>,
   \notag
\end{align}
where $\xi_k = \epsilon_k - \mu_0$.  So
\begin{equation}\label{afle.e:detGinv}
   \Det{ \tilde{\calG}^{-1}[\chi](\bk,n) }
   =
   \omega_k^2
   +
   \omega_n^2 \>.
\end{equation}
where
\begin{equation}\label{afle.e:omegak} 
   \omega_k^2
   =
   [ \, 
      \xi_k + \chi \cosh\theta \,
   ]^2
   -
   | A |^2 \sinh^2\theta \>.
\end{equation}
Stable solutions are possible for $\omega_k^2 \ge 0$.
The trace-log term then becomes
\begin{align}
   &\frac{1}{2V \beta} \,
   \mathrm{Tr}
   \bigl [ \,
      \Ln{  \calG^{-1}[\chi](x,x) } \,
   \bigr ]
   =
   \frac{1}{2V \beta} \,
   \sum_{\bk,n}
   \Ln{ \omega_k^2 + \omega_n^2 }
   \notag \\
   &\qquad 
   =
   \frac{1}{V} \,
   \sum_{\bk}
   \Bigl \{ \,
      \frac{\omega_k}{2}
      +
      \frac{1}{\beta}
      \Ln{ 1 - e^{-\beta \omega_k} } \,
   \Bigr \}
   \notag \\
   &\qquad
   =
   \Intk
   \Bigl \{ \,
      \frac{\omega_k}{2}
      +
      \frac{1}{\beta}
      \Ln{ 1 - e^{-\beta \omega_k} } \,
   \Bigr \} \>.
   \label{aux.e:TrLn}
\end{align}
So from \eqref{afle.e:leading} the effective potential to leading order in the auxiliary field loop expansion (LOAF) is given by
\begin{align}
   &\Veff[\Phi]
   =
   \Veffzero
   +
   | \phi |^2 \, 
   \bigl [ \,
      \chi \cosh\theta - \mu_0 \,
   \bigr ]
   \label{afle.e:GammaIII} \\
   & \qquad
   -
   \frac{1}{2} \, 
   \bigl [ \,
      \phi^{\ast\,2} \, A
      +
      \phi^2 \, A^{\ast} \,
   \bigr ] \,
   \sinh\theta
   \notag \\
   & \qquad
   -
   \frac{\chi^2 - | A |^2}{2\lambda_0}
   +
   \Intk
   \Bigl \{ \,
      \frac{\omega_k}{2}
      +
      \frac{1}{\beta}
      \Ln{ 1 - e^{-\beta \omega_k} } \,
   \Bigr \} \>.
   \notag   
\end{align}
It is useful to introduce new variables $\chi'$ and $A'$ as
\begin{equation}\label{afle.e:redefineAF}
   \chi'
   =
   \chi \cosh\theta - \mu_0 \>,
   \Qquad{and}
   A'
   =
   A \, \sinh\theta \>.
\end{equation}
Then the effective potential \eqref{afle.e:GammaIII} becomes
\begin{align}
   &\Veff[\Phi']
   =
   \Veffzero
   +
   | \phi |^2 \, \chi'
   -
   \frac{1}{2} \, 
   \bigl [ \,
      \phi^{\ast\,2} \, A'
      +
      \phi^2 \, A^{\prime\,\ast} \,
   \bigr ] \,
   \notag \\
   & \qquad
   -
   \frac{ ( \chi' + \mu_0 )^2 }
        { 2 \lambda_0 \cosh^2\theta }
   + 
   \frac{| A' |^2}{2\lambda_0 \sinh^2\theta }
   \label{afle.e:GammaIV} \\
   & \qquad
   +
   \Intk
   \Bigl \{ \,
      \frac{\omega_k}{2}
      +
      \frac{1}{\beta}
      \Ln{ 1 - e^{-\beta \omega_k} } \,
   \Bigr \} \>,
   \notag   
\end{align}
where now $\omega_k^2 =  ( \, \epsilon_k + \chi' \, )^2 - | A' |^2$.  The gap equations \eqref{afle.e:dGammadphidchiII} are now written as
\begin{gather}
   \begin{pmatrix}
      \chi'_0 & -A'_0 \\
      -A^{\prime\,\ast}_0 & \chi'_0
   \end{pmatrix}
   \begin{pmatrix} \phi_0 \\ \phi_0^{\ast} \end{pmatrix}
   =
   0 \>,
   \label{afle.e:gapeqs} \\
   \frac{\chi'_0 + \mu_0}{\lambda_0 \cosh^2\theta}
   =
   | \phi |^2
   +
   \Intk
   \frac{\epsilon_k + \chi'_0}{2\omega_k} \, 
   [ \, 2 n(\beta\omega_k) + 1 \, ] \>,
   \notag \\
   \frac{A'_0}{\lambda_0 \sinh^2 \theta}
   =
   \phi^2
   +
   A'_0
   \Intk \,
   \frac{ [ \, 2 n(\beta\omega_k) + 1 \, ]}{2\omega_k} \>,
   \notag
\end{gather}
where $n(x) = 1 / ( e^x - 1 )$ is the Bose-Einstein particle distribution.  The solutions $\Phi_0^{\prime\,\alpha}$ of Eqs.~\eqref{afle.e:gapeqs} substituted into Eq.~\eqref{afle.e:GammaIV} determine the effective potential.

To calculate the finite temperature effective potential to order $\epsilon$ we need to determine $\Tr{ \Ln{ \calD^{-1}(x,x') }}$.  For the static case in the imaginary time formalism, $\calD_{ij}[\chi](x,x')$ and $\Pi_{ij}[\Phi](x,x')$ are expanded in Fourier series' analogous to Eq.~\eqref{afle.e:Fourierexpand}.  So from Eq.~\eqref{afle.e:Dinverse} we obtain
\begin{equation}\label{BEC.NLOAF.e:tD}
   \tilde{\calD}_{ij}[\Phi](\bk,n)
   =
   \frac{ \eta_{ij} }{\lambda_0}
   +
   \tilde \Pi_{ij}[\Phi](\bk,n) \>.
\end{equation}

%
%%%%%%%%%%%%%%%%%%%%%%%%%%%%%%%%%%%%%%%%%%%%%%%%%%%%%%%%%%%%%%%%%%%%%%
%
\section{\label{s:effpotcon} The effective potential in the condensate phase to leading order}
%
%%%%%%%%%%%%%%%%%%%%%%%%%%%%%%%%%%%%%%%%%%%%%%%%%%%%%%%%%%%%%%%%%%%%%%
%

In the language of broken symmetry, the condensate phase is a phase where the U(1) symmetry of the theory is broken since then $\Expect{\phi} \neq 0$.  From Eq.~\eqref{afle.e:GammaIV}, the minimum of the effective potential is when
\begin{equation}\label{EP.e:brokencase}
   \frac{\delta \Veff[\Phi]}{\delta \phi^{\ast}} \Bigl |_{\phi_0}
   =
   \chi' \, \phi_0
   -
   A' \, \phi_0^{\ast}
   =
   0 \>.
\end{equation}
Because of the $U(1)$ gauge symmetry, we can choose $\phi_0$ to be real, which means that $A$ is also real. Hence, we have the broken symmetry condition $\chi' = A'$, and the dispersion relation reads
\begin{equation}\label{EP.e:disprel}
   \omega_k^2 
   = 
   \epsilon_k \, ( \epsilon_k + 2 A' ) \>, 
\end{equation}
The latter is a consequence of the the Hugenholz-Pines theorem which assures that the dispersion relation does not exhibit a gap. This is equivalent to the Goldstone theorem for a dilute Bose gas with a spontaneously-broken continuos symmetry. This connection is discussed in detail in Ref.~\onlinecite{r:Andersen:2004uq}. In the absence of quantum fluctuations in $\chi'=A'$, one obtains the Bogoliubov dispersion, $\omega_k^2 = \epsilon_k ( \epsilon_k + 2 \lambda \, \phi_0^2 ) $, by setting $A' = \lambda \, \phi_0^2 \, \sinh^2 \theta$ and $\sinh \theta = 1$.

In the spontaneously broken phase, the effective potential is 
\begin{align}
   \Veff[\chi']
   &=
   \Veffzero
   -
   \frac{ ( \chi' + \mu_0 )^2 }
        { 2 \lambda_0 \cosh^2\theta }
   + 
   \frac{\chi^{\prime\,2}}{2 \lambda_0 \sinh^2\theta }
   \label{EP.e:V-i} \\
   & \quad
   +
   \Intk
   \Bigl \{ \,
      \frac{\omega_k}{2}
      +
      \frac{1}{\beta}
      \Ln{ 1 - e^{-\beta \omega_k} } \,
   \Bigr \} \>,
   \notag   
\end{align}
where $\chi'$ is determined by the equation
\begin{align}
   \frac{ \partial \Veff[\chi'] }{ \partial \chi' }
   &=
   \frac{\chi'}{\lambda_0 \sinh^2\theta} 
   -
   \frac{\chi' + \mu}{\lambda_0 \cosh^2\theta} 
   \label{EP.e:dVdchip} \\
   & \qquad
   + 
   \Intk \, \frac{\epsilon_k}{2 \omega_k} \, [ \, 2 n(\beta\omega_k) + 1 \, ]
   =
   0 \>.
   \notag
\end{align}
These equations for $\Veff[\chi']$ and $\chi'$ contain infinite terms that need to be regulated.
In order to regulate the effective potential, we first expand $\omega_k$ in a Laurent series in~$\epsilon_k$
\begin{equation}\label{RR.e:omegakexpand}
   \omega_k
   =
   \sqrt{ ( \, \epsilon_k + \chi' \, )^2 - | A' |^2 }
   =
   \epsilon_k
   +
   \chi'
   -
   \frac{ | A' |^2 }{ 2 \epsilon_k }
   +
   \dotsb \>,
\end{equation}
around $k \rightarrow \infty$.
The first three terms in the series are responsible for the divergences in the integral in Eq.~\eqref{afle.e:GammaIV}.  To regularize the theory, we subtracting these three terms from $\omega_k$ in the integrand, and replace the constant $\Veffzero$ the bare interaction strength $\lambda_0$ and chemical potential $\mu_0$ by regulated ones.  This procedure gives the regulated effective potential
\begin{align}
   &\VeffR[\Phi']
   =
   \VeffzeroR
   +
   | \phi |^2 \, \chi'
   -
   \frac{1}{2} \, 
   \bigl [ \,
      \phi^{\ast\,2} \, A'
      +
      \phi^2 \, A^{\prime\,\ast} \,
   \bigr ] \,
   \notag \\
   & \qquad
   -
   \frac{ ( \chi' + \muR )^2 }
        { 2 \lambdaR \cosh^2\theta }
   + 
   \frac{| A' |^2}{2\lambdaR \sinh^2\theta }
   \label{RR.e:GammaV} \\
   &
   +
   \Intk
   \Bigl \{ \,
      \frac{1}{2}
      \Bigl [ \,
         \omega_k
         -
      \chi'
         +
         \frac{ | A' |^2 }{ 2 \epsilon_k } \,
      \Bigr ]
      +
      \frac{1}{\beta}
      \Ln{ 1 - e^{-\beta \omega_k} } \,
   \Bigr \} \>,
   \notag   
\end{align}
which is now finite.  Similarly, the regulated gap equations for $A'$ and $\chi'$ are now give as
\begin{align}
   \frac{\chi'+ \muR}{\lambdaR \cosh^2 \theta}
   &=
   | \phi |^2 \!
   + \!\!
   \Intk
   \Bigl \{ 
      \frac{\epsilon_k + \chi'}{2\omega_k}
      [ 2 n(\beta\omega_k) + 1 ] \!
      - \!
      \frac{1}{2}
   \Bigr \} \,,
   \notag \\
   \frac{A'}{\lambdaR \sinh^2 \theta}
   &=
   \phi^2 \!
   +\!
   A' \!\!
   \Intk 
   \Bigl \{ 
      \frac{2 n(\beta\omega_k) + 1}{2\omega_k}\!
      -\!
      \frac{1}{2\epsilon_k}
   \Bigr \}
   \label{RR.e:gapeqsA1}
\end{align}
which are also finite.

This regularization scheme is equivalent to dimensional regularization as done for example in Ref.~\onlinecite{r:Papenbrock:1999fk}, or to conventional renormalization of the coupling constant and chemical potential as described in the review article of Andersen and discussed in detail for the LOAF approximation in the App.~\ref{s:renorm}.

%We will show in the next section that this regulation scheme is equivalent to introducing a cutoff, $\Lambda$, and then renormalizing the parameters in the original Lagrangian.

%
%%%%%%%%%%%%%%%%%%%%%%%%%%%%%%%%%%%%%%%%%%%%%%%%%%%%%%%%%%%%%%%%%%%%%%
%
\section{\label{s:theta}Setting the parameter $\theta$}
%
%%%%%%%%%%%%%%%%%%%%%%%%%%%%%%%%%%%%%%%%%%%%%%%%%%%%%%%%%%%%%%%%%%%%%%
%

Up to this point, we considered a one-parameter class  of mean-field approximations governed by the parameter~$\theta$.  The dispersion relation in the condensate phase for the leading-order auxiliary field (LOAF) approximation  is given by  Eq.~\eqref{EP.e:disprel}
\begin{equation}
   \omega_k 
   = 
   \sqrt{ \epsilon_k ( \epsilon_k + 2 A \, \sinh\theta ) } \>,
\end{equation}
where 
$A/\lambda = \Expect{\phi^2} \sinh \theta = [ \, \phi^2 + \hbar \, K(x,x) / i \, ] \, \sinh \theta$.   Now, we will choose $\theta$ by demanding that in the weak coupling limit, when $K(x,x)$ can be ignored, the dispersion relation agrees with the one-loop low-density result obtained by Bogoliubov.  
Using a Hamiltonian formalism, Bogoliubov assumed
\begin{equation}
   \phi = \phi_0 + \psi \>,
   \label{eq:bog}
\end{equation}
subject to the constraint $\Expect{\psi} = 0$.  Realizing that $\phi_0 \approx \sqrt N$, he then wrote the theory in terms of the classical Hamiltonian plus a quadratic fluctuation Hamiltonian, which he diagonalized.  Using Eq.~\eqref{eq:bog} and limiting to at most quadratic fluctuations, one has
\begin{align}
   &[(\phi_0^*+ \psi^*) (\phi_0+\psi)]^2 
   \rightarrow 
   (\phi_0^\ast \phi_0)^2 
   +  
   4 \, \psi^\ast \psi \, (\, \phi_0^\ast\phi_0 \, )
   \notag \\
   & \qquad 
   + 
   \psi \, \psi \, ( \phi_0^\ast \phi_0) 
   + 
   \psi_0^\ast \psi_0^\ast \, (\phi_0^\ast \phi_0) \>.
   \label{bog.e:Bogaves}
\end{align}
The minimum of the classical Hamiltonian defines  $\mu = \lambda \, (\phi_0^\ast \phi_0)$.  
One can reformulate\cite{r:Andersen:2004uq} the Bogoliubov theory in path integral language as the classical approximation plus gaussian fluctuations.  The inverse Green function in the gaussian fluctuation approximation now has 
\begin{equation} 
   V^{a}{}_{b}[\phi](x)
   =
   \lambda 
   \begin{pmatrix}
      2 \, \phi_0^* \phi_0 & \phi_0 \phi_0 \\
      \phi_0^* \phi_0^* & 2 \, \phi_0^* \phi_0 
   \end{pmatrix} \>
\end{equation}
where $V^a{}_b[\phi]$ is defined in Eq.~\eqref{af.e:G0invV0def}.  This leads to the dispersion relation at the minimum: 
\begin{equation}\label{ST.e:bogdisp}
   \omega_k 
   = 
   \sqrt{\epsilon_k ( \epsilon_k + 2 \lambda \, \phi_0^2 ) }
\end{equation} 

We will choose $\theta$ such that our result for $\omega_k$ reduces to the Bogoliubov dispersion relation \eqref{ST.e:bogdisp} when we ignore quantum fluctuations in the anomalous density.  This sets $\sinh\theta = 1$ and $\cosh\theta = \sqrt{2}$.  With our choice of $\theta$, the renormalized effective potential can be written as
\begin{align}
   &\VeffR[\Phi]
   =
   \VeffzeroR
   +
   \chi' \, | \phi |^2
   -
   \frac{1}{2} \,
   \bigl ( 
      A^{\ast} \, \phi^2
      +
      A \, \phi^{\ast\,2} \,
   \bigr ) 
   -
   \frac{ ( \chi' + \mu)^2}{4 \lambdaR}
   \notag \\
   & \qquad
   +
   \frac{| A |^2 }{2 \lambdaR}
   + 
   \Intk \,
   \Bigl \{ \,
      \frac{1}{2}
      \Bigl [
         \omega_k
         -
         \epsilon_k
         -
         \chi'
         +
         \frac{|A|^2}{2 \epsilon_k} 
      \Bigr ]
      \notag \\
      & \qquad\qquad
      +
      \frac{1}{\beta} \,  \Ln{ 1 - e^{-\beta \omega_k} } \,
   \Bigr \} \>,
   \label{BEC.Seff.e:VeffII} 
\end{align}
where now  $\chi' =  \sqrt{2} \, \chi - \mu$ and 
\begin{equation}
\omega_k^2 = ( \epsilon_k + \chi')^2 - |A|^2 \>.
\end{equation}
The  equations for the auxiliary fields are obtained from $\delta \, \VeffR[\Phi] / \delta \chi'_{i} = 0$, as
\begin{subequations}\label{BEC.Seff.e:gapeqs}
\begin{align}
   \frac{A}{\lambdaR}
   &=
   \phi_0^2
   +
   A \!
   \Intk \,
   \Bigl \{ 
      \frac{[ 2 n(\beta\omega_k) + 1 ]}{2\omega_k}
      -
      \frac{1}{2\epsilon_k}
   \Bigr \} \>,
   \label{BEC.Seff.e:gapeqsA} \\
   \frac{\chi'+ \muR}{2 \lambdaR}
   &=
   | \phi_0 |^2 \!
   + \!
   \Intk \,
   \Bigl \{ 
      \frac{\epsilon_k + \chi'}{2\omega_k}
      [ 2 n(\beta\omega_k) + 1 ]
      -
      \frac{1}{2}
   \Bigr \} \>.
   \label{BEC.Seff.e:gapeqsB}
\end{align}
\end{subequations}
From Eq.~\eqref{EP.e:brokencase} we know that at the minimum of the effective potential we have $ (\chi' - A )\, \phi_0 = 0$, and we can replace $\muR$ by the physical density using
\begin{equation}\label{BEC.Seff.e:rhodef}
   \rho 
   = 
   - 
   \frac{\partial \VeffR[\Phi_0]}{\partial \mu} 
   = 
   \frac{\chi' + \muR}{2 \lambdaR} \>.
\end{equation}
In the broken symmetry phase we have $\chi' = A$ in which case Eqs.~\eqref{BEC.Seff.e:gapeqs} become
\begin{subequations}\label{BEC.Seff.e:gapeqsII}
\begin{align}
   \frac{\chi'}{\lambdaR}
   &=
   \rho_0
   +
   \chi'
   \Intk \,
   \Bigl \{ 
      \frac{[ 2 n(\beta\omega_k) + 1 ]}{2\omega_k}
      -
      \frac{1}{2\epsilon_k}
   \Bigr \} \>,
   \label{BEC.Seff.e:gapeqsAII} \\
   \rho
   &=
   \rho_0
   +
   \Intk \,
   \Bigl \{ 
      \frac{\epsilon_k + \chi'}{2\omega_k}
      [ 2 n(\beta\omega_k) + 1 ]
      -
      \frac{1}{2}
   \Bigr \} \>,
   \label{BEC.Seff.e:gapeqsBII}
\end{align}
\end{subequations}
where $\rho_0 = \phi_0^2$ is the condensate density.  

%
%%%%%%%%%%%%%%%%%%%%%%%%%%%%%%%%%%%%%%%%%%%%%%%%%%%%%%%%%%%%%%%%%%%%%%
%
\section{\label{s:meanfield}Related mean-field approximations}
%
%%%%%%%%%%%%%%%%%%%%%%%%%%%%%%%%%%%%%%%%%%%%%%%%%%%%%%%%%%%%%%%%%%%%%%
%

For comparison, we will review next two related mean-field approximations. We will focus on the leading-order large-$N$ approximations, which corresponds to the choice of $\theta=0$ in our formalism, and the Popov approximation that is widely used in the study of BEC condensates. 

%
%%%%%%%%%%%%%%%%%%%%%%%%%%%%%%%%%%%%%%%%%%%%%%%%%%%%%%%%%%%%%%%%%%%%%%
%
\subsection{\label{ss:largeN}Large-$N$ approximation in leading order } 

The large-$N$ approximation corresponds to the value $\theta = 0$.  In obtaining the large-N approximation, one rewrites $\phi^\ast \phi$ in terms of two real components and extends the theory to $N$ real components.  The $O(2)$  [$U(1)$] symmetry is then extended to $O(N)$.  Here the composite field is $\chi = \lambda \, \phi_i \phi_i / N$. With appropriate rescaling, one can show\cite{r:Bender:1977bh} that the composite-field propagator is proportional to $1/N$, so counting loops of bound-state propagators yields the $1/N$ expansion.  In lowest order we find that this approximation in the BEC phase leads to the free-field dispersion relation. %, which is quite different than the Bogoliubov result.  For that reason we feel that the large-$N$ expansion in leading order is not a very good ``mean-field'' approximation.  
A related large-$N$ expansion for the Bose gas at the critical temperature has been used successfully to characterize the behavior near the critical point\cite{r:Baym:2000fk}.  One simplicity of this expansion is that the noninteracting-like dispersion relation simplifies the integrals present in the theory, and one can obtain analytic results even at finite temperatures.  As with the general $\theta$ result, the large-$N$ expansion also provides a complete resummation of the original theory. 

% next-to-leading order, the large-$N$ expansion

The large-$N$ finite-temperature effective potential in leading order is given by 
\begin{equation}\label{ln.e:Vln-I}
   \calV_\text{LN}[\Phi]
   =
   \calV_0
   +
   \chi \, | \phi |^2
   -
   \frac{(\chi + \mu_0)^2}{2 \lambda_0} 
   -
   \frac{1}{2} \, \Tr{ \Ln{ \calG^{-1} } } \>.
\end{equation}
The Matsubara inverse propagator in momentum space is now diagonal:
\begin{equation*}
   \calG^{-1} (\bk, n) 
   = 
   \begin{pmatrix}
      i \omega_n - \epsilon_k - \chi & 0 \\
      0 & -i\omega_n - \epsilon_k - \chi 
   \end{pmatrix}
\end{equation*}
We write the temperature-dependent last term in Eq.~\eqref{ln.e:Vln-I} as
\begin{align}  
   \frac{1}{2} \, \Tr{ \Ln{ \calG^{-1} } }
   &= 
   \frac{1}{2 \beta} 
   \IntkR  \sum_n (\omega_n^2 + \omega_k^2 ) 
   \\
   &=    
   \IntkR \,
   \Bigl \{ \,
      \frac{\omega_k}{2} 
      + 
      \frac{1}{\beta} \,
      \Ln{ 1- e^{-\beta \omega_k} } \,
   \Bigr \} \>,
   \notag
\end{align}
where  $\omega_k = \epsilon_k + \chi$.  Inserting this into Eq.~\eqref{ln.e:Vln-I}, the effective potential for the large-$N$ case is given by
\begin{align}
   \calV_\text{LN}[\Phi]
   &=
   \calV_0
   +
   \chi \, | \phi |^2
   -
   \frac{(\chi + \mu_0)^2}{2 \lambda_0} 
   \label{ln.e:Vln-II} \\
   & \qquad
   +
   \IntkR \,
   \Bigl \{ \,
      \frac{\omega_k}{2} 
      + 
      \frac{1}{\beta} \,
      \Ln{ 1- e^{-\beta \omega_k} } \,
   \Bigr \} \>.
   \notag
\end{align}
Setting the derivative of the effective potential with respect to $\chi$ equal to zero yields the gap equation,
\begin{equation}
   \frac{\chi+\mu_0}{\lambda_0} 
   =
   | \phi |^2 
   +  
   \IntkR \,
   \frac{2 \, n(\beta\omega_k) + 1 }{2} \>.
\end{equation}

The large-$N$ effective potential in leading order is renormalized following the procedure discussed in Ref.~\cite{r:Andersen:2004uq}. We recognize that the infinite constant is related to the renormalization of the chemical potential, i.e
\begin{equation}
   \frac{\mu_0}{\lambda_0} 
   = 
   \frac{\muR}{\lambdaR}
   +
   \IntkR \, \frac{1}{2} \>.
\end{equation}
The renormalization is a consequence of the lack of a normal-ordering step in the the path-integral formalism, in contrast with the usual Hamiltonian formalism.  Performing the renormalization, one obtains the finite gap equation,
\begin{equation}\label{LN.e:gap}
   \rho
   =
   \frac{\chi+\muR}{\lambdaR} 
   =
   \rho_0
   +  
   \Intk \,
   n(\beta\omega_k) \>,
\end{equation}
where we have set the condensate density $\rho_0 = |\phi |^2$.
Eq.~\eqref{LN.e:gap} determines $\chi[\phi]$ implicitly, which is then re-inserted into the expression of $\calV_{\text{LN}}$, so that $\calV_{\text{LN}}[\Phi]$ becomes solely a function of $\rho_0 = |\phi |^2$.  The renormalized potential is now
\begin{align}
   \calV_{\text{LN}}[\phi]  
   &= 
   \chi \, | \phi |^2
   -
   \frac{ (\chi+ \muR)^2}{2 \lambdaR} 
   \notag \\
   & \qquad
   + 
   \Intk \, 
   \frac{1}{\beta} \, \Ln{ 1- e^{-\beta \omega_k} } \>.
\end{align}
The minimum of the effective potential is when
\begin{equation}
   \frac{\partial \, \calV_{\text{LN}}}{\partial \phi^\ast} 
   = 
   0 \>,
   \Qquad{$\Rightarrow$}
   \phi \, \chi = 0 \>.
\end{equation}
So, in the large-$N$ mean-field approximation, the broken-symmetry regime, $\phi \neq 0$, corresponds to the condition
\begin{equation}
   \chi = 0 \>,
\end{equation}
which gives the dispersion relation, $\omega_k = \epsilon_k$, which is the same as the free-field theory dispersion. 
% Thus in the lowest order, the large-$N$ expansion leads to a self-consistent mean-field  approximation that only keeps gaussian fluctuations around the classical limit. 

At finite temperature, the gap equation at the minimum 
\begin{equation}\label{gap_ln}
   \lambdaR \, \rho_0 
   =  
   \muR 
   - 
   \lambdaR   
   \Intk \, n(\beta\omega_k)
\end{equation}
which gives the chemical potential as
\begin{equation}
   \muR
   = 
   \lambdaR \,
   \Bigl \{ \,
      \rho_0 
      + 
      \frac{\sqrt{\pi}}{2} \, T^{3/2} \, \zeta(3/2) \,
   \Bigr \} \>.
\label{mu_ln}
\end{equation}
Correspondingly, the phase transition ($\phi = 0$) takes place at the free-field  critical temperature:
\begin{equation}
   T_c
   = 
   \Bigl [ \,
      \frac{2 \muR}{ \lambdaR \, \zeta(3/2) \, \sqrt{\pi}} \,
   \Bigr ]^{2/3}
   \>.
\end{equation}
At the minimum, $\chi = 0$ so that the value of the effective potential at the minimum as a function of temperature for $T< T_c$ is 
\begin{align}
   \calV_{\text{LN}} 
   & = 
   -
   \frac{\muR^2}{2 \lambdaR} 
   + 
   \Intk \,  \frac{1}{\beta} \, \Ln{ 1- e^{-\beta \epsilon_k} } \>, 
   \\ \notag
   & = 
   -
   \frac{\muR^2}{2 \lambdaR} 
   -
   \frac{T^{3/2}}{16 \pi^2} \,
         \frac{\sqrt{\pi } \, 
         \zeta(5/2)}{2} \>.
\end{align}
The total number density is determined from 
\begin{equation}\label{rho_ln}
   \rho 
   =  
   - \frac{\partial \, \calV_{\text{LN}} }{\partial \muR} 
   = 
   \frac{\muR}{\lambdaR} \>.
\end{equation}
Hence, by combining Eqs.~\eqref{mu_ln} and~\eqref{rho_ln}, we obtain the density of particles in the condensate as
\begin{equation} 
   \rho_0 
   =  
   \rho 
   -  
   \frac {\sqrt{\pi}}{2} \, T^{3/2} \, \zeta(3/2) \>.
\end{equation}

In summary, below $T_c$ the large-$N$ approximation gives essentially the same results as a non-interacting gas since $\chi =0$. Above $T_c$,  the large-$N$ approximation gives rise to a self-consistent correction to the dispersion relation.  The large-$N$ result above $T_c$ is the same as that of the Popov approximation we review below. We will also find that the large-$N$ approximation is equal to the LOAF approximation we are proposing here at high temperatures in the regime where $A=0$. 

%
%%%%%%%%%%%%%%%%%%%%%%%%%%%%%%%%%%%%%%%%%%%%%%%%%%%%%%%%%%%%%%%%%%%%%%
%
\subsection{\label{ss:popov}Hartree and Popov approximations}
%
%%%%%%%%%%%%%%%%%%%%%%%%%%%%%%%%%%%%%%%%%%%%%%%%%%%%%%%%%%%%%%%%%%%%%%
%

The Hartree approximation is a truncation scheme that ignores correlation functions beyond the first two.  Technically this is obtained by setting the third derivative of the generating functional of connected graphs with respect to the external currents to zero, i.e.
\begin{equation}
   \frac {\delta^3 \, W[j]}{ \delta j(x) \, \delta j(y) \, \delta j(z)} 
   \equiv 
   0 \>.
\end{equation}
Then, the vacuum expectation value of the expectation value of the field $\phi[j](x)$ in the presence of external sources is
\begin{align}
   &( h - \mu ) \, \phi(x)
   +
   \lambda_0 \, | \phi(x) |^2 \, \phi(x)
   \label{BEC.hartree.e:funceom} \\
   & \qquad
   +
   2 \lambda_0  \hbar \, G(x,x) \, \phi(x) / i
   \notag \\
   & \qquad
   +
   \lambda_0 \hbar \, K(x,x) \, \phi^{\ast}(x) / i
   =
   j(x) ,
   \notag
\end{align}
where $h$ was defined in Eq.~\eqref{af.e:G0invV0def}.  Here $\phi(x)$, $G(x,x')$ and $K(x,x')$ are considered functionals of the current $j(x)$.  $G(x,x')$ and $K(x,x')$ have the same meaning as the normal and anomalous correlation functions in Eq.~\eqref{af.e:Green}.  Introducing new auxiliary fields $\chi(x)$ and $A(x)$ by the definition,
\begin{subequations}\label{BEC.hartree.e:chiAdefs}
\begin{align}
   \frac{\chi(x) + \mu_0}{2 \lambda_0}
   &=
   | \, \phi(x) \, |^2 + \hbar \, G(x,x) / i  \>,
   \label{BEC.hartree.e:chidef} \\
   \frac{A(x)}{\lambda_0}
   &=
   [ \, \phi(x) \, ]^2 + \hbar \, K(x,x) / i \>.
   \label{BEC.hartree.e:Adef}
\end{align}
\end{subequations}
and setting $j(x) = 0$ in Eq.~\eqref{BEC.hartree.e:funceom} gives an equation for the average field,
\begin{equation}\label{BEC.hartree.e:eomphiII}
   \bigl [ \,
      h
      +
      \chi(x)
      -
      2 \lambda \, | \, \phi(x) \, |^2 \,
   \bigr ] \, \phi(x)
   +
   A(x) \, \phi^{\ast}(x)
   =
   0 \>.
   \notag
\end{equation}
and its complex conjugate.  Functional differentiation of Eq.~\eqref{BEC.hartree.e:funceom} with respect to $j(x')$ and $j^{\ast}(x')$, ignoring third-order functional derivatives leads to equations for the Green functions $G(x,x')$ and $K(x,x')$.  We find
\begin{subequations}\label{BEC.hartree.e:GKeom}
\begin{align}
   \bigl [ \,
      h
      +
      \chi(x) \,
   \bigr ]  \, G(x,x')
   +
   A(x) \, K^{\ast}(x,x')
   &=
   \delta(x,x') \>,
   \label{BEC.hartree.e:Geom} \\
   \bigl [ \,
      h
      +
      \chi(x) \,
   \bigr ] \, K(x,x')
   +
   A(x) \, G^{\ast}(x,x')
   &=
   0 \>,
   \label{BEC.hartree.e:Keom}
\end{align}
\end{subequations}
and the complex conjugates.
 Eqs.~\eqref{BEC.hartree.e:GKeom} can be written in matrix form as
\begin{equation}\label{BEC.hartree.e:Gmatform}
   \int \rd x' \, 
   \calG^{-1}(x,x') \, \calG(x',x'')
   =
   \delta(x,x'') \>,
\end{equation}
where
\begin{subequations}\label{BEC.hartree.e:Gmatdefs}
\begin{align}
   \calG^{-1}(x,x')
   &=
   \delta(x,x') \,
   \begin{pmatrix}
      h + \chi(x) & A(x) \\
      A^{\ast}(x) & h^{\ast} + \chi(x)
   \end{pmatrix} \>.
   \label{BEC.hartree.e:GmatdefsB}
\end{align}
\end{subequations}
The renormalized effective potential for the Hartree approximation can be written as:
\begin{align}
   &\calV_{\text{H}}[\Phi]
   =
   \calV_{\text{R}}
   +
   \chi \, | \phi |^2
   -
   \lambdaR \, | \phi |^4
   -
   \frac{( \chi + \muR )^2}{4 \lambdaR}
   \label{BEC.hartree.e:effVII} \\
   & \qquad
   -
   \frac{| A |^2}{2 \lambdaR}
   +
   \frac{1}{2} \,
   [ \, 
      \phi^2 \, A^{\ast} 
      + 
      \phi^{\ast\,2} \, A \, 
   ] \,
   \notag \\
   &
   +
   \Intk \,
   \Bigl \{ \,
      \frac{1}{2}
      \Bigl [
         \omega_k
         -
         \epsilon_k
         -
         \chi
         +
         \frac{|A|^2}{2 \epsilon_k} 
      \Bigr ]
      +
      \frac{1}{\beta} \,  \Ln{ 1 - e^{-\beta \omega_k} } \,
   \Bigr \} \>.
   \notag
\end{align}
The minimum of the potential is given by
\begin{equation}\label{BEC.hartree.e:dVdphi}
   \frac{\partial V_{\text{H}}}{\partial \phi^{\ast}} \Big |_{\phi_0}
   =
   \chi_0 \, \phi_0
   -
   2 \lambdaR \, | \phi_0 |^2 \, \phi_0
   +
   A_0 \, \phi^{\ast}_0
   =
   0  \>.
\end{equation}
Again, the $U(1)$ gauge symmetry, allows us to choose $\phi_0$ to be real at the minimum.
Then according to Eq.~\eqref{BEC.hartree.e:dVdphi}, $A_0$ is also real.  Hence, Eq.~\eqref{BEC.hartree.e:dVdphi} becomes
\begin{equation}\label{BEC.hartree.e:phimincond}
   [ \,
      \chi_0 - 2 \lambdaR \, | \phi_0 |^2 + A_0 \,
   ] \, \phi_0
   =
   0 \>.
\end{equation}
In the broken symmetry case, $\phi_0 \ne 0$, we have
\begin{equation}\label{BEC.hartree.e:wehave}
   \chi_0 + A_0
   =
   2 \lambdaR \, | \phi_0 |^2 \>,
\end{equation}
and the dispersion relation is
\begin{equation}\label{BEC.hartree.e:omegahartree}
   \omega_k^2
   =
   [ \, \epsilon_k + 2 \lambdaR \, | \phi_0 |^2 \, ] \,
   [ \, 
      \epsilon_k 
      -
      2 \, ( \, A_0 - \lambdaR \, | \phi_0 |^2 \, ) \, ] \>.
\end{equation}

The Hartree approximation has the defect of not being gapless.  This can be fixed by hand by ignoring the fluctuation in the anomalous density, that is by arbitrarily setting 
\begin{equation}
   A_0 - \lambdaR \, | \phi_0 |^2  
   = 
   0 \>.
\end{equation}
This further approximation is known as the ``gapless'' Popov approximation~\cite{r:Popov:1983kx}.  
%Formally, the Popov approximation can also be obtained from Eqs.~\eqref{BEC.Seff.e:gapeqs} by setting $A_0 =\chi_0' = \lambda \rho_0$ and neglecting the quantum fluctuations in the anomalous density. 

The Popov approximation includes the self-consistent fluctuations of $\chi$, but treats $A$ classically.  Below $T_c$, the  Popov approximation has the dispersion relation
\begin{equation}\label{disp:Popov}
   \omega_{k}^{2}
   =
   \epsilon_{k} ( \, \epsilon_{k} + 2 \lambdaR \rho_0 \, ) \>,
\end{equation}
and the chemical potential is $\muR = 2 \lambdaR \rho - \lambdaR \rho_0$.  The condensate density $\rho_0$ is given by
\begin{equation}
   \rho
   =
   \rho_0
   +
   \Intk \,
   \Bigl \{ \,
      \frac{ \epsilon_{k} + \lambdaR \rho_0}
           { \omega_{k}} \,
      [ \, 2n(\beta\omega_{k}) + 1 \, ]
      -
      \frac{1}{2} \,
    \Bigr \} \>.
\end{equation}
In the Popov approximation the critical temperature is the same as in the free-field case, $T_c=T_0$.

The Popov approximation is the most commonly used mean-field theory of weakly interacting bosons at finite temperatures\cite{r:Andersen:2004uq}.  This approximation has a gapless spectrum but is known to produce an artificial first-order phase transition as we shall see below.   Unlike the LOAF and the Hartree approximations, the equations for $\phi, A, \chi$ in the Popov approximation are not derivable from an effective action.

%
%%%%%%%%%%%%%%%%%%%%%%%%%%%%%%%%%%%%%%%%%%%%%%%%%%%%%%%%%%%%%%%%%%%%%%
%
\section{\label{ss:results}Mean-field results and discussions}

%
%%%%%%%%%%%%%%%%%%%%%%%%%%%%%%%%%%%%%%%%%%%%%%%%%%%%%%%%%%%%%%%%%%%%%%
%

We begin by comparing the predictions of the LOAF and Popov approximations for zero-temperature conditions. 
In the broken-symmetry phase at $T=0$, we have $\chi' = A$, and the effective potential is given by
\begin{align} 
   \calV_{\text{eff}} [\chi'] 
   &=  
   -\frac{1}{2 \lambda} \,
   \Bigl [ \,
      \frac{ ( \chi'+ \mu )^2 }{2}
      - 
      \chi'{}^2 \,
   \Bigr ] 
   \\ \notag
   & \qquad
   + 
   \frac{1}{4 \pi^2}  
   \int_{0}^{\infty} \!\! k^2 \, \rd k \, 
   \Bigl [ \,
      \omega_k
      -
      \epsilon_k
      -
      \chi'
      +
      \frac{\chi'{}^2}{2 \epsilon_k} \,
   \Bigr ]
   \\ \notag
   &= 
   -
   \frac{1}{2 \lambda}
   \Bigl [ \,
      \frac{ ( \chi' + \mu )^2 }{2}  
      - 
      \chi'{}^2 \,
    \Bigr ] 
    + 
    \frac { 2 \sqrt{2}} {15 \pi^2} \, \chi'{}^{5/2} \>.
    \label{VNzero}
\end{align}
Setting $\partial \calV_{\text{eff}}/\partial \chi' = 0$, then gives $\chi'$ as a function of $\mu$ at the minimum. We have
\begin{equation}\label{mu1}
   \chi' 
   = 
   \mu
   - 
   \lambda \, \frac{ 2 \sqrt{2}}{3  \pi^2} \, \chi'{}^{3/2}  
   \>.
\end{equation}
The above cubic equation can be solved explicitly. In particular, in the weak coupling limit we obtain
\begin{equation}
   \chi'
   =
   \mu 
   -  
   \lambda \, \frac{2  \sqrt{2} }{3  \pi^2} \, \mu^{3/2} \>.
   \label{chi_weak}
\end{equation}
Using Eq.~\eqref{BEC.Seff.e:rhodef}, in weak coupling we obtain 
\begin{equation}
   \rho 
   = 
   \frac{\mu}{\lambda} 
   \Bigl [ \, 
      1
      -
      \frac{ \lambda \sqrt{2 \mu}}{ 3 \pi^2} \,
    \Bigr ]
    \label{eq:rho_T0}
    \>,
\end{equation}
which agrees with  the one-loop result corresponding to the original Bogoliubov approximation  (see e.g. Eq.~89 in Ref.~\cite{r:Andersen:2004uq}).  By inverting Eq.~\eqref{eq:rho_T0}, we derive $\mu(\rho)$ at weak coupling, as 
\begin{align}
   \mu 
   & = 
   \lambda \rho 
   \Bigl [ \, 
      1
      + 
      \frac{1}{3 \pi^2} \sqrt{  \frac{\lambda \rho} {2} } \,
   \Bigr ] 
   \label{mu_weak}
   \\ \notag 
   &
   =
   8\pi \rho \, a \, 
   \Bigl ( \,
      1
      +
      \frac{32}{3} \, \sqrt{\rho \, a^{3} / \pi } \,
   \Bigr ) \>,
\end{align}
where we have set $\lambdaR = 4\pi \hbar^2 \, a / m$ with $a$ the $s$-wave scattering length.

We can also calculate the condensate depletion, defined as $\rho - \rho_0$. From Eq.~\eqref{BEC.Seff.e:gapeqsBII} at $T=0$, we obtain the exact LOAF result
\begin{equation}
    \rho - \rho_0 = \frac{1}{6\sqrt{2} \, \pi} \, \chi'^{3/2}
    \>.
\end{equation}
Hence, using Eqs.~\eqref{chi_weak} and \eqref{mu_weak} we obtain the weak-coupling result for the fractional depletion (see e.g. Eq. 22.14 in Ref.~\onlinecite{r:Fetter:1971fk})
\begin{equation}
    1 -  \frac{\rho_0}{\rho} = \frac{8}{3} \sqrt{ \frac{\rho a^3}{\pi} }
    \>,
\end{equation}
first obtained by Bogoliubov in 1947~\cite{r:Bogoliubov:1947ys}.

%
%%%%%%%%%%%%%%%%%%%%%%%%%%%%%%%%%%%%%%%%%%%%%%%%%%%%%%%%%%%%%%
\begin{figure}[t!]
   \centering
   \includegraphics[width=0.9\columnwidth]{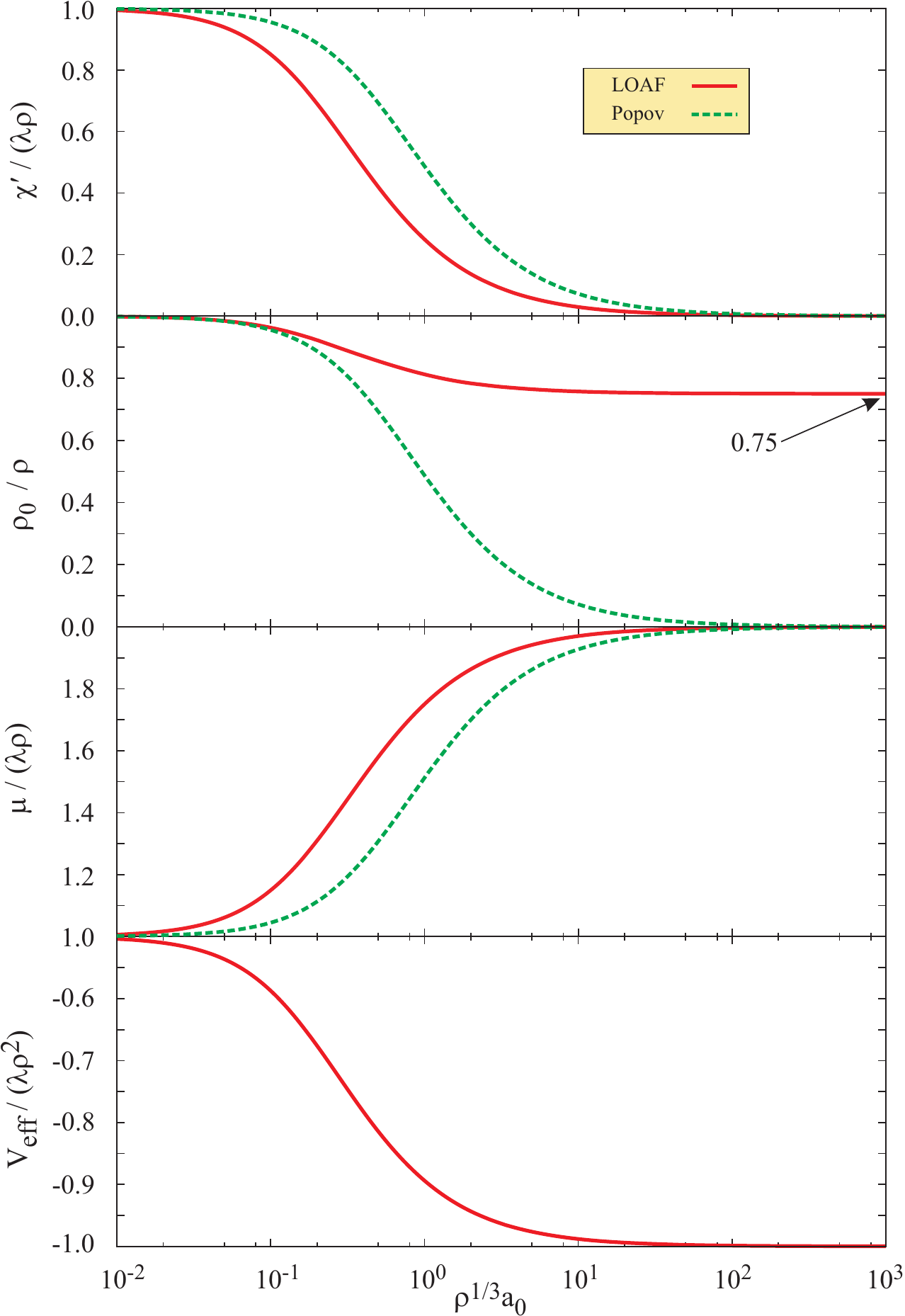}
   \caption{\label{f:Tzero} (Color online) 
   Comparison of the predictions of the Popov and LOAF approximations regarding the zero-temperature values of the densities, $\chi' = A$, condensate fraction, $\rho_0/\rho$, chemical potential, $\mu$, and effective potential, $V_{\mathrm{eff}}$, 
   as a function of dimensionless parameter,  $\rho^{1/3}a$. 
   In the case of the LOAF approximation, the normal and anomalous densities are equal, $\chi'= A$, whereas in the Popov approximation we have $A = \lambda \rho_0$.
   Note that there is no effective potential in the Popov approximation, because this approximation is not derivable from an action.
   }
\end{figure}
%
%%%%%%%%%%%%%%%%%%%%%%%%%%%%%%%%%%%%%%%%%%%%%%%%%%%%%%%%%%%%%%
\begin{figure}[t!]
   \centering
   \includegraphics[width=0.9\columnwidth]{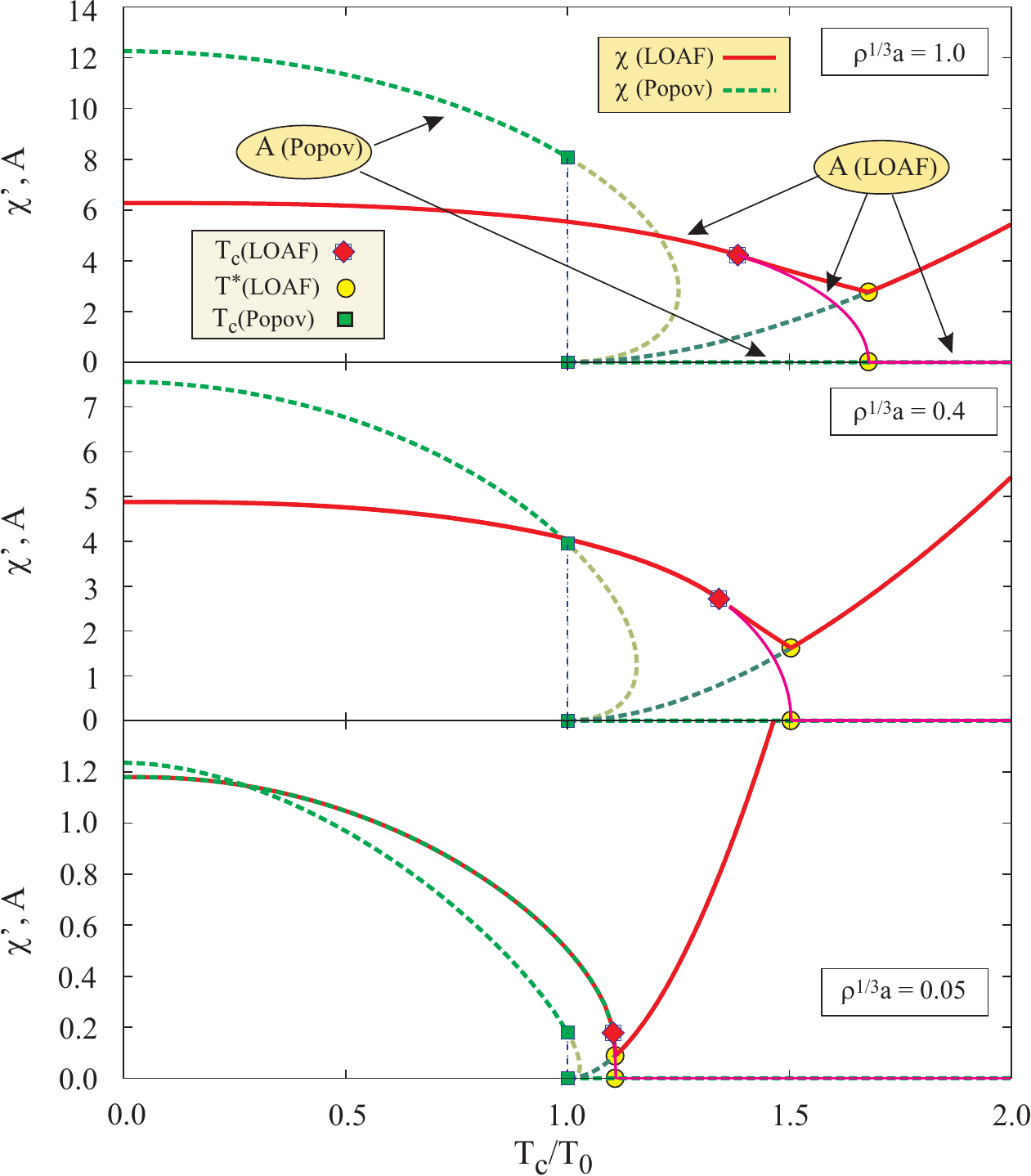}
   \caption{\label{f:densities} (Color online) Normal density, $\chi'$, and anomalous density, $A$, from the LOAF and Popov approximations. The comparison between the LOAF and Popov approximations is carried out for $\rho^{1/3}a = 1$, $\rho^{1/3}a = 0.4$, and $\rho^{1/3}a = 0.05$. 
   $T_c$ and $T^\star$ indicate vanishing condensate density, $\rho_0$, and anomalous density, $A$, respectively. 
   The Popov approximation leads to a first-order phase transition, whereas LOAF predicts a second-order phase transition.
   We have that $T_c = T^\star$ in the Popov approximation but not in LOAF. 
   In LOAF $\chi'$ and $A$ are equal for $T \le T_c$.
   }
\end{figure}
%
%%%%%%%%%%%%%%%%%%%%%%%%%%%%%%%%%%%%%%%%%%%%%%%%%%%%%%%%%%%%%%
\begin{figure}[t!]
   \centering
   \includegraphics[width=0.9\columnwidth]{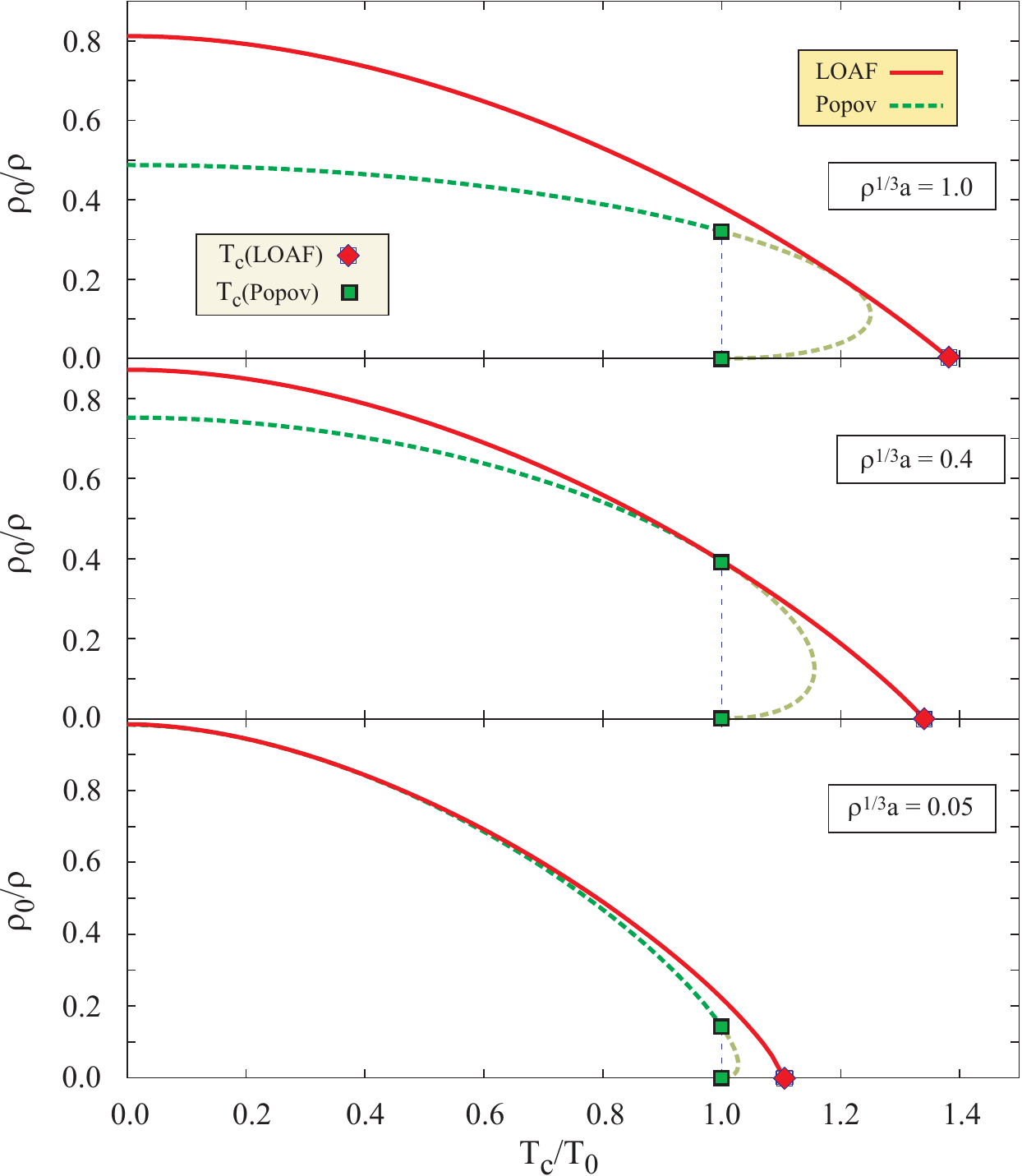}
   \caption{\label{f:rho0} (Color online) Condensate fraction, $\rho_0/\rho$, from the LOAF and Popov approximations. 
   Similarly to Fig.~\ref{f:densities}.
    Because at $T_c$ the Popov approximation and noninteracting dispersion relations are the same, the Popov approximation does not change $T_c$ relative to the noninteracting case.
   LOAF increases $T_c$.
   }
\end{figure}
%
%%%%%%%%%%%%%%%%%%%%%%%%%%%%%%%%%%%%%%%%%%%%%%%%%%%%%%%%%%%%%%
\begin{figure}[t!]
   \centering
   \includegraphics[width=0.9\columnwidth]{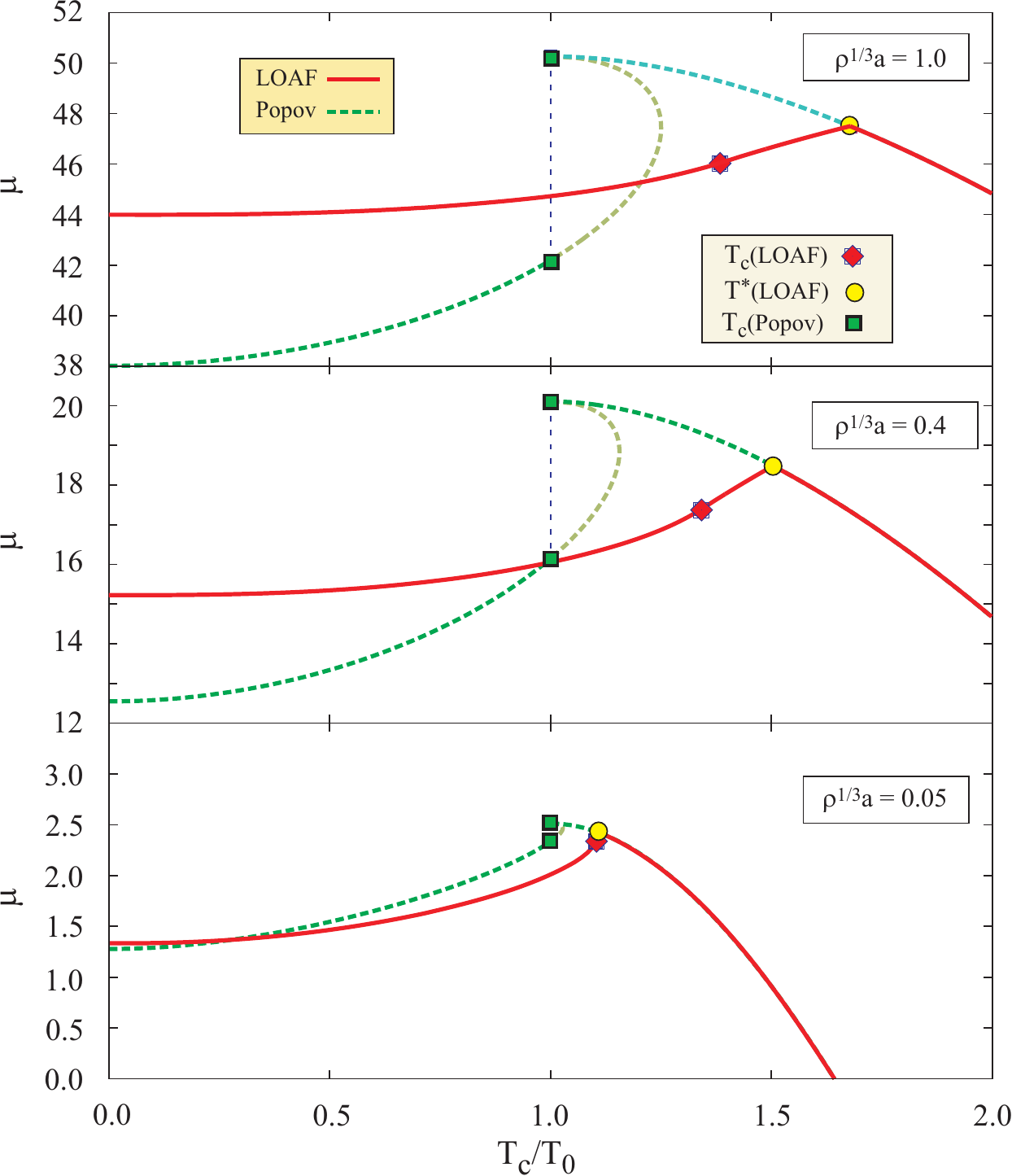}
   \caption{\label{f:mu} (Color online) Chemical potential, $\mu$, from the LOAF and Popov approximations. 
   Similarly to Fig.~\ref{f:densities}.
   }
\end{figure}
%
%%%%%%%%%%%%%%%%%%%%%%%%%%%%%%%%%%%%%%%%%%%%%%%%%%%%%%%%%%%%%%
\begin{figure}[t!]
   \centering
   \includegraphics[width=0.9\columnwidth]{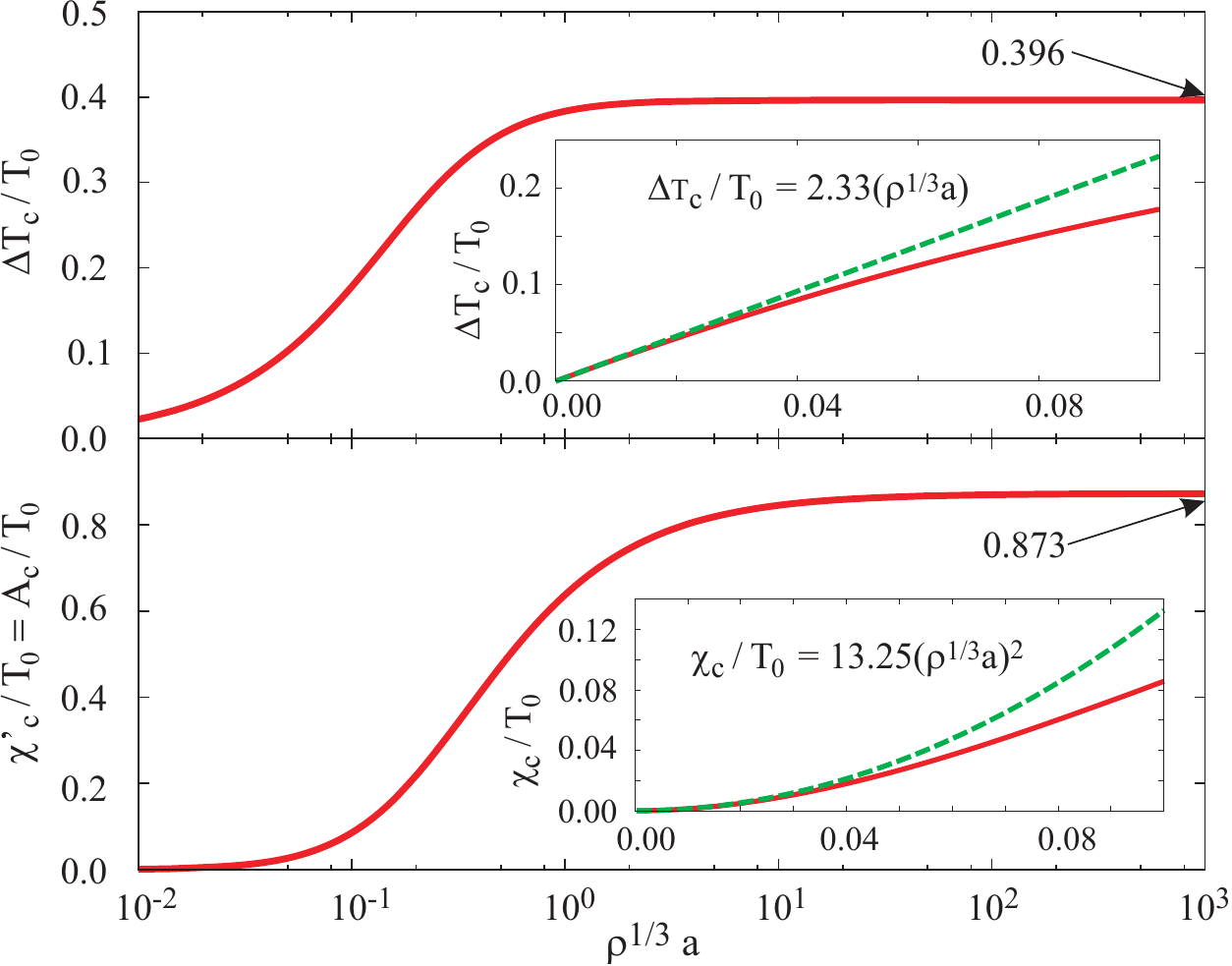}
   \caption{\label{f:unity} (Color online) 
   LOAF predictictions for the critical regime, $T=T_c$, as a function of $\rho^{1/3}a$:
   (Top panel) Relative change in $T_c$ with respect to the noninteracting critical temperature, $T_0$.
   (Bottom panel) Critical value of the normal and anomalous densities, $\chi'_c = A_c$.
   The insets illustrate the $\rho^{1/3}a$ dependence of $\Delta T_c/T_0 = (T_c - T_0)/T_0$ and $\chi'_c$ in the weak-coupling regime. 
   }
\end{figure}
%
%%%%%%%%%%%%%%%%%%%%%%%%%%%%%%%%%%%%%%%%%%%%%%%%%%%%%%%%%%%%%%%%%%%%%%
%

In Fig.~\ref{f:Tzero} we depict the coupling constant dependence of the  zero-temperature values of the normal densities, $\chi'$, condensate fraction, $\rho_0/\rho$, chemical potential, $\mu$, and effective potential, $V_{\mathrm{eff},0}$. The coupling constant depends linearly of the dimensionless parameter,  $\rho^{1/3}a$. We note that in the case of the LOAF approximation, the normal and anomalous densities are equal, $\chi' = A$, whereas in the Popov approximation we have $A = \lambda \rho_0$. Also, in the Popov approximation there is no effective potential, because this approximation is not derivable from an action. At zero temperature, LOAF predicts that the condensate fraction in the unitarity limit is 3/4, whereas in the Popov approximation the condensate fraction approaches zero asymptotically. 

Turning now to the discussion of results in the finite temperature regime, we note that throughout this section, the temperature is scaled by its noninteracting critical value, $T_0 = (2\pi \hbar^2 / m) [\rho/\zeta(3/2)]^{2/3}$, where $\zeta(x)$ is the Riemann zeta function. 
In Fig.~\ref{f:densities} we depict the temperature dependence of the normal density $\chi'$, and anomalous density, $A$,  at constant $\rho^{1/3}a$. We compare the results derived using the LOAF and and the Popov approximations. For illustrative purposes, we show results for $\rho^{1/3}a = 1$, $\rho^{1/3}a = 0.4$, and $\rho^{1/3}a = 0.05$.  Similarly, in Figs.~\ref{f:rho0} and~\ref{f:mu}, we depict the temperature dependence of the condensate fraction, $\rho_0/\rho$,  and chemical potential, $\mu$, respectively, for different interaction strengths. 

We identify two special temperatures, at $T_c$ where the condensate density vanishes, and at $T^\star$ where the anomalous density, $A$, vanishes. These temperatures are the same in the Popov aproximation formalism, but they are different in the LOAF approximation. The existence of a temperature range, $T_c < T < T^\star$, for which the anomalous density, $A$, is nonzero despite a zero condensate fraction, $\rho_0/\rho$, is a fundamental prediction of LOAF. In this temperature range, the dispersion relation departs from the quadratic form predicted by the Popov approximation for $T > T_c$. Above $T_c$ the solution of the Popov-approximation equations becomes multivalued, indicating that the system undergoes a first-order phase transition at $T_c$. In contrast, LOAF predicts a second-order transition. Because at the critical temperature, $T_c$, in the Popov approximations and the noninteracting gas case, the dispersion relations are the same, the Popov approximation does not change $T_c$ relative to the noninteracting case. The LOAF formalism predicts a higher critical temperature than in the noninteracting case, $T_c \ge T_0$. In the weak coupling limit, we wave $T_c \rightarrow T_0$, as $\rho^{1/3}a \rightarrow 0$. 

As illustrated in Figs.~\ref{f:densities},  \ref{f:rho0}, and~\ref{f:mu}, the LOAF and Popov approximations results become qualitatively similar in the weak coupling limit, even though the order of the phase transitions remains different. However, strengthening the interaction between particles in the Bose gas results in enhanced differences between the LOAF and Popov predictions, even for temperatures, $T \ll T_c$.  A larger value of $\rho^{1/3}a$ indicates stronger coupling. 

We also note for comparison purposes, that below $T_c$ the large-$N$ approximation gives the same results as a non-interacting gas. Above $T_c$,  the large-$N$ result above $T_c$ is the same as that of the Popov approximation. Also, above $T^\star$,  where $A=0$ in the LOAF approximation, the large-$N$, Popov and LOAF approximation give the same results. 

In Fig.~\ref{f:unity} we depict the relative change in $T_c$ with respect to $T_0$, $\Delta T_c/T_0 = (T_c - T_0)/T_0$, and the critical value of the normal and anomalous densities, $\chi'_c = A_c$, predicted by LOAF, as a function of  the interaction strength characterized by the dimensionless parameter $\rho^{1/3}a $. The insets show the weak-coupling limit of LOAF results, emphasizing the departure from the noninteracting result in lowest order. 

The leading-order auxiliary formalism, LOAF,  produces a more realistic set of observables away from the weak-coupling limit  because of its non-perturbative character. In contrast,  the Popov approximation is appropriate only in the case of a weakly-interacting gas of bosons. The former is made explicit by studying the LOAF prediction for the relative change, $\Delta T_c/T_0 = (T_c - T_0)/T_0$, as a function of $\rho^{1/3}a $. The inset in the top panel in Fig.~\ref{f:unity} illustrates that in the weak-coupling regime, $\rho^{1/3}a \ll 1$, LOAF produces the same slope, 2.33, for the linear departure as that derived by Baym \emph{et al.}\cite{r:Baym:2000fk} using the large-N expansion, but at next-to-leading order (i.e. they include density fluctuations in their calculation).  The LOAF corrections to the critical temperature are due to the inclusion of self-consistent fluctuations effects in the mean-field $\chi'$ and $A$ densities. We note that carrying that approach to the next order, the slope is reduced to $1.71$~\cite{PhysRevA.62.063604}, and is approaching the Monte Carlo estimates of $1.32\pm0.02$ \cite{PhysRevLett.87.120401,PhysRevE.64.066113,PhysRevE.68.049902}, and $1.29\pm 0.05$~\cite{PhysRevLett.87.120402}.  It will be interesting to see how our next to leading order calculation compares to these results. 
A summary of other  $\Delta T_c/T_0$ theoretical predictions is found in Ref.~\onlinecite{r:Andersen:2004uq}.

As the system approaches the unitarity limit,  LOAF predicts that $\Delta T_c/T_0  \rightarrow 0.396$ and $\chi'_c/T_0 = A_c/T_0  \rightarrow 0.873$ for $\rho^{1/3}a \gg 1$.

%
%
%%%%%%%%%%%%%%%%%%%%%%%%%%%%%%%%%%%%%%%%%%%%%%%%%%%%%%%%%%%%%%%%%%%%%%
%
\section{\label{s:conclusions}Conclusions}
%
%%%%%%%%%%%%%%%%%%%%%%%%%%%%%%%%%%%%%%%%%%%%%%%%%%%%%%%%%%%%%%%%%%%%%%
%

In this paper we discussed in detail a new auxiliary-field formulation for the BEC problem that was first introduced in Ref.~\onlinecite{r:Cooper:2010fk}. At mean-field level this approach
meets three very important criteria  \cite{r:Andersen:2004uq} for a
satisfactory mean-field theory for weakly interacting bosons:
(1) the excitation spectrum should be gapless (Goldstone theorem),
(2) at $T=0$ and weak coupling, it reproduces the known results from Bogoliubov theory, and
(3) it has a smooth second-order phase transition.
The commonly used theories violate those criteria: the Hartree approximation
violates (1), the Bogoliubov and Popov theories violate (3), and the $T$-matrix Popov
theory violates (2).  Also at mean-field level, we obtain a result for $\Delta T_c/T_0 = (T_c - T_0)/T_0$ which was obtain only at next-to-leading order in a large-$N$ expansion, showing that  including the anamolous density in our auxiliary-field formulation is quite important.
This approach will be useful to study both the static and dynamic properties of dilute Bose gases.  %To compare with real experiments where atoms are trapped, one can generalize this method using the local density approximation\cite{r:Pethick:2002fk}.

As described above, one can systematically improve upon the LOAF approximation  discussed here
by  calculating the 1-PI action order-by-order in~$\epsilon$. The broken $U(1)$ symmetry Ward identities guarantee Goldstone's theorem order-by-order in $\epsilon$~\cite{r:Bender:1977bh}. 
For time-dependent problems, however, this expansion is secular\cite{r:MCD01}, and a further resummation is required. The latter is performed using the two-particle irreducible (2-PI) formalism\cite{r:Baym62,r:CJT}. 
The corresponding Schwinger-Dyson (SD) equations for the scalar field and the two-particle correlation functions are simplified dramatically because all vertices are trilinear. 
A~practical implementation of this approach is the bare-vertex approximation (BVA)\cite{r:BCDM01}. The BVA is an energy-momentum and particle-number conserving truncation of the SD infinite hierarchy of equations obtained by ignoring the derivatives of the self-energy, similarly to the Migdal's theorem\cite{r:Migdal:1958uq} approach in condensed matter physics. The BVA proved effective in the case of classical and quantum $\lambda \phi^4$ field theory problems\cite{r:CDM02,r:CDM02ii,r:Mihaila:2003ys} and can be applied to the BEC case. In this context, we note that a related approximation is the 2PI-1/N expansion which has been used in particle theory to study thermalization of various quantum field theories \cite{r:AB01,r:B02,r:AABBS}.  Its use for studying dilute Bose gases was discussed by Calzetta and Hu~\cite{r:Calzetta:2008pb}.  The 2-PI approach has been used also to study the quantum dynamics in the Bose-Hubbard model~\cite{r:ReyHuCalzettaRouraClark03,PhysRevA.75.013613}.

%
%%%%%%%%%%%%%%%%%%%%%%%%%%%%%%%%%%%%%%%%%%%%%%%%%%%%%%%%%%%%%%%%%%%%%%
%
\begin{acknowledgments}
This work was performed in part under the auspices of the U.~S.~Dept.~of Energy.  The authors would like to thank E. Mottola and P.~B.~Littlewood for useful discussions.
\end{acknowledgments}

%
%%%%%%%%%%%%%%%%%%%%%%%%%%%%%%%%%%%%%%%%%%%%%%%%%%%%%%%%%%%%%%%%%%%%%%
%
\appendix

%
%%%%%%%%%%%%%%%%%%%%%%%%%%%%%%%%%%%%%%%%%%%%%%%%%%%%%%%%%%%%%%%%%%%%%%
%
\section{\label{s:renorm}Regularization and renormalization}

Unlike the case of an operator formalism where one can remove vacuum energies by normal ordering, in the path integral method we have to subtract an infinite zero-point vacuum energy $\Veffzero$.  In addition the interaction strength $\lambda_0$ needs to be renormalized to obtain the physical scattering amplitude, as in the Bogoliubov theory for a $\delta$-function interaction.  This is accomplished by summing the Born series to find the physical $s$-wave scattering amplitude.  We will find that regularizing by subtracting the leading divergences in the expression for the potential for the broken symmetry case is equivalent to dimensional regularization, which is known to preserve the Ward identities.  It is also equivalent to renormalizing the vacuum energy, chemical potential, and coupling constant.  

%
%%%%%%%%%%%%%%%%%%%%%%%%%%%%%%%%%%%%%%%%%%%%%%%%%%%%%%%%%%%%%%%%%%%%%%
%
\subsection{\label{ss:DR}Dimensional regularization}
%
%%%%%%%%%%%%%%%%%%%%%%%%%%%%%%%%%%%%%%%%%%%%%%%%%%%%%%%%%%%%%%%%%%%%%%
%

Our regularization scheme of subtracting the leading divergence is equivalent to a  dimensional regularization procedure, which guarantees that the Ward identities of the unrenormalized theory are preserved.  Dimensional regularization consists of evaluating a generalization of the integral in a regime  where it is defined and then analytically continuing to the original ill-defined integral.

In the broken symmetry phase, we need to evaluate an integral of the form 
\begin{equation}\label{dr.e:dint}
   I[M^2]
   =  
   \frac{1}{4 \pi^2} \int_0^\infty  \!\! k^2 \, \rd k \,
   \sqrt{ k^2 ( k^2 + M^2 ) }
   \>.
\end{equation}
If we consider instead the integral
\begin{align}
   &\frac{1}{4 \pi^2} 
   \int_0^\infty  \!\! \rd k \, k^{-\alpha} \,
   ( k^2 + M^2 )^\gamma 
   \\ \notag & =   
   \frac{ [ \, M^2 \, ]^{\gamma +1}}{[ \, M^2 \, ]^{ (\alpha+1)/2} }
   \frac{ 
      \Gamma [ \, ( \, 1 - \alpha \, ) / 2 \, ] \,
      \Gamma [ \, ( \, \alpha - 2 \gamma - 1 \, ) / 2 \, ]
         }{ 8 \pi^2 \Gamma (-\gamma) } \>,
 \end{align}
and then analytically continue this expression to $\alpha=-3$ and $\gamma = 1/2$, we obtain the dimensionally-regularized value of the integral in Eq.~\eqref{dr.e:dint} as
\begin{equation}
  I[M^2]
  = 
  \frac{1}{30 \pi^2} \, [ \, M^2 \, ]^{5/2} \>.
\end{equation}
This is exactly what we obtained by regulating the integral by subtracting the leading divergences, i.e. 
\begin{align*}
   I_{\text{R}}[M^2] 
   &=
   \frac{1}{4 \pi^2} \int_0^\infty \!\!\! k^2 \, \rd k \,
   \Bigl \{ 
      \sqrt{ k^2 ( k^2+M^2 )}
      \\
      & \qquad
      -
      k^2
      -
      \frac{M^2}{2}
      +
      \frac{M^4}{8 k^2} 
   \Bigr \} 
   =  
   \frac{1}{30 \pi^2} \,  [ \, M^2 \, ]^{5/2} \>,
\end{align*}
because the terms we subtracted are formally zero in the dimensional regularization scheme. 

%
%%%%%%%%%%%%%%%%%%%%%%%%%%%%%%%%%%%%%%%%%%%%%%%%%%%%%%%%%%%%%%%%%%%%%%
%
\subsection{\label{ss:R}Renormalization}
%
%%%%%%%%%%%%%%%%%%%%%%%%%%%%%%%%%%%%%%%%%%%%%%%%%%%%%%%%%%%%%%%%%%%%%%
%

In the broken symmetry phase, our regularization scheme of subtracting the leading divergence is also equivalent to renormalizing the vacuum energy, chemical potential, and coupling constant. 

Introducing a cutoff $\Lambda$ in the momentum integrals in Eq.~\eqref{EP.e:V-i}, the effective potential in the broken symmetry case is given by
\begin{align}
   &\Veff[\chi']
   =
   \Veffzero
   -
   \frac{ \mu_0^2 }
        { 2 \lambda_0 \cosh^2\theta }
   -
   \frac{ \mu_0 \chi' }
        { \lambda_0 \cosh^2\theta }
   \label{Re.e:V-i} \\
   & \quad
   +
   \frac{2 \chi^{\prime\,2}}{\lambda_0 \sinh 2\theta }
   + 
   \IntkR
   \Bigl \{ \,
      \frac{\omega_k}{2}
      +
      \frac{1}{\beta}
      \Ln{ 1 - e^{-\beta \omega_k} } \,
   \Bigr \} \>,
   \notag   
\end{align}
and $\omega_k = \sqrt{ \epsilon_k \, ( \epsilon_k + 2 \chi' ) }$.  We first renormalize the interaction strength $\lambda_0$ by setting
\begin{equation}\label{Re.e:lambdaRdef}
   \frac{2}{\lambda_0 \sinh 2\theta} 
   = 
   \frac{2}{\lambdaR \sinh 2\theta} 
   + 
   \IntkR \, \frac{1}{4\epsilon_k} \>.
\end{equation}
Next we renormalize the chemical potential $\mu_0$ by setting
\begin{equation}\label{Re.e:muRdef}
   \frac{\mu_0}{\lambda_0 \cosh^2 \theta}
   = 
   \frac{\muR}{\lambdaR \cosh^2 \theta} 
   +
   \IntkR \, \frac{1}{2} \>.
\end{equation}
The renormalized vacuum energy is then defined by the equation
\begin{equation}\label{Re.e:VzeroRdef}
   \Veffzero
   -
   \frac{ \mu_0^2 }
        { 2 \lambda_0 \cosh^2\theta }
   =
   \VeffzeroR
   -
   \frac{ \muR^2 }
        { 2 \lambdaR \cosh^2\theta }   
   + 
   \frac{1}{2} \IntkR \, \epsilon_k \>,
\end{equation}
so that the effective potential \eqref{Re.e:V-i} becomes
\begin{align}
   &\VeffR[\chi']
   =
   \VeffzeroR
   -
   \frac{ ( \chi' + \muR )^2 }
        { 2 \lambdaR \cosh^2\theta }
   + 
   \frac{\chi^{\prime\,2}}{2 \lambdaR \sinh^2\theta }
   \label{Re.e:VeffR} \\
   & \quad
   +
   \Intk
   \Bigl \{ \,
      \frac{1}{2}
      \Bigl [
         \omega_k
         -
         \chi'
         +
         \frac{\chi^{\prime\,2}}{2 \epsilon_k} \,
      \Bigr ]
      +
      \frac{1}{\beta}
      \Ln{ 1 - e^{-\beta \omega_k} } \,
   \Bigr \} \>,
   \notag   
\end{align}
where we have taken the limit $\Lambda \rightarrow \infty$ since the integral is now finite.
For completeness, we note that the renormalized gap equation \eqref{EP.e:dVdchip} for $\chi'$ is
\begin{align}
   &\frac{\partial \VeffR[\chi']}{\partial \chi'} 
   = 
   \frac{\chi'}{\lambdaR \sinh^2 \theta} 
   -
   \frac{\chi' + \muR}{\lambdaR \cosh^2 \theta} 
   \label{Re.e:dVdchip} \\
   & \qquad
   + 
   \Intk \, 
   \Bigl \{ \,
      \frac{\epsilon_k}{2 \omega_k} \,
      [ \, 2 \, n(\beta\omega_k) + 1 \, ] 
      - 
      \frac{1}{2} 
      + 
      \frac{\chi'}{2 \, \omega_k} \,
   \Bigr \}
   =
   0 \>.
   \notag
\end{align}

%
%%%%%%%%%%%%%%%%%%%%%%%%%%%%%%%%%%%%%%%%%%%%%%%%%%%%%%%%%%%%%%%%%%%%%%
%

\section{\label{s:building}Building blocks for graphs}
%
%%%%%%%%%%%%%%%%%%%%%%%%%%%%%%%%%%%%%%%%%%%%%%%%%%%%%%%%%%%%%%%%%%%%%%
%

Mean-field perturbation theory is an expansion around the stationary point of the effective action and uses the propagators and vertices of the stationary point to construct all the graphs.  The propagators that enter into the loop expansion are the mean-field propagators $\calG^a{}_b[\chi]$, where $\calG^{-1}{}^a{}_b[\chi]$ is given by Eq.~\eqref{af.e:G0invV0def}, and $\calD_{ij}[\Phi]$, where $\calD_{ij}[\Phi]^{-1}$ is defined by Eq.~\eqref{afle.e:Dinverse}. The basic local vertices are the three-point vertex $\calV^i_{ab}$, which connects $\chi_i$ with a $\phi_a$ and $\phi_b$, and the two-point vertex, $\calV^i_{ab} \phi^b$, which changes a $\phi_a$  into a $\chi^i$.  The lowest-order theory also consists of the nonlocal 1-PI  vertices for $N$-$\chi$ lines, namely
\begin{equation}
   \Gamma_N^{i_1,i_2, \ldots i_N} 
   = 
   \frac{\delta^N \Tr{ \Ln{ \calG^{-1}[\chi] } } }
   {\delta \chi_{i_1} \delta \chi_{i _2}  \ldots \delta \chi_{i_N}}
\end{equation}
These nonlocal $N$-$\chi$ vertices are polygons made up of $N$ mean-field propagators $\calG[\chi]$.
Once we have  $\Gamma[\phi_a, \chi_i]$ to some order in $\epsilon$,  we can determine the equations for $\phi$ and $\chi$ from $\delta \Gamma/ \delta \phi_a = j^a$ and $\delta \Gamma/ \delta \chi_i = s^i$. Subsequently, all higher-order 1-PI vertex functions can be obtained by knowing what happens when we differentiate either $\calG$ with respect to $\chi_i$ or $\calD$ with respect to both $\chi_i$ and $\phi_a$. Because we know both $\calG^{-1}$ and $\calD^{-1}$ explicitly, one uses the identity
\begin{equation}
   \frac{\delta A}{\delta \Phi} 
   = 
   - A \circ \frac{\delta A^{-1}}{\delta \Phi} \circ A \,
\end{equation}
to obtain the rules for how to functionally differentiate $\calG$ and $\calD$ in a graph. Here the $\circ$ symbol stands for both an integration and a matrix product.   Using the notation of Eq.~\eqref{af.e:chiSupperdefs} with $\chi_i = \eta_{ij} \, \chi^j$,  we note that
\begin{gather}
   \frac{\delta \chi^i(x)}{\delta \chi^j(x')}
   =
   \delta^i{}_j \, \delta(x,x') \>,
   \quad
   \frac{\delta \chi^i(x)}{\delta \chi_j(x')}
   =
   \eta^{ij} \, \delta(x,x') \>,
   \label{BEC.NLOAF.e:chidervs} \\
   \frac{\delta [ \, \chi_i(x) \chi^i(x) \, ]}{\delta \chi^j(x')}
   =
   2 \, \chi_j(x) \, \delta(x,x') \>.
   \notag
\end{gather}
% $\calG^{-1}[\chi]$ is given by
%\begin{align}
%   &\calG^{-1}[\chi](x,x')
%   \label{BEC.NLOAF.e:Ginvdef} \\
%   &=
%   \delta(x,x')
%   \begin{pmatrix}
%      h + \chi(x) \cosh\theta - \mu & 
%      - A(x) \sinh\theta \\[3pt]
%      - A^{\ast}(x) \sinh\theta & 
%      h^{\ast} + \chi(x) \cosh\theta - \mu 
%   \end{pmatrix} \>,
%   \notag
%\end{align}
%where $h$ is given by Eq.~\eqref{BEC.aux.e:G0invV0def}.
Functional derivatives of $\calG^{-1}[\chi]$ with respect to $\chi^i$ are given in terms of
\begin{equation}\label{BEC.Seff.e:dGinvdchi}
   \frac{ \delta \calG^{-1}[\chi] }{ \delta \chi_i }
   =
   \calV^{i}(\theta) \>,
\end{equation}
where
\begin{subequations}\label{BEC.NLOAF.e:Vimats}
\begin{align}
   \calV^{1}(\theta)
   &=
   \cosh\theta
   \begin{pmatrix}
      1 & 0 \\
      0 & 1
   \end{pmatrix} \>,
   \\
   \calV^{2}(\theta)
   &=
   \sqrt{2} \, \sinh\theta
   \begin{pmatrix}
      0 & 0 \\
      1 & 0
   \end{pmatrix} \>,
   \\
   \calV^{3}(\theta)
   &=
   \sqrt{2} \, \sinh\theta
   \begin{pmatrix}
      0 & 1 \\
      0 & 0
   \end{pmatrix} \>. 
\end{align}
\end{subequations}
In addition, we have
\begin{equation}\label{e:dGdchis}
   \frac{\delta \calG[\chi]}{\delta \chi_i}
   =
   -
   \calG[\chi] \circ \calV^{i}(\theta) \circ \calG[\chi] \>.
\end{equation}
In this notation, the inverse composite-field propagator  $\calD_{ij}^{-1}[\Phi](x,x')$ defined in Eq.~\eqref{afle.e:Dinverse} is given by
\begin{equation}\label{BEC.NLOAF.e:DinverseII}
   \calD_{ij}^{-1}[\Phi](x,x')
   =
   \frac{\eta_{ij} }{\lambda} \, \delta(x,x')
   +
   \Pi_{ij}[\Phi](x,x') \>,
\end{equation}
where the polarization $\Pi{}^{ij}[ \Phi ]$ is
\begin{align}
   \Pi^{ij}[\Phi]
   &=
   \phi \circ \calV^{ij}[\chi](\theta) \circ \phi
   \label{BEC.NLOAF.e:Sigmadef} \\
   & \qquad
   -
   \frac{\hbar}{2i} \,
   \Tr{\calG [\chi] \circ \calV^i(\theta) \circ 
       \calG [\chi] \circ \calV^j(\theta) } \>,
   \notag   
\end{align}
with
\begin{align}
   &\calV^{ij}[\chi](\theta)
   \label{BEC.NLOAF.e:Vijdef} \\
   &=
   \frac{1}{2} \,
   \bigl [ \,    
      \calV^i(\theta) \circ \calG[\chi] \circ \calV^j(\theta) 
      + 
      \calV^j(\theta) \circ \calG[\chi] \circ \calV^i(\theta) \,
   \bigr ] \>.
   \notag
\end{align}
Another quantity we will need for obtaining the graphical rules is  $\calR^{ijk}[\chi]$ defined by  by
\begin{align}
   &\calR^{ijk}[\chi](\theta)
   =
   -\frac{\delta \calV^{ij}[\chi](\theta)}{\delta \chi_k}
   \label{BEC.NLOAF.e:Rijkdef} \\
   &=
   \frac{1}{2} \,
   \bigl \{ \, 
      \calV^i(\theta) \circ \calG[\chi] \circ \calV^k(\theta) 
      \circ \calG[\chi] \circ \calV^j(\theta)
      \notag \\
      & \qquad
      +
      \calV^j(\theta) \circ \calG[\chi] \circ \calV^k(\theta) 
      \circ \calG[\chi] \circ \calV^i(\theta) \,
   \bigr \} \>.
   \notag
\end{align}
Similarly 
\begin{align}
   &\calR^{ijkl}[\chi]
   =
   \frac{\delta \calV^{ij}[\chi]}{\delta \chi_k \, \delta \chi_l}
   \label{BEC.NLOAF.e:Rijkldef} \\
   &=
   \frac{1}{2} \,
   \bigl \{ \, 
      \calV^i \circ \calG[\chi]  \circ \calV^l 
      \circ \calG[\chi] \circ \calV^k \circ \calG\circ \calV^j 
      \notag \\
      & \qquad
      + 
      \calV^i  \circ \calG[\chi] \circ \calV^k\circ \calG\circ \calV^l 
      \circ \calG[\chi] \circ \calV^j  \,   
   \bigr \} 
   \notag \\
   &
   +
   \frac{1}{2} \,
     \bigl \{ \,    \calV^j \circ \calG[\chi] \circ \calV^l \circ \calG\circ \calV^k 
      \circ \calG[\chi] \circ \calV^i
      \notag \\
      & \qquad
      +
      \calV^j \circ \calG[\chi] \circ \calV^k 
      \circ \calG[\chi] \circ \calV^l \circ \calG[\chi] \circ \calV^i \,
   \bigr \} \>.
   \notag
\end{align}
We also define the leading-order 3-$\chi$ 1-PI vertex function, $Q_3^{ijk}[\chi]$, as
\begin{align}
   &Q_3^{ijk}[\chi]
   =
   \frac{\delta \calD^{-1}{}^{ij}[\chi] }{ \delta \chi_{k} }
   =
   \frac{\delta \Pi^{ij}[\chi] }{ \delta \chi_{k} }
   \label{BEC.NLOAF.e:Qdef1} \\
   &=
   {}-
   \phi \circ \calR^{ijk}[\chi](\theta) \circ \phi
   \notag \\
   &
   {}+
   \frac{\hbar}{2i} \,
   \Tr{ \calG[\chi] \circ \calV^k(\theta) \circ 
        \calG[\chi] \circ \calV^i(\theta) \circ
        \calG[\chi] \circ \calV^j(\theta)
        \notag \\
        & \qquad 
        +
        \calG[\chi] \circ \calV^k(\theta) \circ \calG[\chi] 
        \circ \calV^j(\theta) \circ
        \calG[\chi] \circ \calV^i(\theta) } \>.
   \notag   
\end{align} 
The 4-$\chi$ vertex is then given by
\begin{align}
   &Q_4^{ijkl}  
   =  
   \phi \circ \calR^{ijkl}[\chi](\theta) \circ \phi 
   -  
   \frac{\hbar}{2i} \, 
   \Tr{\calG  \circ \calV^i(\theta)  
   \label{e:Q4} \\
   & \qquad
   \circ \calG  \circ  \calV^j(\theta) \circ \calG \circ \calV^k(\theta)  
   \circ \calG  \circ \calV^l(\theta)}  + \text{perms.}
   \notag
\end{align}

With the above definitions we can construct the rules for inserting a $\phi$ or $\chi$ vertex into a graph: Inserting a $\chi$ line into $\calG[\chi]$, we obtain:
\begin{align}
   \frac{\delta \calG^a{}_b[\chi] (x_1,x_2)} {\delta \chi_i(z_1)}
   &= 
   -
   \calG^a{}_c (x_1,z_1) \, \calV^i{}^c{}_d(\theta) \, \calG^d{}_b(z_1,x_2) 
   \label{BB.e:dGdchi} \\
   &= 
   - 
   \calG \circ \calV^i(\theta) \circ \calG \>.
   \notag
\end{align}
Inserting  a $\chi$ line into $\calD[\Phi]$, we obtain
\begin{align}
   &\frac{\delta \calD[\chi \phi]^{i,j}(z_1,z_2)} {\delta \chi_k(z_3 )}
   \label{BB.e:dDdchi} \\
   &=  
   - 
   \int [ \rd z_4 ] \, [ \rd z_5 ] \,
   \calD^{im}(z_1,z_4)  \,
   \frac{\delta \calD^{-1}_{mn}(z_4,z_5) } 
        {\delta \chi_k(z_3)} \,
   \calD^{nj}(z_5, z_2)  
   \notag \\
   &=
   - \calD^{im}  \circ Q_{mn}{}^k \circ  \calD^{nj}
   \notag
\end{align}
We also need to insert a $\phi$ line into $\calD$.  The $2$-$\chi$ $1$-$\phi$ vertex is given by
\begin{align}
   \Gamma^3{}^{ij,a} 
   &=
   \frac {\delta \calD^{-1}{}^{ij} (z_1,z_2)} 
         {\delta  \phi_a(x_1)} 
   = 
   \delta (z_1,x_1) \, \calV^{ij}{}^a{}_d (z_1,z_2) \, \phi^d(z_2) 
   \notag \\
   & 
   + 
   \phi^c(z_1) \, \calV^{ij\,a}{}_c (z_1,z_2) \, \delta (z_2,x_1) \>,
   \notag
\end{align}
and for the $2$-$\chi_i$ $2$-$\phi_a$ vertex we find
\begin{align}
   \Gamma^{4\,ij,ab} 
   &=
   \frac{\delta^2 \calD^{-1}{}^{ij} (z_1,z_2)} 
        {\delta  \phi_a(x_1) \, \delta \phi_b(x_2) }
   \\
   &
   = 
   \delta (z_1,x_1) \, \calV^{ijab}(z_1,z_2) \, \delta(z_2,x_2)
   \notag \\
   & \qquad
   + 
   \delta (z_1,x2) \, \calV^{ijab}(z_1,z_2) \, \delta (z_2,x_1) \>.
   \notag
\end{align}
Thus we obtain
\begin{align}
   &\frac{\delta \calD^{i,j}[\Phi](z_1,z_2)} 
        {\delta \phi_a(x_1)}
   \\ 
   &=  
   - 
   \int  [ \rd z_3 ] \, [ \rd z_4 ] \,
   \calD^{im}(z_1,z_3) \, \Gamma^3_{mn}{}^{a}(z_3,z_4,x_1) \,
   \calD^{nj} (z_4, z_2) 
   \notag \\
   &=  
   -
   \calD^{im}  \circ \Gamma^3_{mn}{}^a \circ  \calD^{nj} \>.
   \notag
\end{align}

%
%%%%%%%%%%%%%%%%%%%%%%%%%%%%%%%%%%%%%%%%%%%%%%%%%%%%%%%%%%%%%%%%%%%%%%
% 
\subsection{\label{ss:invprop}Inverse propagators to order $\epsilon$ }
%
%%%%%%%%%%%%%%%%%%%%%%%%%%%%%%%%%%%%%%%%%%%%%%%%%%%%%%%%%%%%%%%%%%%%%%
%

Using the above rules, we derive the one and two-point vertex function to order $\epsilon$. 
For the one-point function in the presence of sources we have the following two equations:  For $\phi_c^a$ we have
\begin{align}
   j^a 
   &= 
   \frac{\delta \, \Gamma[\Phi]}{\delta \phi_a} 
   \\
   &= 
   \frac{1}{2} \, 
   \bigl [ \,
      \phi \circ \calG^{-1} \,
   \bigr ]^a 
   +
   \frac{1}{2} \,
   \bigl [ \,
      \calG^{-1} \circ \phi \,
   \bigr ]^a
   + 
   \frac{\epsilon}{2} \Tr{ \calD \, \Gamma^{3\,..a} }
   \>,
   \notag
\end{align}
whereas for $\chi_i$ we find
\begin{align}
   s^i 
   &= 
   \frac{\delta \, \Gamma[\Phi]}{\delta \chi_i}
   = 
   \frac{1}{2} \, \phi \circ \calV^i \circ \phi 
   \\
   & \qquad
   - 
   \frac{ \chi^i}{\lambda} +\frac{1}{2i} \Tr{ \calG \circ  \calV^i } 
   +
   \frac{\epsilon}{2i} \Tr{ \calD \circ Q^3 {}^{i..}} \>.
   \notag
\end{align}

\begin{figure}[t]
   \centering
   \includegraphics[width=0.9\columnwidth]{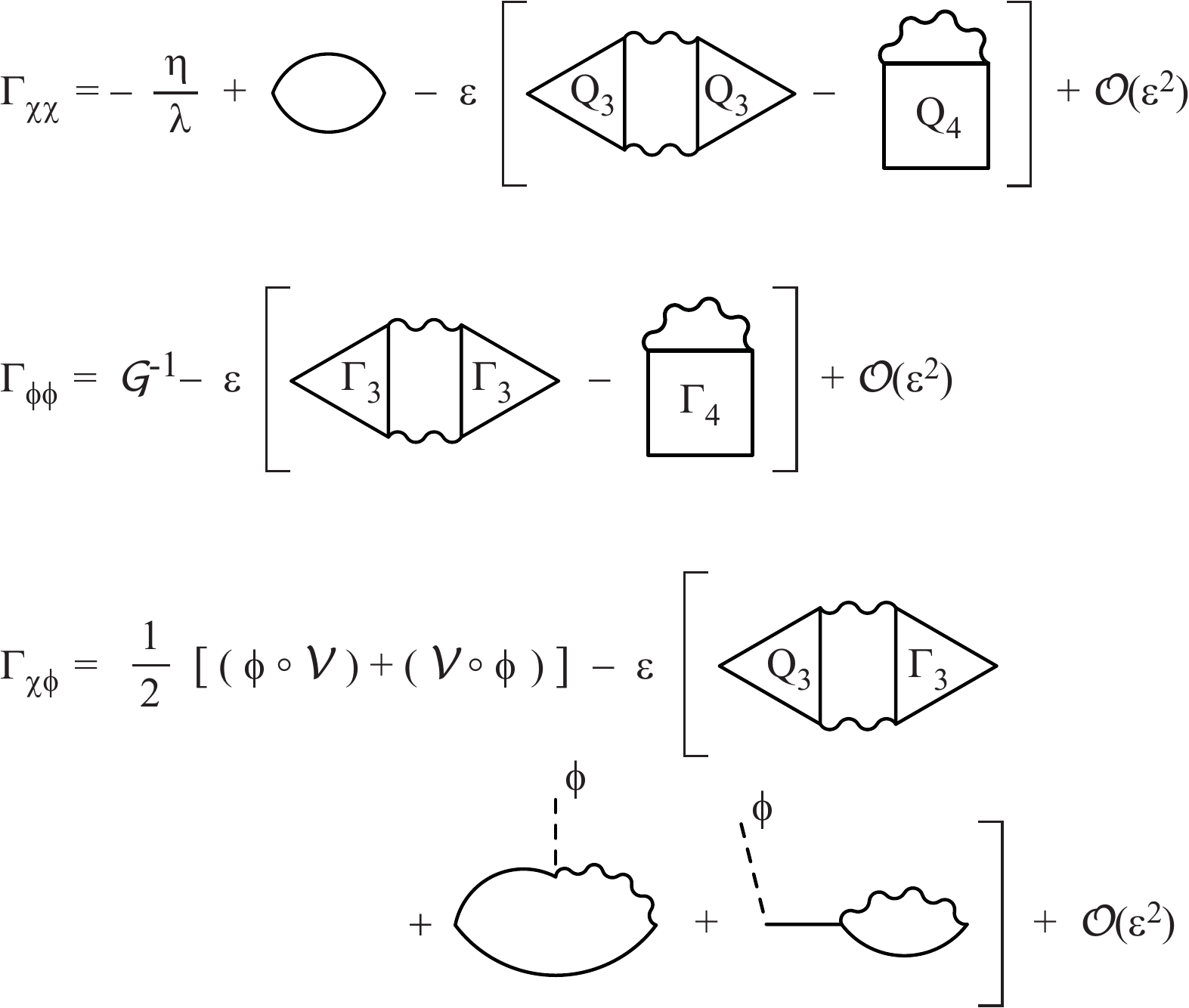}
   \caption{\label{f:dGamma} Feynman diagrams for the second derivatives of~$\Gamma$. Solid and wavy lines correspond to the propagators of~$\phi$ and~$\chi$. Dashed lines denote $\phi$.}
\end{figure}

In turn, for the inverse propagator matrix we have:
\begin{align}
   &\frac{\delta^2 \, \Gamma[\Phi]}{ \delta \chi_i \, \delta \chi_j } 
   = 
   - 
   \calD^{-1}{}^{ij} [\chi, \phi=0] 
   \\
   & \qquad
   - 
   \frac{\epsilon}{2i} \Tr{ \calD \circ Q_3^{i..} \circ \calD \circ Q_3^{j..} }
   +
   \frac{\epsilon}{2i} \Tr{ \calD \circ Q_4^{i...}} \>,
   \notag
\end{align}
and
\begin{align}
   &\frac{\delta \, \Gamma[\Phi]} {\delta \phi_a \, \delta \phi_b} 
   = 
   G_0^{-1}{}^{ab} 
   \\
   & \qquad
   - 
   \frac{\epsilon}{2i} \Tr {D \circ \Gamma_3^{b..} \circ D \circ  \Gamma_3^{a..}} 
   + 
   \frac{\epsilon}{2i} \Tr {D \circ \Gamma_4^{ab..}} \>.
   \notag
\end{align}
The term that mixes $\phi$ and $\chi$ is 
\begin{align}
   &\frac{\delta \, \Gamma[\Phi]} {\delta \phi_a \, \delta \chi_i} 
   = 
   \frac{1}{2} \left[ \phi \circ \calV^i \right]^a
   +  
   \frac{1}{2} \left[ \calV^i \circ \phi \right]^a
   \\
   & \quad
   -
   \frac{\epsilon}{2i} \Tr{ \calD \circ Q_3^{i ..} \circ \calD \circ  \Gamma_3^{a..}} 
   - 
   \frac{\epsilon}{2i} \Tr{ \calD \circ [R_3^{i..} \circ \phi]^a} \>.
   \notag
\end{align}
The propagators for the theory are obtained by inverting the $5 \times 5$ inverse propagator matrix.  Expanding the propagators in a power series in $\epsilon$ and keeping terms to order~$\epsilon$ gives the graphs for the propagators that one would have obtained by working directly with $\ln Z$ to order $\epsilon$.  The Feynman diagrams for the second derivatives of $\Gamma[\Phi]$ are shown in Fig.~\ref{f:dGamma}.
%
%%%%%%%%%%%%%%%%%%%%%%%%%%%%%%%%%%%%%%%%%%%%%%%%%%%%%%%%%%%%%%%%%%%%%%
%
%
%%%%%%%%%%%%%%%%%%%%%%%%%%%%%%%%%%%%%%%%%%%%%%%%%%%%%%%%%%%%%%%%%%%%%%
%

%
%%%%%%%%%%%%%%%%%%%%%%%%%%%%%%%%%%%%%%%%%%%%%%%%%%%%%%%%%%%%%%%%%%%%%%
% 
\subsection{\label{ss:pi0}$\Pi^{ij}[\Phi](x,x')$}
%
%%%%%%%%%%%%%%%%%%%%%%%%%%%%%%%%%%%%%%%%%%%%%%%%%%%%%%%%%%%%%%%%%%%%%%
%

To conclude this section, we complete the calculation of $\Pi^{ij}[\Phi](x,x')$ introduced first in Eq.~\eqref{BEC.NLOAF.e:DinverseII} and explicitly evaluated in Eq.~\eqref{BEC.NLOAF.e:Sigmadef} above.
In the imaginary-time formalism, we first introduce
\begin{align}
   &\calM^{ij}[\chi](x,x')
   \label{BEC.NLOAF.e:trGVGV} \\
   & 
   =
   \frac{1}{2} \,
   \Tr{ \calG[\chi](x',x) \circ \calV^i(\theta) \circ 
        \calG[\chi](x,x') \circ \calV^j(\theta) }
   \notag \\
   & 
   =
   \frac{1}{\beta^2} 
   \iint \frac{\rd^3 k_1 \, \rd^3 k_2}{(2\pi)^6} 
   \sum_{n_1,n_2=-\infty}^{+\infty}
   \notag \\
   & \quad \times
   \frac{1}{2} \,
   \Tr{ \tilde{\calG}[\chi](\bk_2,n_2) \circ \calV^i(\theta) \circ 
        \tilde{\calG}[\chi](\bk_1,n_1) \circ \calV^j(\theta) }
   \notag \\
   & \qquad\qquad
   \times
   e^{i [ \,
      (\bk_1 - \bk_2 ) \cdot ( \br - \br' ) 
      - 
      (\omega_{n_1} - \omega_{n_2} ) ( \tau - \tau' ) \, ] }
   \notag \\
   &=
   \frac{1}{\beta} \int \frac{\rd^3 k}{(2\pi)^3} 
   \sum_{n=-\infty}^{+\infty}
   e^{i [ \, \bk \cdot ( \br - \br' ) - \omega_{n} ( \tau - \tau' ) \, ] }
   \notag \\
   & \qquad
   \frac{1}{\beta} 
   \iint \frac{\rd^3 k_1 \, \rd^3 k_2}{(2\pi)^6} \,
   \sum_{n_1,n_2}
   (2\pi)^3 \, \delta( \bk - \bk_1 + \bk_2 ) \,
   \delta_{n,n_1 - n_2} \, 
   \notag \\
   &\times
   \frac{1}{2} \,
   \Tr{ \tilde{\calG}[\chi](\bk_2,n_2) \circ \calV^i(\theta) \circ 
        \tilde{\calG}[\chi](\bk_1,n_1) \circ \calV^j(\theta) } \>.
   \notag 
\end{align}
Expanding $\calM^{ij}[\chi](x,x')$ in a Fourier series,
\begin{align}
   & \calM^{ij}[\chi](x,x')
   \label{BEC.NLOAF.e:trGVGVxx} \\ 
   &=
   \frac{1}{\beta} \int \frac{\rd^3 k}{(2\pi)^3} 
   \sum_{n=-\infty}^{+\infty}
   \tilde{\calM}^{ij}[\chi](\bk,n) \,
   e^{i [ \, \bk \cdot ( \br - \br' ) - \omega_{n} ( \tau - \tau' ) \, ] } \>,
   \notag 
\end{align}
where from \eqref{BEC.NLOAF.e:trGVGV}, $\tilde{\calM}^{ij}[\chi](\bk,n)$ is given by the convolution integral,
\begin{align}
   &\tilde{\calM}^{ij}[\chi](\bk,n)
   =
   \frac{1}{2\beta} \int \frac{\rd^3 k'}{(2\pi)^3} 
   \sum_{n'=-\infty}^{+\infty}
   \label{BEC.NLOAF.e:tMdef} \\
   & \times
   \Tr{ \tilde{\calG}[\chi](\bk - \bk',n - n') \circ \calV^i(\theta) \circ 
        \tilde{\calG}[\chi](\bk',n') \circ \calV^j(\theta) } \>.
   \notag
\end{align}
From Eqs.~\eqref{BEC.NLOAF.e:Sigmadef}, \eqref{BEC.NLOAF.e:Vijdef}, and \eqref{BEC.NLOAF.e:trGVGV}, we then have
\begin{equation}\label{BEC.NLOAF.e:tSigma}
   \tilde{\Pi}^{ij}[\Phi](\bk,n)
   =
   \phi \circ \tilde{\calV}^{ij}[\chi](\bk,n) \circ \phi
   -
   \tilde{\calM}^{ij}[\chi](\bk,n) \>,
\end{equation}
with
\begin{align}
   &\tilde{\calV}^{ij}[\chi](\bk,n)
   =
   \frac{1}{2} \,
   \bigl [ \,    
      \calV^i(\theta) \circ \tilde{\calG}[\chi](\bk,n) \circ \calV^j(\theta)
      \label{BEC.NLOAF.e:tcalVij} \\
      & \qquad\qquad
      + 
      \calV^j(\theta) \circ \tilde{\calG}[\chi](-\bk,-n) \circ \calV^i(\theta) \,
   \bigr ] \>.   
   \notag
\end{align}

%
%%%%%%%%%%%%%%%%%%%%%%%%%%%%%%%%%%%%%%%%%%%%%%%%%%%%%%%%%%%%%%%%%%%%%%
%
\bibliography{johns}

\begin{thebibliography}{62}%
\makeatletter
\providecommand \@ifxundefined [1]{%
 \@ifx{#1\undefined}
}%
\providecommand \@ifnum [1]{%
 \ifnum #1\expandafter \@firstoftwo
 \else \expandafter \@secondoftwo
 \fi
}%
\providecommand \@ifx [1]{%
 \ifx #1\expandafter \@firstoftwo
 \else \expandafter \@secondoftwo
 \fi
}%
\providecommand \natexlab [1]{#1}%
\providecommand \enquote  [1]{``#1''}%
\providecommand \bibnamefont  [1]{#1}%
\providecommand \bibfnamefont [1]{#1}%
\providecommand \citenamefont [1]{#1}%
\providecommand \href@noop [0]{\@secondoftwo}%
\providecommand \href [0]{\begingroup \@sanitize@url \@href}%
\providecommand \@href[1]{\@@startlink{#1}\@@href}%
\providecommand \@@href[1]{\endgroup#1\@@endlink}%
\providecommand \@sanitize@url [0]{\catcode `\\12\catcode `\$12\catcode
  `\&12\catcode `\#12\catcode `\^12\catcode `\_12\catcode `\%12\relax}%
\providecommand \@@startlink[1]{}%
\providecommand \@@endlink[0]{}%
\providecommand \url  [0]{\begingroup\@sanitize@url \@url }%
\providecommand \@url [1]{\endgroup\@href {#1}{\urlprefix }}%
\providecommand \urlprefix  [0]{URL }%
\providecommand \Eprint [0]{\href }%
\providecommand \doibase [0]{http://dx.doi.org/}%
\providecommand \selectlanguage [0]{\@gobble}%
\providecommand \bibinfo  [0]{\@secondoftwo}%
\providecommand \bibfield  [0]{\@secondoftwo}%
\providecommand \translation [1]{[#1]}%
\providecommand \BibitemOpen [0]{}%
\providecommand \bibitemStop [0]{}%
\providecommand \bibitemNoStop [0]{.\EOS\space}%
\providecommand \EOS [0]{\spacefactor3000\relax}%
\providecommand \BibitemShut  [1]{\csname bibitem#1\endcsname}%
\let\auto@bib@innerbib\@empty
%</preamble>
\bibitem [{\citenamefont {Onnes}(1911)}]{r:Kamerling:1911fk}%
  \BibitemOpen
  \bibfield  {author} {\bibinfo {author} {\bibfnamefont {H.~K.}\ \bibnamefont
  {Onnes}},\ }\href@noop {} {\bibfield  {journal} {\bibinfo  {journal} {Proc.
  Roy. Acad. Amsterdam}\ }\textbf {\bibinfo {volume} {13}},\ \bibinfo {pages}
  {1903} (\bibinfo {year} {1911})}\BibitemShut {NoStop}%
\bibitem [{\citenamefont {Kapitza}(1938)}]{r:Kapitza:1938kx}%
  \BibitemOpen
  \bibfield  {author} {\bibinfo {author} {\bibfnamefont {P.~L.}\ \bibnamefont
  {Kapitza}},\ }\href@noop {} {\bibfield  {journal} {\bibinfo  {journal}
  {Nature}\ }\textbf {\bibinfo {volume} {141}},\ \bibinfo {pages} {74}
  (\bibinfo {year} {1938})}\BibitemShut {NoStop}%
\bibitem [{\citenamefont {Allen}\ and\ \citenamefont
  {Misener}(1938)}]{r:Allen:1938vn}%
  \BibitemOpen
  \bibfield  {author} {\bibinfo {author} {\bibfnamefont {J.~F.}\ \bibnamefont
  {Allen}}\ and\ \bibinfo {author} {\bibfnamefont {A.~D.}\ \bibnamefont
  {Misener}},\ }\href@noop {} {\bibfield  {journal} {\bibinfo  {journal}
  {Nature}\ }\textbf {\bibinfo {volume} {141}},\ \bibinfo {pages} {75}
  (\bibinfo {year} {1938})}\BibitemShut {NoStop}%
\bibitem [{\citenamefont {London}(1938{\natexlab{a}})}]{r:London:1938ys}%
  \BibitemOpen
  \bibfield  {author} {\bibinfo {author} {\bibfnamefont {F.}~\bibnamefont
  {London}},\ }\href@noop {} {\bibfield  {journal} {\bibinfo  {journal}
  {Nature}\ }\textbf {\bibinfo {volume} {141}},\ \bibinfo {pages} {643}
  (\bibinfo {year} {1938}{\natexlab{a}})}\BibitemShut {NoStop}%
\bibitem [{\citenamefont {London}(1938{\natexlab{b}})}]{r:London:1938zr}%
  \BibitemOpen
  \bibfield  {author} {\bibinfo {author} {\bibfnamefont {F.}~\bibnamefont
  {London}},\ }\href@noop {} {\bibfield  {journal} {\bibinfo  {journal} {Phys.
  Rev.}\ }\textbf {\bibinfo {volume} {54}},\ \bibinfo {pages} {947} (\bibinfo
  {year} {1938}{\natexlab{b}})}\BibitemShut {NoStop}%
\bibitem [{\citenamefont {Bogoliubov}(1947)}]{r:Bogoliubov:1947ys}%
  \BibitemOpen
  \bibfield  {author} {\bibinfo {author} {\bibfnamefont {N.~N.}\ \bibnamefont
  {Bogoliubov}},\ }\href@noop {} {\bibfield  {journal} {\bibinfo  {journal} {J.
  Phys. USSR}\ }\textbf {\bibinfo {volume} {11}},\ \bibinfo {pages} {23}
  (\bibinfo {year} {1947})}\BibitemShut {NoStop}%
\bibitem [{\citenamefont {Landau}(1941)}]{r:Landau:1941ly}%
  \BibitemOpen
  \bibfield  {author} {\bibinfo {author} {\bibfnamefont {L.~D.}\ \bibnamefont
  {Landau}},\ }\href@noop {} {\bibfield  {journal} {\bibinfo  {journal} {J.
  Phys. USSR}\ }\textbf {\bibinfo {volume} {5}},\ \bibinfo {pages} {71}
  (\bibinfo {year} {1941})}\BibitemShut {NoStop}%
\bibitem [{\citenamefont {Lee}\ \emph {et~al.}(1957)\citenamefont {Lee},
  \citenamefont {Huang},\ and\ \citenamefont {Yang}}]{r:Lee:1957ve}%
  \BibitemOpen
  \bibfield  {author} {\bibinfo {author} {\bibfnamefont {T.~D.}\ \bibnamefont
  {Lee}}, \bibinfo {author} {\bibfnamefont {K.}~\bibnamefont {Huang}}, \ and\
  \bibinfo {author} {\bibfnamefont {C.~N.}\ \bibnamefont {Yang}},\ }\href@noop
  {} {\bibfield  {journal} {\bibinfo  {journal} {Phys. Rev.}\ }\textbf
  {\bibinfo {volume} {106}},\ \bibinfo {pages} {1135} (\bibinfo {year}
  {1957})}\BibitemShut {NoStop}%
\bibitem [{\citenamefont {Fedichev}\ \emph {et~al.}(1996)\citenamefont
  {Fedichev}, \citenamefont {Reynolds},\ and\ \citenamefont
  {Shlyapnikov}}]{r:Fedichev:1996hc}%
  \BibitemOpen
  \bibfield  {author} {\bibinfo {author} {\bibfnamefont {P.~O.}\ \bibnamefont
  {Fedichev}}, \bibinfo {author} {\bibfnamefont {M.~W.}\ \bibnamefont
  {Reynolds}}, \ and\ \bibinfo {author} {\bibfnamefont {G.~V.}\ \bibnamefont
  {Shlyapnikov}},\ }\href@noop {} {\bibfield  {journal} {\bibinfo  {journal}
  {Phys. Rev. Lett.}\ }\textbf {\bibinfo {volume} {77}},\ \bibinfo {pages}
  {2921} (\bibinfo {year} {1996})}\BibitemShut {NoStop}%
\bibitem [{\citenamefont {Esry}\ \emph {et~al.}(1999)\citenamefont {Esry},
  \citenamefont {Greene},\ and\ \citenamefont {Burke}}]{r:Esry:1991ij}%
  \BibitemOpen
  \bibfield  {author} {\bibinfo {author} {\bibfnamefont {B.~D.}\ \bibnamefont
  {Esry}}, \bibinfo {author} {\bibfnamefont {C.~H.}\ \bibnamefont {Greene}}, \
  and\ \bibinfo {author} {\bibfnamefont {J.~P.}\ \bibnamefont {Burke}},\
  }\href@noop {} {\bibfield  {journal} {\bibinfo  {journal} {Phys. Rev. Lett.}\
  }\textbf {\bibinfo {volume} {83}},\ \bibinfo {pages} {1751} (\bibinfo {year}
  {1999})}\BibitemShut {NoStop}%
\bibitem [{\citenamefont {Shin}\ \emph {et~al.}(2007)\citenamefont {Shin},
  \citenamefont {Schunck}, \citenamefont {Schirotzek},\ and\ \citenamefont
  {Ketterle}}]{r:Shin:2007oq}%
  \BibitemOpen
  \bibfield  {author} {\bibinfo {author} {\bibfnamefont {Y.}~\bibnamefont
  {Shin}}, \bibinfo {author} {\bibfnamefont {C.~H.}\ \bibnamefont {Schunck}},
  \bibinfo {author} {\bibfnamefont {A.}~\bibnamefont {Schirotzek}}, \ and\
  \bibinfo {author} {\bibfnamefont {W.}~\bibnamefont {Ketterle}},\ }\href@noop
  {} {\bibfield  {journal} {\bibinfo  {journal} {Phys. Rev. Lett.}\ }\textbf
  {\bibinfo {volume} {99}},\ \bibinfo {pages} {090403} (\bibinfo {year}
  {2007})}\BibitemShut {NoStop}%
\bibitem [{\citenamefont {Daley}\ \emph {et~al.}(2009)\citenamefont {Daley},
  \citenamefont {Taylor}, \citenamefont {Diehl}, \citenamefont {Baranov},\ and\
  \citenamefont {Zoller}}]{r:Daley:2009bs}%
  \BibitemOpen
  \bibfield  {author} {\bibinfo {author} {\bibfnamefont {A.~J.}\ \bibnamefont
  {Daley}}, \bibinfo {author} {\bibfnamefont {J.~M.}\ \bibnamefont {Taylor}},
  \bibinfo {author} {\bibfnamefont {S.}~\bibnamefont {Diehl}}, \bibinfo
  {author} {\bibfnamefont {M.}~\bibnamefont {Baranov}}, \ and\ \bibinfo
  {author} {\bibfnamefont {P.}~\bibnamefont {Zoller}},\ }\href@noop {}
  {\bibfield  {journal} {\bibinfo  {journal} {Phys. Rev. Lett.}\ }\textbf
  {\bibinfo {volume} {102}},\ \bibinfo {pages} {040402} (\bibinfo {year}
  {2009})}\BibitemShut {NoStop}%
\bibitem [{\citenamefont {Henderson}\ \emph {et~al.}(2006)\citenamefont
  {Henderson}, \citenamefont {Kelkar}, \citenamefont {Lee}, \citenamefont
  {Gutirez-Medina},\ and\ \citenamefont {Raizen}}]{r:Henderson:2006fv}%
  \BibitemOpen
  \bibfield  {author} {\bibinfo {author} {\bibfnamefont {K.}~\bibnamefont
  {Henderson}}, \bibinfo {author} {\bibfnamefont {H.}~\bibnamefont {Kelkar}},
  \bibinfo {author} {\bibfnamefont {T.~C.}\ \bibnamefont {Lee}}, \bibinfo
  {author} {\bibfnamefont {B.}~\bibnamefont {Gutirez-Medina}}, \ and\ \bibinfo
  {author} {\bibfnamefont {M.~G.}\ \bibnamefont {Raizen}},\ }\href@noop {}
  {\bibfield  {journal} {\bibinfo  {journal} {Europhys. Lett.}\ }\textbf
  {\bibinfo {volume} {75}},\ \bibinfo {pages} {392} (\bibinfo {year}
  {2006})}\BibitemShut {NoStop}%
\bibitem [{\citenamefont {Henderson}\ \emph {et~al.}(2009)\citenamefont
  {Henderson}, \citenamefont {Ryu}, \citenamefont {Mac{C}ormic},\ and\
  \citenamefont {Boshier}}]{r:Henderson:2009dz}%
  \BibitemOpen
  \bibfield  {author} {\bibinfo {author} {\bibfnamefont {K.}~\bibnamefont
  {Henderson}}, \bibinfo {author} {\bibfnamefont {C.}~\bibnamefont {Ryu}},
  \bibinfo {author} {\bibfnamefont {C.}~\bibnamefont {Mac{C}ormic}}, \ and\
  \bibinfo {author} {\bibfnamefont {M.}~\bibnamefont {Boshier}},\ }\href@noop
  {} {\bibfield  {journal} {\bibinfo  {journal} {New J. Phys.}\ }\textbf
  {\bibinfo {volume} {11}},\ \bibinfo {pages} {043030} (\bibinfo {year}
  {2009})}\BibitemShut {NoStop}%
\bibitem [{\citenamefont {Hohenberg}\ and\ \citenamefont
  {Martin}(1965)}]{r:Hohenberg:1965fu}%
  \BibitemOpen
  \bibfield  {author} {\bibinfo {author} {\bibfnamefont {P.~C.}\ \bibnamefont
  {Hohenberg}}\ and\ \bibinfo {author} {\bibfnamefont {P.~C.}\ \bibnamefont
  {Martin}},\ }\href@noop {} {\bibfield  {journal} {\bibinfo  {journal} {Ann.
  Phys.}\ }\textbf {\bibinfo {volume} {34}},\ \bibinfo {pages} {291} (\bibinfo
  {year} {1965})}\BibitemShut {NoStop}%
\bibitem [{\citenamefont {Andersen}(2004)}]{r:Andersen:2004uq}%
  \BibitemOpen
  \bibfield  {author} {\bibinfo {author} {\bibfnamefont {J.~O.}\ \bibnamefont
  {Andersen}},\ }\href@noop {} {\bibfield  {journal} {\bibinfo  {journal}
  {Revs. Mod. Phys.}\ }\textbf {\bibinfo {volume} {76}},\ \bibinfo {pages}
  {599} (\bibinfo {year} {2004})}\BibitemShut {NoStop}%
\bibitem [{\citenamefont {Toyoda}(1982)}]{r:Toyoda:1982pi}%
  \BibitemOpen
  \bibfield  {author} {\bibinfo {author} {\bibfnamefont {T.}~\bibnamefont
  {Toyoda}},\ }\href@noop {} {\bibfield  {journal} {\bibinfo  {journal} {Ann.
  Phys.}\ }\textbf {\bibinfo {volume} {141}},\ \bibinfo {pages} {154} (\bibinfo
  {year} {1982})}\BibitemShut {NoStop}%
\bibitem [{\citenamefont {Huang}(1999)}]{r:Huang:1999qa}%
  \BibitemOpen
  \bibfield  {author} {\bibinfo {author} {\bibfnamefont {K.}~\bibnamefont
  {Huang}},\ }\href@noop {} {\bibfield  {journal} {\bibinfo  {journal} {Phys.
  Rev. Lett.}\ }\textbf {\bibinfo {volume} {83}},\ \bibinfo {pages} {3770}
  (\bibinfo {year} {1999})}\BibitemShut {NoStop}%
\bibitem [{\citenamefont {Baym}\ \emph {et~al.}(1999)\citenamefont {Baym},
  \citenamefont {Blaizot}, \citenamefont {Holzmann}, \citenamefont {Laloe},\
  and\ \citenamefont {Vautherin}}]{r:Baym:1999mi}%
  \BibitemOpen
  \bibfield  {author} {\bibinfo {author} {\bibfnamefont {G.}~\bibnamefont
  {Baym}}, \bibinfo {author} {\bibfnamefont {J.-P.}\ \bibnamefont {Blaizot}},
  \bibinfo {author} {\bibfnamefont {M.}~\bibnamefont {Holzmann}}, \bibinfo
  {author} {\bibfnamefont {F.}~\bibnamefont {Laloe}}, \ and\ \bibinfo {author}
  {\bibfnamefont {D.}~\bibnamefont {Vautherin}},\ }\href@noop {} {\bibfield
  {journal} {\bibinfo  {journal} {Phys. Rev. Lett.}\ }\textbf {\bibinfo
  {volume} {83}},\ \bibinfo {pages} {1703} (\bibinfo {year}
  {1999})}\BibitemShut {NoStop}%
\bibitem [{\citenamefont {Baym}\ \emph {et~al.}(2000)\citenamefont {Baym},
  \citenamefont {Blaizot},\ and\ \citenamefont {Zinn-Justin}}]{r:Baym:2000fk}%
  \BibitemOpen
  \bibfield  {author} {\bibinfo {author} {\bibfnamefont {G.}~\bibnamefont
  {Baym}}, \bibinfo {author} {\bibfnamefont {J.-P.}\ \bibnamefont {Blaizot}}, \
  and\ \bibinfo {author} {\bibfnamefont {J.}~\bibnamefont {Zinn-Justin}},\
  }\href@noop {} {\bibfield  {journal} {\bibinfo  {journal} {Europhys. Lett.}\
  }\textbf {\bibinfo {volume} {49}},\ \bibinfo {pages} {150} (\bibinfo {year}
  {2000})}\BibitemShut {NoStop}%
\bibitem [{\citenamefont {Cooper}\ \emph {et~al.}(2010)\citenamefont {Cooper},
  \citenamefont {Chien}, \citenamefont {Mihaila}, \citenamefont {Dawson},\ and\
  \citenamefont {Timmermans}}]{r:Cooper:2010fk}%
  \BibitemOpen
  \bibfield  {author} {\bibinfo {author} {\bibfnamefont {F.}~\bibnamefont
  {Cooper}}, \bibinfo {author} {\bibfnamefont {C.-C.}\ \bibnamefont {Chien}},
  \bibinfo {author} {\bibfnamefont {B.}~\bibnamefont {Mihaila}}, \bibinfo
  {author} {\bibfnamefont {J.~F.}\ \bibnamefont {Dawson}}, \ and\ \bibinfo
  {author} {\bibfnamefont {E.~M.}\ \bibnamefont {Timmermans}},\ }\href@noop {}
  {\bibfield  {journal} {\bibinfo  {journal} {Phys. Rev. Lett.}\ }\textbf
  {\bibinfo {volume} {105}},\ \bibinfo {pages} {240402} (\bibinfo {year}
  {2010})}\BibitemShut {NoStop}%
\bibitem [{\citenamefont {Bender}\ \emph {et~al.}(1977)\citenamefont {Bender},
  \citenamefont {Cooper},\ and\ \citenamefont {Guralnik}}]{r:Bender:1977bh}%
  \BibitemOpen
  \bibfield  {author} {\bibinfo {author} {\bibfnamefont {C.}~\bibnamefont
  {Bender}}, \bibinfo {author} {\bibfnamefont {F.}~\bibnamefont {Cooper}}, \
  and\ \bibinfo {author} {\bibfnamefont {G.}~\bibnamefont {Guralnik}},\
  }\href@noop {} {\bibfield  {journal} {\bibinfo  {journal} {Ann. Phys.}\
  }\textbf {\bibinfo {volume} {109}},\ \bibinfo {pages} {165} (\bibinfo {year}
  {1977})}\BibitemShut {NoStop}%
\bibitem [{\citenamefont {Hubbard}(1959)}]{r:Hubbard:1959kx}%
  \BibitemOpen
  \bibfield  {author} {\bibinfo {author} {\bibfnamefont {J.}~\bibnamefont
  {Hubbard}},\ }\href@noop {} {\bibfield  {journal} {\bibinfo  {journal} {Phys.
  Rev. Lett.}\ }\textbf {\bibinfo {volume} {3}},\ \bibinfo {pages} {77}
  (\bibinfo {year} {1959})}\BibitemShut {NoStop}%
\bibitem [{\citenamefont {Stratonovich}(1958)}]{r:Stratonovich:1958vn}%
  \BibitemOpen
  \bibfield  {author} {\bibinfo {author} {\bibfnamefont {R.~L.}\ \bibnamefont
  {Stratonovich}},\ }\href@noop {} {\bibfield  {journal} {\bibinfo  {journal}
  {Doklady}\ }\textbf {\bibinfo {volume} {2}},\ \bibinfo {pages} {416}
  (\bibinfo {year} {1958})}\BibitemShut {NoStop}%
\bibitem [{\citenamefont {Coleman}\ \emph {et~al.}(1974)\citenamefont
  {Coleman}, \citenamefont {Jackiw},\ and\ \citenamefont
  {Politzer}}]{r:Coleman:1974ve}%
  \BibitemOpen
  \bibfield  {author} {\bibinfo {author} {\bibfnamefont {S.}~\bibnamefont
  {Coleman}}, \bibinfo {author} {\bibfnamefont {R.}~\bibnamefont {Jackiw}}, \
  and\ \bibinfo {author} {\bibfnamefont {H.~D.}\ \bibnamefont {Politzer}},\
  }\href@noop {} {\bibfield  {journal} {\bibinfo  {journal} {Phys. Rev. D}\
  }\textbf {\bibinfo {volume} {10}},\ \bibinfo {pages} {2491} (\bibinfo {year}
  {1974})}\BibitemShut {NoStop}%
\bibitem [{\citenamefont {Root}(1974)}]{r:Root:1974qf}%
  \BibitemOpen
  \bibfield  {author} {\bibinfo {author} {\bibfnamefont {R.}~\bibnamefont
  {Root}},\ }\href@noop {} {\bibfield  {journal} {\bibinfo  {journal} {Phys.
  Rev. D}\ }\textbf {\bibinfo {volume} {10}},\ \bibinfo {pages} {3322}
  (\bibinfo {year} {1974})}\BibitemShut {NoStop}%
\bibitem [{\citenamefont {{S{\'a} de Melo}}\ \emph {et~al.}(1993)\citenamefont
  {{S{\'a} de Melo}}, \citenamefont {Randeria},\ and\ \citenamefont
  {Engelbrecht}}]{r:Melo:1993vn}%
  \BibitemOpen
  \bibfield  {author} {\bibinfo {author} {\bibfnamefont {C.~A.~R.}\
  \bibnamefont {{S{\'a} de Melo}}}, \bibinfo {author} {\bibfnamefont
  {M.}~\bibnamefont {Randeria}}, \ and\ \bibinfo {author} {\bibfnamefont
  {J.~R.}\ \bibnamefont {Engelbrecht}},\ }\href@noop {} {\bibfield  {journal}
  {\bibinfo  {journal} {Phys. Rev. Lett.}\ }\textbf {\bibinfo {volume} {71}},\
  \bibinfo {pages} {3202} (\bibinfo {year} {1993})}\BibitemShut {NoStop}%
\bibitem [{\citenamefont {Engelbrecht}\ \emph {et~al.}(1997)\citenamefont
  {Engelbrecht}, \citenamefont {Randeria},\ and\ \citenamefont {{S{\'a} de
  Melo}}}]{r:Engelbrecht:1997fk}%
  \BibitemOpen
  \bibfield  {author} {\bibinfo {author} {\bibfnamefont {J.~R.}\ \bibnamefont
  {Engelbrecht}}, \bibinfo {author} {\bibfnamefont {M.}~\bibnamefont
  {Randeria}}, \ and\ \bibinfo {author} {\bibfnamefont {C.~A.~R.}\ \bibnamefont
  {{S{\'a} de Melo}}},\ }\href@noop {} {\bibfield  {journal} {\bibinfo
  {journal} {Phys. Rev. B}\ }\textbf {\bibinfo {volume} {55}},\ \bibinfo
  {pages} {15153} (\bibinfo {year} {1997})}\BibitemShut {NoStop}%
\bibitem [{\citenamefont {Floerchinger}\ \emph {et~al.}(2008)\citenamefont
  {Floerchinger}, \citenamefont {Scherer}, \citenamefont {Diehl},\ and\
  \citenamefont {Wetterich}}]{r:Floerchinger:2008kxx}%
  \BibitemOpen
  \bibfield  {author} {\bibinfo {author} {\bibfnamefont {S.}~\bibnamefont
  {Floerchinger}}, \bibinfo {author} {\bibfnamefont {M.}~\bibnamefont
  {Scherer}}, \bibinfo {author} {\bibfnamefont {S.}~\bibnamefont {Diehl}}, \
  and\ \bibinfo {author} {\bibfnamefont {C.}~\bibnamefont {Wetterich}},\
  }\href@noop {} {\bibfield  {journal} {\bibinfo  {journal} {Phys. Rev. B}\
  }\textbf {\bibinfo {volume} {78}},\ \bibinfo {pages} {174528} (\bibinfo
  {year} {2008})}\BibitemShut {NoStop}%
\bibitem [{\citenamefont {Negele}\ and\ \citenamefont
  {Orland}(1988)}]{r:Negele:1988fk}%
  \BibitemOpen
  \bibfield  {author} {\bibinfo {author} {\bibfnamefont {J.~W.}\ \bibnamefont
  {Negele}}\ and\ \bibinfo {author} {\bibfnamefont {H.}~\bibnamefont
  {Orland}},\ }\href@noop {} {\emph {\bibinfo {title} {Quantum Many-Particle
  Systems}}}\ (\bibinfo  {publisher} {Addison-Wesley},\ \bibinfo {address} {New
  York, NY},\ \bibinfo {year} {1988})\BibitemShut {NoStop}%
\bibitem [{\citenamefont {Braaten}\ and\ \citenamefont
  {Nieto}(1997)}]{r:Braaten:1997uq}%
  \BibitemOpen
  \bibfield  {author} {\bibinfo {author} {\bibfnamefont {E.}~\bibnamefont
  {Braaten}}\ and\ \bibinfo {author} {\bibfnamefont {A.}~\bibnamefont
  {Nieto}},\ }\href@noop {} {\bibfield  {journal} {\bibinfo  {journal} {Phys.
  Rev. B}\ }\textbf {\bibinfo {volume} {56}},\ \bibinfo {pages} {14745}
  (\bibinfo {year} {1997})}\BibitemShut {NoStop}%
\bibitem [{\citenamefont {Rey}\ \emph {et~al.}(2004)\citenamefont {Rey},
  \citenamefont {Hu}, \citenamefont {Calzetta}, \citenamefont {Roura},\ and\
  \citenamefont {Clark}}]{r:ReyHuCalzettaRouraClark03}%
  \BibitemOpen
  \bibfield  {author} {\bibinfo {author} {\bibfnamefont {A.~M.}\ \bibnamefont
  {Rey}}, \bibinfo {author} {\bibfnamefont {B.~L.}\ \bibnamefont {Hu}},
  \bibinfo {author} {\bibfnamefont {E.}~\bibnamefont {Calzetta}}, \bibinfo
  {author} {\bibfnamefont {A.}~\bibnamefont {Roura}}, \ and\ \bibinfo {author}
  {\bibfnamefont {C.~W.}\ \bibnamefont {Clark}},\ }\href@noop {} {\bibfield
  {journal} {\bibinfo  {journal} {Phys. Rev. A}\ }\textbf {\bibinfo {volume}
  {69}},\ \bibinfo {pages} {033610} (\bibinfo {year} {2004})}\BibitemShut
  {NoStop}%
\bibitem [{\citenamefont {Gasenzer}\ \emph {et~al.}(2005)\citenamefont
  {Gasenzer}, \citenamefont {Berges}, \citenamefont {Schmidt},\ and\
  \citenamefont {Seco}}]{r:gasenzer:2005}%
  \BibitemOpen
  \bibfield  {author} {\bibinfo {author} {\bibfnamefont {T.}~\bibnamefont
  {Gasenzer}}, \bibinfo {author} {\bibfnamefont {J.}~\bibnamefont {Berges}},
  \bibinfo {author} {\bibfnamefont {M.~G.}\ \bibnamefont {Schmidt}}, \ and\
  \bibinfo {author} {\bibfnamefont {M.}~\bibnamefont {Seco}},\ }\href {\doibase
  10.1103/PhysRevA.72.063604} {\bibfield  {journal} {\bibinfo  {journal} {Phys.
  Rev. A}\ }\textbf {\bibinfo {volume} {72}},\ \bibinfo {pages} {063604}
  (\bibinfo {year} {2005})}\BibitemShut {NoStop}%
\bibitem [{\citenamefont {Temme}\ and\ \citenamefont
  {Gasenzer}(2006)}]{r:gasenzer:2006}%
  \BibitemOpen
  \bibfield  {author} {\bibinfo {author} {\bibfnamefont {K.}~\bibnamefont
  {Temme}}\ and\ \bibinfo {author} {\bibfnamefont {T.}~\bibnamefont
  {Gasenzer}},\ }\href {\doibase 10.1103/PhysRevA.74.053603} {\bibfield
  {journal} {\bibinfo  {journal} {Phys. Rev. A}\ }\textbf {\bibinfo {volume}
  {74}},\ \bibinfo {pages} {053603} (\bibinfo {year} {2006})}\BibitemShut
  {NoStop}%
\bibitem [{\citenamefont {Berges}\ and\ \citenamefont
  {Gasenzer}(2007)}]{r:gasenzer:2007}%
  \BibitemOpen
  \bibfield  {author} {\bibinfo {author} {\bibfnamefont {J.}~\bibnamefont
  {Berges}}\ and\ \bibinfo {author} {\bibfnamefont {T.}~\bibnamefont
  {Gasenzer}},\ }\href {\doibase 10.1103/PhysRevA.76.033604} {\bibfield
  {journal} {\bibinfo  {journal} {Phys. Rev. A}\ }\textbf {\bibinfo {volume}
  {76}},\ \bibinfo {pages} {033604} (\bibinfo {year} {2007})}\BibitemShut
  {NoStop}%
\bibitem [{\citenamefont {Friederich}\ \emph {et~al.}(2010)\citenamefont
  {Friederich}, \citenamefont {Krahl},\ and\ \citenamefont
  {Wetterich}}]{PhysRevB.81.235108}%
  \BibitemOpen
  \bibfield  {author} {\bibinfo {author} {\bibfnamefont {S.}~\bibnamefont
  {Friederich}}, \bibinfo {author} {\bibfnamefont {H.~C.}\ \bibnamefont
  {Krahl}}, \ and\ \bibinfo {author} {\bibfnamefont {C.}~\bibnamefont
  {Wetterich}},\ }\href {\doibase 10.1103/PhysRevB.81.235108} {\bibfield
  {journal} {\bibinfo  {journal} {Phys. Rev. B}\ }\textbf {\bibinfo {volume}
  {81}},\ \bibinfo {pages} {235108} (\bibinfo {year} {2010})}\BibitemShut
  {NoStop}%
\bibitem [{\citenamefont {Floerchinger}\ and\ \citenamefont
  {Wetterich}(2008)}]{r:Floerchinger:2008kx}%
  \BibitemOpen
  \bibfield  {author} {\bibinfo {author} {\bibfnamefont {S.}~\bibnamefont
  {Floerchinger}}\ and\ \bibinfo {author} {\bibfnamefont {C.}~\bibnamefont
  {Wetterich}},\ }\href@noop {} {\bibfield  {journal} {\bibinfo  {journal}
  {Phys. Rev. A}\ }\textbf {\bibinfo {volume} {77}},\ \bibinfo {pages} {053603}
  (\bibinfo {year} {2008})}\BibitemShut {NoStop}%
\bibitem [{\citenamefont {Calzetta}\ and\ \citenamefont
  {Hu}(2008)}]{r:Calzetta:2008pb}%
  \BibitemOpen
  \bibfield  {author} {\bibinfo {author} {\bibfnamefont {E.~A.}\ \bibnamefont
  {Calzetta}}\ and\ \bibinfo {author} {\bibfnamefont {B.-L.~B.}\ \bibnamefont
  {Hu}},\ }\href@noop {} {\emph {\bibinfo {title} {Nonequilibrium quantum field
  theory}}}\ (\bibinfo  {publisher} {Camb. U. Press},\ \bibinfo {address}
  {{Cam\-bridge, Eng\-land}},\ \bibinfo {year} {2008})\BibitemShut {NoStop}%
\bibitem [{\citenamefont {Brezin}\ and\ \citenamefont
  {Wada}(1993)}]{r:brezin:1993}%
  \BibitemOpen
  \bibinfo {editor} {\bibfnamefont {E.}~\bibnamefont {Brezin}}\ and\ \bibinfo
  {editor} {\bibfnamefont {S.~R.}\ \bibnamefont {Wada}},\ eds.,\ \href@noop {}
  {\emph {\bibinfo {title} {The large-{N} expansion in quantum field theory and
  statistical physics}}}\ (\bibinfo  {publisher} {{World Scien\-tif\-ic}},\
  \bibinfo {address} {Singapore},\ \bibinfo {year} {1993})\BibitemShut
  {NoStop}%
\bibitem [{\citenamefont {Moshe}\ and\ \citenamefont
  {Zinn-Justin}(2003)}]{r:Moshe:2003uq}%
  \BibitemOpen
  \bibfield  {author} {\bibinfo {author} {\bibfnamefont {M.}~\bibnamefont
  {Moshe}}\ and\ \bibinfo {author} {\bibfnamefont {J.}~\bibnamefont
  {Zinn-Justin}},\ }\href@noop {} {\bibfield  {journal} {\bibinfo  {journal}
  {Phys. Rept.}\ }\textbf {\bibinfo {volume} {385}},\ \bibinfo {pages} {69}
  (\bibinfo {year} {2003})}\BibitemShut {NoStop}%
\bibitem [{\citenamefont {Arnold}\ and\ \citenamefont
  {Tomasik}(2000)}]{PhysRevA.62.063604}%
  \BibitemOpen
  \bibfield  {author} {\bibinfo {author} {\bibfnamefont {P.}~\bibnamefont
  {Arnold}}\ and\ \bibinfo {author} {\bibfnamefont {B.}~\bibnamefont
  {Tomasik}},\ }\href {\doibase 10.1103/PhysRevA.62.063604} {\bibfield
  {journal} {\bibinfo  {journal} {Phys. Rev. A}\ }\textbf {\bibinfo {volume}
  {62}},\ \bibinfo {pages} {063604} (\bibinfo {year} {2000})}\BibitemShut
  {NoStop}%
\bibitem [{\citenamefont {Itzykson}\ and\ \citenamefont
  {Zuber}(1980)}]{ref:ItzyksonZuber}%
  \BibitemOpen
  \bibfield  {author} {\bibinfo {author} {\bibfnamefont {C.}~\bibnamefont
  {Itzykson}}\ and\ \bibinfo {author} {\bibfnamefont {J.-B.}\ \bibnamefont
  {Zuber}},\ }\href@noop {} {\emph {\bibinfo {title} {Quantum Field Theory}}}\
  (\bibinfo  {publisher} {{McGraw-Hill}},\ \bibinfo {address} {New York, NY},\
  \bibinfo {year} {1980})\BibitemShut {NoStop}%
\bibitem [{\citenamefont {Luttinger}\ and\ \citenamefont {Ward}(1960)}]{r:LW}%
  \BibitemOpen
  \bibfield  {author} {\bibinfo {author} {\bibfnamefont {J.~M.}\ \bibnamefont
  {Luttinger}}\ and\ \bibinfo {author} {\bibfnamefont {J.~C.}\ \bibnamefont
  {Ward}},\ }\href@noop {} {\bibfield  {journal} {\bibinfo  {journal} {Phys.
  Rev.}\ }\textbf {\bibinfo {volume} {118}},\ \bibinfo {pages} {1417} (\bibinfo
  {year} {1960})}\BibitemShut {NoStop}%
\bibitem [{\citenamefont {Baym}(1962)}]{r:Baym62}%
  \BibitemOpen
  \bibfield  {author} {\bibinfo {author} {\bibfnamefont {G.}~\bibnamefont
  {Baym}},\ }\href@noop {} {\bibfield  {journal} {\bibinfo  {journal} {Phys.
  Rev.}\ }\textbf {\bibinfo {volume} {127}},\ \bibinfo {pages} {1391} (\bibinfo
  {year} {1962})}\BibitemShut {NoStop}%
\bibitem [{\citenamefont {Cornwall}\ \emph {et~al.}(1974)\citenamefont
  {Cornwall}, \citenamefont {Jackiw},\ and\ \citenamefont {Tomboulis}}]{r:CJT}%
  \BibitemOpen
  \bibfield  {author} {\bibinfo {author} {\bibfnamefont {J.~M.}\ \bibnamefont
  {Cornwall}}, \bibinfo {author} {\bibfnamefont {R.}~\bibnamefont {Jackiw}}, \
  and\ \bibinfo {author} {\bibfnamefont {E.}~\bibnamefont {Tomboulis}},\
  }\href@noop {} {\bibfield  {journal} {\bibinfo  {journal} {Phys. Rev. D}\
  }\textbf {\bibinfo {volume} {10}},\ \bibinfo {pages} {2428} (\bibinfo {year}
  {1974})}\BibitemShut {NoStop}%
\bibitem [{\citenamefont {Papenbrock}\ and\ \citenamefont
  {Bertsch}(1999)}]{r:Papenbrock:1999fk}%
  \BibitemOpen
  \bibfield  {author} {\bibinfo {author} {\bibfnamefont {T.}~\bibnamefont
  {Papenbrock}}\ and\ \bibinfo {author} {\bibfnamefont {G.~F.}\ \bibnamefont
  {Bertsch}},\ }\href@noop {} {\bibfield  {journal} {\bibinfo  {journal} {Phys.
  Rev. C}\ }\textbf {\bibinfo {volume} {59}},\ \bibinfo {pages} {2052}
  (\bibinfo {year} {1999})}\BibitemShut {NoStop}%
\bibitem [{\citenamefont {Popov}(1983)}]{r:Popov:1983kx}%
  \BibitemOpen
  \bibfield  {author} {\bibinfo {author} {\bibfnamefont {V.~N.}\ \bibnamefont
  {Popov}},\ }\href@noop {} {\emph {\bibinfo {title} {Functional integrals in
  quantum field theory and statistical physics}}}\ (\bibinfo  {publisher}
  {Reidel},\ \bibinfo {address} {Dordrecht},\ \bibinfo {year}
  {1983})\BibitemShut {NoStop}%
\bibitem [{\citenamefont {Fetter}\ and\ \citenamefont
  {Walecka}(1971)}]{r:Fetter:1971fk}%
  \BibitemOpen
  \bibfield  {author} {\bibinfo {author} {\bibfnamefont {A.~L.}\ \bibnamefont
  {Fetter}}\ and\ \bibinfo {author} {\bibfnamefont {J.~D.}\ \bibnamefont
  {Walecka}},\ }\href@noop {} {\emph {\bibinfo {title} {Quantum theory of
  many-particle systems}}}\ (\bibinfo  {publisher} {{McGraw-Hill}},\ \bibinfo
  {address} {{New York, NY}},\ \bibinfo {year} {1971})\BibitemShut {NoStop}%
\bibitem [{\citenamefont {Arnold}\ and\ \citenamefont
  {Moore}(2001{\natexlab{a}})}]{PhysRevLett.87.120401}%
  \BibitemOpen
  \bibfield  {author} {\bibinfo {author} {\bibfnamefont {P.}~\bibnamefont
  {Arnold}}\ and\ \bibinfo {author} {\bibfnamefont {G.~D.}\ \bibnamefont
  {Moore}},\ }\href {\doibase 10.1103/PhysRevLett.87.120401} {\bibfield
  {journal} {\bibinfo  {journal} {Phys. Rev. Lett.}\ }\textbf {\bibinfo
  {volume} {87}},\ \bibinfo {pages} {120401} (\bibinfo {year}
  {2001}{\natexlab{a}})}\BibitemShut {NoStop}%
\bibitem [{\citenamefont {Arnold}\ and\ \citenamefont
  {Moore}(2001{\natexlab{b}})}]{PhysRevE.64.066113}%
  \BibitemOpen
  \bibfield  {author} {\bibinfo {author} {\bibfnamefont {P.}~\bibnamefont
  {Arnold}}\ and\ \bibinfo {author} {\bibfnamefont {G.~D.}\ \bibnamefont
  {Moore}},\ }\href {\doibase 10.1103/PhysRevE.64.066113} {\bibfield  {journal}
  {\bibinfo  {journal} {Phys. Rev. E}\ }\textbf {\bibinfo {volume} {64}},\
  \bibinfo {pages} {066113} (\bibinfo {year} {2001}{\natexlab{b}})}\BibitemShut
  {NoStop}%
\bibitem [{\citenamefont {Arnold}\ and\ \citenamefont
  {Moore}(2003)}]{PhysRevE.68.049902}%
  \BibitemOpen
  \bibfield  {author} {\bibinfo {author} {\bibfnamefont {P.}~\bibnamefont
  {Arnold}}\ and\ \bibinfo {author} {\bibfnamefont {G.~D.}\ \bibnamefont
  {Moore}},\ }\href {\doibase 10.1103/PhysRevE.68.049902} {\bibfield  {journal}
  {\bibinfo  {journal} {Phys. Rev. E}\ }\textbf {\bibinfo {volume} {68}},\
  \bibinfo {pages} {049902(E)} (\bibinfo {year} {2003})}\BibitemShut {NoStop}%
\bibitem [{\citenamefont {Kashurnikov}\ \emph {et~al.}(2001)\citenamefont
  {Kashurnikov}, \citenamefont {Prokof'ev},\ and\ \citenamefont
  {Svistunov}}]{PhysRevLett.87.120402}%
  \BibitemOpen
  \bibfield  {author} {\bibinfo {author} {\bibfnamefont {V.~A.}\ \bibnamefont
  {Kashurnikov}}, \bibinfo {author} {\bibfnamefont {N.~V.}\ \bibnamefont
  {Prokof'ev}}, \ and\ \bibinfo {author} {\bibfnamefont {B.~V.}\ \bibnamefont
  {Svistunov}},\ }\href {\doibase 10.1103/PhysRevLett.87.120402} {\bibfield
  {journal} {\bibinfo  {journal} {Phys. Rev. Lett.}\ }\textbf {\bibinfo
  {volume} {87}},\ \bibinfo {pages} {120402} (\bibinfo {year}
  {2001})}\BibitemShut {NoStop}%
\bibitem [{\citenamefont {Mihaila}\ \emph {et~al.}(2001)\citenamefont
  {Mihaila}, \citenamefont {Dawson},\ and\ \citenamefont {Cooper}}]{r:MCD01}%
  \BibitemOpen
  \bibfield  {author} {\bibinfo {author} {\bibfnamefont {B.}~\bibnamefont
  {Mihaila}}, \bibinfo {author} {\bibfnamefont {J.~F.}\ \bibnamefont {Dawson}},
  \ and\ \bibinfo {author} {\bibfnamefont {F.}~\bibnamefont {Cooper}},\
  }\href@noop {} {\bibfield  {journal} {\bibinfo  {journal} {Phys. Rev. D}\
  }\textbf {\bibinfo {volume} {63}},\ \bibinfo {pages} {096003} (\bibinfo
  {year} {2001})}\BibitemShut {NoStop}%
\bibitem [{\citenamefont {Blagoev}\ \emph {et~al.}(2001)\citenamefont
  {Blagoev}, \citenamefont {Cooper}, \citenamefont {Dawson},\ and\
  \citenamefont {Mihaila}}]{r:BCDM01}%
  \BibitemOpen
  \bibfield  {author} {\bibinfo {author} {\bibfnamefont {K.~B.}\ \bibnamefont
  {Blagoev}}, \bibinfo {author} {\bibfnamefont {F.}~\bibnamefont {Cooper}},
  \bibinfo {author} {\bibfnamefont {J.~F.}\ \bibnamefont {Dawson}}, \ and\
  \bibinfo {author} {\bibfnamefont {B.}~\bibnamefont {Mihaila}},\ }\href@noop
  {} {\bibfield  {journal} {\bibinfo  {journal} {Phys. Rev. D}\ }\textbf
  {\bibinfo {volume} {64}},\ \bibinfo {pages} {125003} (\bibinfo {year}
  {2001})}\BibitemShut {NoStop}%
\bibitem [{\citenamefont {Migdal}(1958)}]{r:Migdal:1958uq}%
  \BibitemOpen
  \bibfield  {author} {\bibinfo {author} {\bibfnamefont {A.~B.}\ \bibnamefont
  {Migdal}},\ }\href@noop {} {\bibfield  {journal} {\bibinfo  {journal} {Sov.
  Phys. JETP}\ }\textbf {\bibinfo {volume} {7}},\ \bibinfo {pages} {996}
  (\bibinfo {year} {1958})}\BibitemShut {NoStop}%
\bibitem [{\citenamefont {Cooper}\ \emph
  {et~al.}(2003{\natexlab{a}})\citenamefont {Cooper}, \citenamefont {Dawson},\
  and\ \citenamefont {Mihaila}}]{r:CDM02}%
  \BibitemOpen
  \bibfield  {author} {\bibinfo {author} {\bibfnamefont {F.}~\bibnamefont
  {Cooper}}, \bibinfo {author} {\bibfnamefont {J.~F.}\ \bibnamefont {Dawson}},
  \ and\ \bibinfo {author} {\bibfnamefont {B.}~\bibnamefont {Mihaila}},\
  }\href@noop {} {\bibfield  {journal} {\bibinfo  {journal} {Phys. Rev. D}\
  }\textbf {\bibinfo {volume} {67}},\ \bibinfo {pages} {051901R} (\bibinfo
  {year} {2003}{\natexlab{a}})}\BibitemShut {NoStop}%
\bibitem [{\citenamefont {Cooper}\ \emph
  {et~al.}(2003{\natexlab{b}})\citenamefont {Cooper}, \citenamefont {Dawson},\
  and\ \citenamefont {Mihaila}}]{r:CDM02ii}%
  \BibitemOpen
  \bibfield  {author} {\bibinfo {author} {\bibfnamefont {F.}~\bibnamefont
  {Cooper}}, \bibinfo {author} {\bibfnamefont {J.~F.}\ \bibnamefont {Dawson}},
  \ and\ \bibinfo {author} {\bibfnamefont {B.}~\bibnamefont {Mihaila}},\
  }\href@noop {} {\bibfield  {journal} {\bibinfo  {journal} {Phys. Rev. D}\
  }\textbf {\bibinfo {volume} {67}},\ \bibinfo {pages} {056003} (\bibinfo
  {year} {2003}{\natexlab{b}})}\BibitemShut {NoStop}%
\bibitem [{\citenamefont {Mihaila}(2003)}]{r:Mihaila:2003ys}%
  \BibitemOpen
  \bibfield  {author} {\bibinfo {author} {\bibfnamefont {B.}~\bibnamefont
  {Mihaila}},\ }\href@noop {} {\bibfield  {journal} {\bibinfo  {journal} {Phys.
  Rev. D}\ }\textbf {\bibinfo {volume} {68}},\ \bibinfo {pages} {036002}
  (\bibinfo {year} {2003})}\BibitemShut {NoStop}%
\bibitem [{\citenamefont {Aarts}\ and\ \citenamefont {Berges}(2001)}]{r:AB01}%
  \BibitemOpen
  \bibfield  {author} {\bibinfo {author} {\bibfnamefont {G.}~\bibnamefont
  {Aarts}}\ and\ \bibinfo {author} {\bibfnamefont {J.}~\bibnamefont {Berges}},\
  }\href@noop {} {\bibfield  {journal} {\bibinfo  {journal} {Phys. Rev. D}\
  }\textbf {\bibinfo {volume} {64}},\ \bibinfo {pages} {105010} (\bibinfo
  {year} {2001})}\BibitemShut {NoStop}%
\bibitem [{\citenamefont {Berges}(2002)}]{r:B02}%
  \BibitemOpen
  \bibfield  {author} {\bibinfo {author} {\bibfnamefont {J.}~\bibnamefont
  {Berges}},\ }\href@noop {} {\bibfield  {journal} {\bibinfo  {journal} {Nuc.
  Phys. A}\ }\textbf {\bibinfo {volume} {699}},\ \bibinfo {pages} {847}
  (\bibinfo {year} {2002})}\BibitemShut {NoStop}%
\bibitem [{\citenamefont {Aarts}\ \emph {et~al.}(2002)\citenamefont {Aarts},
  \citenamefont {Ahrensmeier}, \citenamefont {Baier}, \citenamefont {Berges},\
  and\ \citenamefont {Serreau}}]{r:AABBS}%
  \BibitemOpen
  \bibfield  {author} {\bibinfo {author} {\bibfnamefont {G.}~\bibnamefont
  {Aarts}}, \bibinfo {author} {\bibfnamefont {D.}~\bibnamefont {Ahrensmeier}},
  \bibinfo {author} {\bibfnamefont {R.}~\bibnamefont {Baier}}, \bibinfo
  {author} {\bibfnamefont {J.}~\bibnamefont {Berges}}, \ and\ \bibinfo {author}
  {\bibfnamefont {J.}~\bibnamefont {Serreau}},\ }\href@noop {} {\bibfield
  {journal} {\bibinfo  {journal} {Phys. Rev. D}\ }\textbf {\bibinfo {volume}
  {66}},\ \bibinfo {pages} {045008} (\bibinfo {year} {2002})}\BibitemShut
  {NoStop}%
\bibitem [{\citenamefont {Tikhonenkov}\ \emph {et~al.}(2007)\citenamefont
  {Tikhonenkov}, \citenamefont {Anglin},\ and\ \citenamefont
  {Vardi}}]{PhysRevA.75.013613}%
  \BibitemOpen
  \bibfield  {author} {\bibinfo {author} {\bibfnamefont {I.}~\bibnamefont
  {Tikhonenkov}}, \bibinfo {author} {\bibfnamefont {J.~R.}\ \bibnamefont
  {Anglin}}, \ and\ \bibinfo {author} {\bibfnamefont {A.}~\bibnamefont
  {Vardi}},\ }\href@noop {} {\bibfield  {journal} {\bibinfo  {journal} {Phys.
  Rev. A}\ }\textbf {\bibinfo {volume} {75}},\ \bibinfo {pages} {013613}
  (\bibinfo {year} {2007})}\BibitemShut {NoStop}%
\end{thebibliography}%

%
%%%%%%%%%%%%%%%%%%%%%%%%%%%%%%%%%%%%%%%%%%%%%%%%%%%%%%%%%%%%%%%%%%%%%%%
%
%\vfill
%
%%%%%%%%%%%%%%%%%%%%%%%%%%%%%%%%%%%%%%%%%%%%%%%%%%%%%%%%%%%%%%%%%%%%%%
%
\end{document}